\documentclass[a4paper,12pt,preprint, notitlepage,aps]{article}

\usepackage{diagbox}
\usepackage{authblk}
\usepackage{tikz}
\usepackage{amsmath,amssymb,graphicx,multirow,xspace,slashed,array,booktabs}
\usepackage[colorlinks=true,urlcolor=blue,anchorcolor=blue,citecolor=blue,filecolor=blue,linkcolor=blue,menucolor=blue]{hyperref}
\usepackage{cite}
\usepackage{subcaption}
\usepackage{booktabs}
\usepackage{blindtext, rotating}
\usepackage{afterpage}
\usepackage{enumitem}
\usepackage{ marvosym }
\usepackage{authblk} 
\usepackage{verbatim}
\usepackage{soul}
\usepackage[normalem]{ulem}
\usepackage{pifont}
\usepackage{booktabs}
\usepackage{bm}
\usepackage{cleveref}
\usepackage{siunitx}
\usepackage{tikz-feynman}
\usepackage{graphicx}
\usepackage{orcidlink}
\tikzfeynmanset{compat=1.1.1}

\newcommand{\slg}[1]{#1\hspace{-0.45em}/}

\usepackage{floatrow}
\newfloatcommand{capbtabbox}{table}[][\FBwidth]

\usepackage[font=footnotesize,labelfont=bf]{caption}

\usepackage{lineno}

\allowdisplaybreaks

\long\def\symbolfootnote[#1]#2{\begingroup%
\def\thefootnote{\fnsymbol{footnote}}\footnote[#1]{#2}\endgroup}

\allowdisplaybreaks

\makeatletter
	
	\@addtoreset{equation}{section}
\makeatother

\newcommand{\newc}{\newcommand}
\newc{\gsim}{\lower.7ex\hbox{$\;\stackrel{\textstyle>}{\sim}\;$}}
\newc{\lsim}{\lower.7ex\hbox{$\;\stackrel{\textstyle<}{\sim}\;$}}
\newc{\gev}{\,{\rm GeV}}
\newc{\mev}{\,{\rm MeV}}
\newc{\ev}{\,{\rm eV}}
\newc{\kev}{\,{\rm keV}}
\newc{\tev}{\,{\rm TeV}}
\newc{\MHT}{$H_T^{\text{miss}}$}
\newc{\MET}{$\slashed{E}_T$}
\newc{\MTT}{$M_{T2}$}

\def\ln{\mathop{\rm ln}}

\newc{\mz}{M_Z}
\newc{\mpl}{M_*}
\newc{\mw}{m_{\rm weak}}
\newc{\nr}[1]{N^c_R{}_{#1}}

\usepackage{accents}
\newlength{\dhatheight}

\def\beq{\begin{equation}}
\def\eeq{\end{equation}}
\newcommand{\bea}{\begin{eqnarray}\begin{aligned}}
\newcommand{\eea}{\end{aligned}\end{eqnarray}}
\def\bitem{\begin{itemize}}
\def\eitem{\end{itemize}}

\begin{document}

\title{Probing CP Violation in Dark Sector through \\ the Electron Electric Dipole Moment
}

\author[1,2]{\orcidlink{0000-0001-7386-0253} Jia Liu \thanks{jialiu@pku.edu.cn}}
\author[3,4]{Yuichiro Nakai \thanks{ynakai@sjtu.edu.cn}}
\author[3,4]{\\ Yoshihiro Shigekami \thanks{shigekami@sjtu.edu.cn}}
\author[2,1]{Muyuan Song \thanks{muyuansong@pku.edu.cn}}
\affil[1]{ School of Physics and State Key Laboratory of Nuclear Physics and Technology, \protect \\
Peking University, Beijing 100871, China } 
\affil[2]{Center for High Energy Physics, Peking University, Beijing 100871, China }
\affil[3]{Tsung-Dao Lee Institute, Shanghai Jiao Tong University, \protect \\
520 Shengrong Road, Shanghai 201210, China }
\affil[4]{School of Physics and Astronomy, Shanghai Jiao Tong University, \protect \\ 
800 Dongchuan Road, Shanghai 200240, China}

\date{}

\maketitle

\begin{abstract}
The Two Higgs Doublet Model (2HDM) stands as a promising framework for exploring physics beyond the Standard Model (SM). Within this context, we explore the possibility that the two Higgs doublets may serve as a window into CP-violating dark sectors, neutral under the SM gauge groups. Specifically, our focus is on investigating the electric dipole moment (EDM) of the electron, generated solely by CP violation in the dark sector. We present a general formula for the electron EDM, without specifying the structure of the dark sectors, and discuss the current constraints on various dark sector models.
It is noteworthy that even in the case of a CP-conserving 2HDM, the resulting electron EDM is capable of reaching the current experimental limit, with CP violation arising exclusively from the dark sectors.
Furthermore, we introduce a heavy dark sector (HDS) approximation for the analytic calculation of the EDM, assuming that the dark sector particles are much heavier than the physical states in the 2HDM. This approximation yields simplified analytic results that are consistent with the full numerical calculations.
\end{abstract}

\clearpage

\tableofcontents

\section{Introduction}
\label{intro}

Identifying the structure of the Higgs sector to spontaneously break the electroweak symmetry is one of the most important missions in modern particle physics. 
Despite that the discovery of the Higgs boson at the Large Hadron Collider (LHC)~\cite{ATLAS:2012yve, CMS:2012qbp} largely advances our understanding of the Higgs sector, the current experimental data still allow a lot of freedom for its structure. 
Among others, two Higgs doublet models (2HDMs)~\cite{Lee:1973iz} are classified into the simplest and viable extensions of the Standard Model (SM) Higgs sector (see $e.g.$ ref.~\cite{Branco:2011iw} for a review). 
2HDMs are motivated by various phenomenological reasons. 
One of them is supersymmetry, which requires two Higgs doublets to give masses to both up-type and down-type fermions~\cite{Inoue:1982Kenzo, fayet:1975supergauge, Gunion:1984yn, flores:1983higgs}. 
Another motivation is dark matter (DM), which can be provided by a stable neutral scalar
in some versions of 2HDMs, such as the inert doublet model~\cite{Deshpande:1977rw, LopezHonorez:2006gr, Dolle:2009fn, LopezHonorez:2010eeh, Blinov:2015qva, Krawczyk:2015xhl}. 
Therefore, the paradigm of 2HDMs has been considered one of the top candidates of physics beyond SM. 
In the present paper, we point out a new interesting feature of 2HDMs: two Higgs doublets become a window with a good view of CP-violating dark sectors that are neutral under the SM gauge groups but interact with SM particles through two Higgs doublet portal couplings. 

Exploration of dark sector models has received a lot of attention recently. 
In particular, if a dark sector contains a new source of CP violation, it may be able to address the matter-antimatter asymmetry of the Universe. 
In fact, refs.~\cite{Carena:2018cjh, Carena:2019xrr} have demonstrated a connection between CP violation in a dark sector and electroweak baryogenesis, which is responsible for the observed baryon asymmetry during the electroweak phase transition (EWPT). 
While new light particles in a dark sector are often probed by high-intensity accelerator-based experiments, a CP-violating dark sector may generate electric dipole moments (EDMs) of nucleons, atoms, and molecules that are precisely measured. 
Among others, the measurement of the electron EDM is fascinating. 
The recent results of the ACME~\cite{ACME:2018yjb} and JILA~\cite{Roussy:2022cmp, Caldwell:2022xwj} experiments have reached $\mathcal{O}(10^{-29}-10^{-30}) \, e\rm cm$. 
Since the electron EDM is immune to QCD effects and its precise calculation is possible, it seems to provide a sensitive probe of a CP-violating dark sector. 
The authors of ref.~\cite{Okawa:2019arp} have initiated to study of this issue by considering various dark sectors feebly interacting with the SM sector. 
They found that the effect of dark sector CP violation appears only from higher loops and is too tiny to be observed unless a neutrino portal coupling contains a nonzero complex phase. 
Ref.~\cite{Chen:2022vac} has extended the Higgs sector by a complex singlet coupling to a dark sector fermion. 
The authors showed that such a complex singlet extension does not generate the electron EDM up to at least the two-loop level. 
Moreover, various researchers have explored the implications of the new dark gauge symmetry coupled with the scalar, generating CP-violation at two-loop levels as well~\cite{deLima:2020lzl}. 
This implies that EDM originating from the dark sector can be either highly suppressed at higher loops or, if present at the tree level, plays a vital role in understanding the CP violation. 

In the present paper, we explore the effect of dark sector CP violation within the framework of 2HDMs and calculate molecule EDMs that constrain the electron EDM.\footnote{
Ref.~\cite{Azevedo:2018fmj} has discussed a similar setup in a different context and studied a dark sector contribution to anomalous gauge couplings.} 
A new CP-violating phase, which can be $\mathcal{O}(1)$, exists in an interaction involving dark sector particles, while the ordinary 2HDM sector does not contain CP violation. 
We find that the produced electron EDM reaches the current experimental limit in some parameter space. 
The 2HDM portal enhances the sensitivity of the electron EDM to a CP-violating dark sector. 

The rest of the paper is organized as follows. 
In section~\ref{sec:2HDMandDS}, we introduce our setup. 
Section~\ref{sec:EDMfrom2HDM} then presents a general formula for the electron EDM without specifying the structure of a dark sector. 
To discuss the current constraints numerically, we specify the dark sector structure and consider two cases of a dark complex scalar and a dark vector-like fermion in section~\ref{sec:NumChecks}. 
Section~\ref{sec:conclusion} is devoted to conclusions and discussions. 
Some calculational details are summarized in appendices.

\section{The framework}
\label{sec:2HDMandDS}

Let us start with an overview of the features of our CP-conserving 2HDMs and CP-violating dark sectors under consideration in the present paper. 
The 2HDM is a well-studied case of physics beyond the SM, involving an extra Higgs doublet scalar field. 
Here, our objective is to provide a brief description of the relevant physical modes and interactions. 
More details are summarized in appendix~\ref{app:2HDM}. 
While the current section does not specify a concrete dark sector model, dark sector CP violation will be encoded through a (radiatively generated) mixing term between CP-even and CP-odd scalars of the 2HDM.

\subsection{Two Higgs Doublet Model}
\label{sec:2HDM}

The potential of two Higgs doublets $H_{1,2}$ can be expressed as
\begin{align}
V_H &= \sum_{i = 1, 2} \Bigl[ m_i^2 H_i^{\dagger} H_i + \frac{\lambda_i}{2} \left( H_i^{\dagger} H_i \right)^2 \Bigr] + \lambda_3 \left( H_1^{\dagger} H_1 \right) \left( H_2^{\dagger} H_2 \right) \nonumber \\[0.3ex]
&\hspace{1.2em} + \lambda_4 \left( H_1^{\dagger} H_2 \right) \left( H_2^{\dagger} H_1 \right) + \left[ - m_{12}^2 H_1^{\dagger} H_2 + \frac{\lambda_5}{2} \left( H_1^{\dagger} H_2 \right)^2 + {\rm h.c.} \right] \, , \label{eq:V2HDM}
\end{align}
where $m_{12}^2$ and $\lambda_5$ are generally complex, while the other parameters are real. 
By field redefinition of either $H_1$ or $H_2$, we can set $m_{12}^2$ or $\lambda_5$ to be real, while $\arg [\lambda_5^* (m_{12}^2)^2]$ remains invariant. 
This phase is severely constrained by the electron EDM measurement, as previously studied in refs.~\cite{Egana-Ugrinovic:2018fpy, Chun:2019oix, Cheung:2020ugr, Kanemura:2020ibp, Chen:2020soj, Low:2020iua, Kanemura:2021atq, Frank:2021pkc, Kanemura:2023jbz}. 
Since our focus is on dark sector CP violation and how it is transferred into the 2HDM sector, we assume that the pure 2HDM sector is CP-conserving, $\arg [\lambda_5^* (m_{12}^2)^2] = 0$, and both $\lambda_5$ and $m_{12}^2$ are real\footnote{Even we consider the CP-conserving 2HDM, CP is not a symmetry of our study, because of the CP violation in the dark sector.}. 
The two $SU(2)_L$ doublets $H_{1,2}$ are parametrized as
\begin{align}
H_i = \begin{pmatrix}
H_i^+ \\
\left( v_i + H_i^0 + i A_i^0 \right) / \sqrt{2}
\end{pmatrix} ~~ (i = 1, 2) \, .
\label{eq:Hiparam}
\end{align}
Here, $v_{1}$ and $v_{2}$ are nonzero vacuum expectation values (VEVs) for the neutral components of $H_{1}$ and $H_{2}$, respectively. 
They are related by $v_1^2 + v_2^2 = v_{\rm SM}^2 = (246.22 \, \rm GeV)^2$. 
We define their ratio as
\begin{align}
\tan \beta \equiv \frac{v_2}{v_1} \, .
\label{eq:tanbeta}
\end{align}
Upon diagonalizing the mass matrices derived from the scalar potential in Eq.~\eqref{eq:V2HDM}, we obtain the mass eigenstates for the neutral scalars:
\begin{align}
H_1^0 &= H^0 c_{\alpha} - h^0 s_{\alpha} \, , &&H_2^0 = H^0 s_{\alpha} + h^0 c_{\alpha} \, , \label{eq:h0H0} \\[0.5ex]
A_1^0 &= G^0 c_{\beta} - A^0 s_{\beta} \, , &&A_2^0 = G^0 s_{\beta} + A^0 c_{\beta} \, , \label{eq:A0G0}
\end{align}
with mixing angle $\alpha$ whose expression is provided in Eq.~\eqref{eq:sin2alpha}. 
Here, $G^0$ represents the Nambu-Goldstone (NG) field, while $h^0$ and $H^0$ denote the two CP-even scalars, with the former being the SM Higgs, and $A^0$ represents the CP-odd heavy scalar. 

\begin{table}[!t]
\centering
\begin{tabular}{|c|c|}
\hline
Type & Yukawa terms \\ \hline
Type-I ($d_R: +$, $e_R: +$) & $- Y_u^{(2)} \overline{Q_L} \widetilde{H_2} u_R - Y_d^{(2)} \overline{Q_L} H_2 d_R - Y_e^{(2)} \overline{L_L} H_2 e_R + {\rm h.c.}$ \\
Type-II ($d_R: -$, $e_R: -$) & $- Y_u^{(2)} \overline{Q_L} \widetilde{H_2} u_R - Y_d^{(1)} \overline{Q_L} H_1 d_R - Y_e^{(1)} \overline{L_L} H_1 e_R + {\rm h.c.}$ \\
Type-X ($d_R: +$, $e_R: -$) & $- Y_u^{(2)} \overline{Q_L} \widetilde{H_2} u_R - Y_d^{(2)} \overline{Q_L} H_2 d_R - Y_e^{(1)} \overline{L_L} H_1 e_R + {\rm h.c.}$ \\
Type-Y ($d_R: -$, $e_R: +$) & $- Y_u^{(2)} \overline{Q_L} \widetilde{H_2} u_R - Y_d^{(1)} \overline{Q_L} H_1 d_R - Y_e^{(2)} \overline{L_L} H_2 e_R + {\rm h.c.}$ \\ \hline
\end{tabular}
\vspace{0.1cm}
\caption{The four types of Yukawa couplings in the 2HDM, showing the $Z_2$ charges of $d_R$ and $e_R$ along with the type names in the first column and the allowed Yukawa couplings in the second column. 
Note that $Q_L$ and $L_L$ are assumed to be even under the $Z_2$.}
\label{tab:YukawaTypes}
\end{table}

The Yukawa couplings for SM fermions in the 2HDM are represented by the following Lagrangian term:
\begin{align}
\mathcal{L}_{\rm Yuk} = - Y_u^{(i)} \overline{Q_L} \widetilde{H}_i u_R - Y_d^{(i)} \overline{Q_L} H_i d_R - Y_e^{(i)} \overline{L_L} H_i e_R + {\rm h.c.} \, ,
\end{align}
where $\widetilde{H}_i = i \sigma_2 H_i^*$, $Q_L$ and $L_L$ denote the left-handed $SU(2)_L$ doublet quark and lepton, and $u_R$, $d_R$ and $e_R$ are the right-handed $SU(2)_L$ singlet up-type quark, down-type quark and charged lepton. 
In general, there are Yukawa couplings associated with both $H_1$ and $H_2$ for each SM fermion. 
However, including both Yukawa couplings easily leads to dangerous flavor-changing neutral current (FCNC) processes. 
To avoid such a situation, one can assume different symmetry properties for $H_1$ and $H_2$~\cite{Glashow:1976nt}. 
For example, $H_1$ behaves as odd while $H_2$ as even under a discrete $Z_2$ symmetry.\footnote{
Under the $Z_2$ symmetry, the scalar potential term $m_{12}^2 H_1^{\dagger} H_2$ in Eq.~\eqref{eq:V2HDM} can be considered as a soft $Z_2$ breaking term.} 
Without loss of generality, one can select a scenario in which $\overline{Q_L} u_R$ couples exclusively to $H_2$. 
Then, there are four types of Yukawa coupling arrangements, namely Type-I, II, X, and Y, depending on the charges of $d_R$ and $e_R$, which are summarized in Table~\ref{tab:YukawaTypes}. 
Due to the mixings in Eqs.~\eqref{eq:h0H0} and \eqref{eq:A0G0}, the Yukawa interactions between the SM fermions and the physical neutral scalars are given by
\begin{align}
\mathcal{L}_{\rm Yuk} \supset - \sum_{f = u, d, e} \frac{m_f}{v_{\rm SM}} \left( \xi_h^f \bar{f} f h^0 + \xi_H^f \bar{f} f H^0 - \xi_A^f \bar{f} i \gamma^5 f A^0 \right) \, ,
\label{eq:SMYukawain2HDM}
\end{align}
where $m_f$ is the SM fermion mass, and the coefficients $\xi_{h, H, A}^f$ are summarized in Table~\ref{tab:xif}. 
It is notable that the lepton couplings to the heavy scalars are significantly enhanced by a large $\tan \beta$ for Type-II and Type-X 2HDMs, leading to a larger effect on the electron EDM. 

\begin{table}[!t]
\centering
\begin{tabular}{|c||ccc|ccc|ccc|}
\hline
 & $\xi_h^u$ & $\xi_H^u$ & $\xi_A^u$ & $\xi_h^d$ & $\xi_H^d$ & $\xi_A^d$ & $\xi_h^e$ & $\xi_H^e$ & $\xi_A^e$ \\ \hline
Type-I & $c_{\alpha}/s_{\beta}$  & $s_{\alpha}/s_{\beta}$ & $\cot \beta$ & $c_{\alpha}/s_{\beta}$  & $s_{\alpha}/s_{\beta}$ & $- \cot \beta$ & $c_{\alpha}/s_{\beta}$  & $s_{\alpha}/s_{\beta}$ & $- \cot \beta$ \\
Type-II & $c_{\alpha}/s_{\beta}$  & $s_{\alpha}/s_{\beta}$ & $\cot \beta$ & $- s_{\alpha}/c_{\beta}$  & $c_{\alpha}/c_{\beta}$ & $\tan \beta$ & $- s_{\alpha}/c_{\beta}$  & $c_{\alpha}/c_{\beta}$ & $\tan \beta$ \\
Type-X & $c_{\alpha}/s_{\beta}$  & $s_{\alpha}/s_{\beta}$ & $\cot \beta$ & $c_{\alpha}/s_{\beta}$  & $s_{\alpha}/s_{\beta}$ & $- \cot \beta$ & $- s_{\alpha}/c_{\beta}$  & $c_{\alpha}/c_{\beta}$ & $\tan \beta$ \\
Type-Y & $c_{\alpha}/s_{\beta}$  & $s_{\alpha}/s_{\beta}$ & $\cot \beta$ & $- s_{\alpha}/c_{\beta}$  & $c_{\alpha}/c_{\beta}$ & $\tan \beta$ & $c_{\alpha}/s_{\beta}$  & $s_{\alpha}/s_{\beta}$ & $- \cot \beta$ \\ \hline
\end{tabular}
\vspace{0.1cm}
\caption{The coefficients between SM fermions and physical neutral scalars.}
\label{tab:xif}
\end{table}

Due to the additional Higgs doublet, the $W$ boson couplings to CP-even scalars are modified as
\begin{align}
\mathcal{L} \supset \frac{2 m_W^2}{v_{\rm SM}} s_{\beta - \alpha} W_{\mu}^+ W^{- \, \mu} h^0 + \frac{2 m_W^2}{v_{\rm SM}} c_{\beta - \alpha} W_{\mu}^+ W^{- \, \mu} H^0 \, . \label{eq:WWhandWWHin2HDM}
\end{align}
Note that if the model does not mix between CP-even and CP-odd scalars, there is no $W$ boson coupling to the CP-odd scalar, $A^0$~\cite{Djouadi:2005gj}.

\subsection{2HDM portal to the dark sector}
\label{sec:DS}

To couple the 2HDM to a dark sector, which is neutral under the SM gauge symmetry, we make the following assumptions:
\begin{itemize}
\item[(i)] The dark sector incorporates physical CP phase(s), and it does not contain any new scalar field that acquires a nonzero VEV. 

\item[(ii)] New scalar potential terms that couple $H_{1, 2}$ to the dark sector are expressed as
\begin{align}
V_{H \mathcal{O}} = \lambda_{ij} H_i^{\dagger} H_j \mathcal{O}_{ij}^{D} + {\rm h.c.} \, , \label{eq:Vmix}
\end{align}
where $\mathcal{O}^D_{ij}$ represents operators composed of dark sector fields. 
\end{itemize}
The assumption (i) implies the absence of tree-level mass mixing between CP-even and CP-odd scalars in $H_{1,2}$. 
Loop corrections through the interactions~\eqref{eq:Vmix} generate effective mixing terms for those scalars:
\begin{equation}
\begin{tikzpicture}
\begin{feynman}[large]
\vertex (a1) {\({h^0, H^0}\)};
\vertex[blob] [right=2.8cm of a1] (b1) {};
\vertex [right=2.5cm of b1] (c1) {\(A^0\)};
\diagram [medium] {
(a1) -- [scalar, momentum=\(k\)] (b1) -- [scalar] (c1)
};
\end{feynman}
\end{tikzpicture} ,
\label{eq:effmix}
\end{equation}
where the gray shaded blob represents loops of dark sector particles. 
Nonzero mass mixing between CP-even ($h^0, ~H^0$) and CP-odd ($A^0$) scalars in the 2HDM is induced by dark sector CP violation. 

We have comment on a renormalization group (RG) running of couplings in 2HDM sector. 
When we consider a UV complete model that incorporates our dark sector, the RG running induces a phase for $\lambda_5$ as well as $m_{12}^2$, due to the coupling between 2HDM and dark sectors in Eq.~\eqref{eq:Vmix}. 
In addition, there is the possibility of generating terms such as $\lambda_6 (H_1^{\dagger} H_1) (H_1^{\dagger} H_2) + \lambda_7 (H_2^{\dagger} H_2) (H_1^{\dagger} H_2) + {\rm h.c.}$ through the RG running, which have complex parameters of $\lambda_{6, 7}$. 
These terms and $\lambda_5$ are additional sources of CP violation, inducing a CP-violating nature in our 2HDM sector where CP-even and CP-odd scalars mix. 
As mentioned in section~\ref{sec:2HDM}, a CP-violating 2HDM is severely constrained by the bound on $d_e$, and the current focus is electron EDM prediction from dark sector CP violation. 
Therefore, we specifically consider a scenario in which the phases of $\lambda_{5, 6, 7}$ induced by the RG running are sufficiently small. 
This ensures that contributions from these CP phases to the electron EDM are sub-dominant, allowing our EDM prediction to be predominantly influenced by CP violation in the dark sector. 
The details of effect of this RG running will be performed in the future work.

\section{EDMs from general dark sector}
\label{sec:EDMfrom2HDM}

We now summarize the current experimental limits on molecule EDMs which constrain the electron EDM, and derive analytical expressions for dark sector contributions to the EDMs without specifying the detailed structure of the dark sector. 
As mentioned in the previous section, the mixing between CP-even and CP-odd scalars is induced via loops of CP-violating dark sector particles.

\subsection{Current limits on EDMs}

Measuring EDMs of leptons, nucleons, atoms, and molecules stands as one of the most sensitive probes of new physics with CP violation, and numerous experiments set upper limits on their magnitudes~\cite{ACME:2018yjb, Roussy:2022cmp, Muong-2:2008ebm, Abel:2020pzs}, which put stringent constraints on models incorporating CP violation, such as the complex 2HDM and supersymmetry (see e.g. refs.~\cite{Nakai:2016atk, Cesarotti:2018huy}). 
For an in-depth understanding of EDMs and their measurement techniques, along with theoretical aspects, one can refer to comprehensive reviews such as refs.~\cite{Pospelov:2005pr, Engel:2013lsa, Chupp:2017rkp}. 

One intriguing approach in EDM experiments is to utilize paramagnetic atoms or molecules. 
A particularly stringent constraint on the electron EDM from this type of experiment was reported by the ACME collaboration before 2022~\cite{ACME:2018yjb}. 
They conducted a measurement on the EDM of the paramagnetic ThO molecule, establishing an upper bound of $| d_{\rm ThO} | < 1.1 \times 10^{-29} \, e \, {\rm cm}$ at $90\%$ confidence level (CL). 
For paramagnetic molecules, the EDM measurement offers sensitivity not only to the electron EDM $d_e$ but also to the CP-odd electron-nucleon interaction $C_S$. 
The relevant dimension-five and dimension-six operators are described by the following Lagrangian:
\begin{align}
\mathcal{L}_{{\rm EDM}, eN} \supset - \frac{i}{2} d_e \bar{e} \sigma^{\mu \nu} \gamma^5 e F_{\mu \nu} - C_S \bar{e} i \gamma^5 e \bar{N} N \, , \label{eq:LagEDMs}
\end{align}
where $\sigma^{\mu \nu} = \frac{i}{2} [\gamma^{\mu}, \gamma^{\nu}]$, and $N$ represents the nucleon. 
These CP-violating interactions contribute to the EDM of the paramagnetic molecule, and the bound from the ACME collaboration is expressed as~\cite{ACME:2018yjb}
\begin{align}
\left| d_e + k_{\rm ThO} C_S \right| < 1.1 \times 10^{-29} \, e \, {\rm cm} \, , \label{eq:deACME}
\end{align}
with $k_{\rm ThO} \simeq 1.8 \times 10^{-15} \, {\rm GeV}^2 e \, {\rm cm}$, where we have used $\mathcal{E}_{\rm eff}^{\rm ThO} \approx 78 \, {\rm GV cm}^{-1}$ for the effective electric field and $W_S^{\rm ThO} = - 282 \, {\rm kHz} \times h$ for the molecule specific constant. 

Towards the end of 2022, the JILA experiments presented a new upper bound on the electron EDM, utilizing the HfF$^+$ molecule ion~\cite{Roussy:2022cmp, Caldwell:2022xwj}. 
The result is expressed as $| d_{{\rm HfF}^+} | < 4.1 \times 10^{-30} \, e \, {\rm cm}$ at 90\% CL, which can be interpreted as
\begin{align}
\left| d_e + k_{{\rm HfF}^+} C_S \right| < 4.1 \times 10^{-30} \, e \, {\rm cm} \, , \label{eq:deJILA}
\end{align}
with $k_{{\rm HfF}^+} \simeq 1.1 \times 10^{-15} \, {\rm GeV}^2 e \, {\rm cm}$, where we have used $\mathcal{E}_{\rm eff}^{{\rm HfF}^+} \approx 23 \, {\rm GV cm}^{-1}$ for the effective electric field and $W_S^{{\rm HfF}^+} = - 51 \, {\rm kHz} \times h$ for the molecule specific constant. 

In addition to EDMs of paramagnetic molecules, there are experiments to measure EDMs of diamagnetic atoms, such as the mercury EDM~\cite{Graner:2016ses} and the neutron EDM~\cite{Abel:2020pzs}. 
These experiments, in principle, can impose constraints on the 2HDM or supersymmetry, and the details can be found in refs.~\cite{Ellis:2008zy, Jung:2013hka}. 
However, upon careful examination, we have verified that the electron EDM constraint in the JILA experiment is considerably stronger than the bounds stemming from the other EDM experiments. 
Consequently, we focus on the constraint obtained by the JILA experiment in the following discussion.

\subsection{The CP-violating diagrams}
\label{sec:CPVdiagrams}

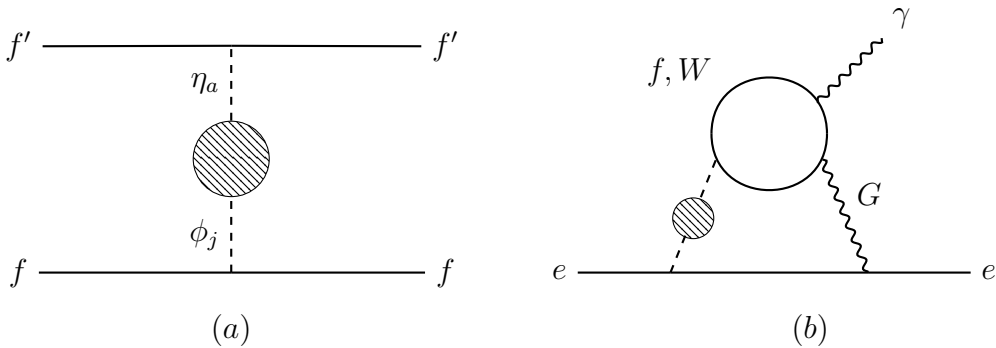
\begin{figure}[t]
\begin{center}
\begin{tikzpicture}
\begin{feynman}[large]
\vertex (a1) {\(f\)};
\vertex [right=2.8cm of a1] (b1);
\vertex [right=2.55cm of b1] (c1) {\(f\)};
\vertex[blob] [above=1.0cm of b1] (o1) {};
\vertex [above=1.5cm of o1] (e1);
\vertex [above=3.0cm of a1] (d1) {\(f'\)};
\vertex [above=3.0cm of c1] (f1) {\(f'\)};
\vertex [below=0.4cm of b1] (h1) {\((a)\)};
\diagram [medium] {
(a1) -- (b1) -- (c1),
(b1) -- [scalar, edge label=\(\phi_j\)] (o1) -- [scalar, edge label=\(\eta_a\)] (e1),
(d1) -- (e1) -- (f1),
};
\vertex [right=1.5cm of c1] (a3) {\(e\)};
\vertex [right=1.46cm of a3] (b3);
\vertex [right=2.6cm of b3] (c3);
\vertex [right=1.35cm of c3] (d3) {\(e\)};
\vertex [above=1.5cm of b3] (bc1);
\vertex [right=0.6cm of bc1] (e3);
\vertex [above=1.0cm of b3] (bc2);
\vertex [right=0.4cm of bc2] (s1);
\vertex [above=0.5cm of b3] (bc3);
\vertex [right=0.2cm of bc3] (s2);
\vertex [above=0.72cm of b3] (bl);
\vertex [right=0.3cm of bl] (blb);
\vertex [above=1.5cm of c3] (cc);
\vertex [left=0.60cm of cc] (f3);
\vertex [above=1.84cm of b3] (lc1);
\vertex [above=1.84cm of c3] (lc2);
\vertex [right=0.54cm of lc1] (lc3);
\vertex [left=0.54cm of lc2] (lc4);
\vertex [above=0.75cm of f3] (pp);
\vertex [left=0.07cm of pp] (p3);
\vertex [above right=1.2cm of p3] (q3) {\(\gamma\)};
\vertex [right=3.3cm of a3] (cc1);
\vertex [below=0.4cm of cc1] (i3) {\((b)\)};
\diagram [medium] {
(a3) -- (b3) -- (c3) -- (d3),
(b3) -- [scalar] (s2),
(s1) -- [scalar] (e3),
(c3) -- [photon, edge label'=\(G\)] (f3),
(lc3) -- [half left, looseness=1.68, pos=0.2, edge label=\({f, W}\)] (lc4),
(lc3) -- [half right, looseness=1.68] (lc4),
(p3) -- [photon] (q3)
};
\draw [pattern=north west lines, pattern color=black, draw=black] (blb) circle [radius=0.27];
\end{feynman}
\end{tikzpicture}
\end{center}
\vspace{-0.5cm}
\caption{The two diagrams relevant to the CP-violating Lagrangian in Eq.~\eqref{eq:LagEDMs}. 
\textit{Left}: The four-fermion operators mediated by the scalars. 
\textit{Right}: The Barr-Zee diagram contributing to the electron EDM, with $G$ denoting the photon or $Z$ boson. 
In both diagrams, the scalar lines represent the CP-even scalars $\phi_j$ or CP-odd scalars $\eta_a$, and the gray shaded blob symbolizes the effective mixing between $\phi_j$ and $\eta_a$ arising from the CP-violating dark sector contribution, which corresponds to $F_{\rm DS}^{\phi_j \eta_a} (k^2, M^2)$ in Eq.~\eqref{eq:LeffFDS}.}
\label{fig:EDMdiagrams}
\end{figure}

Let us now consider dark sector contributions to the CP-violating Lagrangian in Eq.~\eqref{eq:LagEDMs}. 
The relevant contributions are shown in Fig.~\ref{fig:EDMdiagrams}. 
The left diagram leads to the four-fermion operator $C_S \bar{e} i \gamma^5 e \bar{N} N$, while the right corresponds to the Barr-Zee (BZ) diagram that results in the electron EDM operator $d_e \bar{e} \sigma^{\mu \nu} \gamma^5 e F_{\mu \nu}$~\cite{BarrZee:1990}. 
We can also consider one-loop diagrams induced by loops of neutral scalars, but when compared with the diagrams $(a)$, $(b)$ in Fig.~\ref{fig:EDMdiagrams}, these one-loop diagrams are suppressed by the cubic power of the electron mass $m_e^3$: one arises from a chirality flip on the electron line and the others stem from electron-neutral scalar couplings. 
Therefore, for the current study, we can neglect these one-loop diagrams. 

We keep our calculation as general as possible by incorporating a general CP-violating dark sector contribution into the two diagrams of Fig.~\ref{fig:EDMdiagrams} as propagator corrections. 
Moreover, we consider a general scalar model with $N_E$ CP-even and $N_O$ CP-odd neutral scalars contributing to the diagrams, denoted as $\phi_j$ and $\eta_a$, respectively. 
For example, the 2HDM can be identified with $(N_E, N_O) = (2, 2)$ and $(\phi_1, \phi_2, \eta_1, \eta_2) = (h^0, H^0, G^0, A^0)$. 
Note that we assume $\eta_1$ always represents the NG field $G^0$, and thus, the sum of the index $a$ should start from $2$ for $\eta_a$ scalars. 
$\phi_j$ and $\eta_a$ are considered as mass eigenstates, and hence, masses for these scalars are described as
\begin{align}
V_{\rm scl} \supset \sum_{j = 1}^{N_E} m_{\phi_j}^2 \phi_j^2 + \sum_{a = 2}^{N_O} m_{\eta_a}^2 \eta_a^2 \, .
\end{align}
Here, we assume $m_{\phi_1}^2 < m_{\phi_2}^2 < \cdots < m_{\phi_{N_E}}^2$ and $m_{\eta_2}^2 < m_{\eta_3}^2 < \cdots < m_{\eta_{N_O}}^2$. 

The neutral scalars $\phi_j, \eta_a$ couple to the SM fermions through the Yukawa terms in the 2HDM. 
In addition, CP-even scalars also couple to the $W$ boson. 
We then have the following Lagrangian terms:
\begin{align}
\mathcal{L} &\supset - \sum_{j = 1}^{N_E} \sum_{a = 2}^{N_O} \sum_{f = u, d, e} \frac{m_f}{v_{\rm SM}} \left( \xi_{\phi_j}^f \bar{f} f \phi_j - \xi_{\eta_a}^f \bar{f} i \gamma^5 f \eta_a \right) + \sum_{j = 1}^{N_E} \frac{2 m_W^2}{v_{\rm SM}} \xi_{\phi_j}^W W_{\mu}^+ W^{- \, \mu} \phi_j \, , \label{eq:YukawaandWGenaral}
\end{align}
where $\xi_{\phi_j, \eta_a}^f$ and $\xi_{\phi_j}^W$ are related to $\tan \beta$ and mixing angles for $\phi_j$ and $\eta_a$ in the visible sector (for the 2HDM, see Table~\ref{tab:xif} and Eq.~\eqref{eq:WWhandWWHin2HDM}). 
Note that there is no coupling between a CP-odd scalar and the $W$ boson because our assumption forbids any tree-level mixing between CP-even and CP-odd scalars in the visible sector. 
Furthermore, since we do not extend the fermion and gauge sectors, couplings among these sectors are unchanged from the SM ones. 
Finally, we have the effective mixing between $\phi_j$ and $\eta_a$, which is generated in the same manner as in Eq.~\eqref{eq:effmix}. 
In our analytical calculation, we define its mixing as the following term in the effective Lagrangian:
\begin{align}
\mathcal{L}_{\rm eff} \supset F_{\rm DS}^{\phi_j \eta_a} (k^2, M^2) \phi_j \eta_a \, ,
\label{eq:LeffFDS}
\end{align}
where $k$ is four-momentum passing through this effective mixing, and $M$ is a typical mass scale of a dark sector. 
The gray shaded blobs in Fig.~\ref{fig:EDMdiagrams}
correspond to $F_{\rm DS}^{\phi_j \eta_a} (k^2, M^2)$.

\subsection{The four-fermion diagram}
\label{sec:diagram(a)}

In contrast to the nucleon operator in Eq.~\eqref{eq:LagEDMs}, we introduce a new four-fermion operator at the quark level,
\begin{align}
\mathcal{L}_{4 \mathchar`- {\rm fermi}} = C_{f f'} \left( \bar{f} f \right) \left( \bar{f'} i \gamma^5 f' \right) \, . \label{eq:Lag4fermi}
\end{align}
The coefficient $C_{f f'}$ stemming from the diagram $(a)$ can be readily calculated. 
Moreover, we can set the momentum of the scalar to be zero for the current purpose, i.e., $F_{\rm DS}^{\phi_j \eta_a} (k^2, M^2) = F_{\rm DS}^{\phi_j \eta_a} (0, M^2)$. 
Consequently, we obtain
\begin{align}
C_{f f'} = \frac{m_f m_{f'}}{v_{\rm SM}^2} \sum_{j = 1}^{N_E} \sum_{a = 2}^{N_O} \frac{\xi_{\phi_j}^f}{m_{\phi_j}^2} \frac{\xi_{\eta_a}^{f'}}{m_{\eta_a}^2} F_{\rm DS}^{\phi_j \eta_a} (0, M^2) \, , \label{eq:CffpGen}
\end{align}
where $\xi_{\phi_j}^f$ and $\xi_{\eta_a}^{f'}$ are defined in Eq.~\eqref{eq:YukawaandWGenaral}. 

The $C_{f f'}$ term contributes to $C_S$ in Eq.~\eqref{eq:LagEDMs}. 
The coefficient of the electron-quark interaction $C_{q e}$ leads to
\begin{align}
C_S \approx \sum_q C_{q e} \langle N | \bar{q} q | N \rangle &\approx C_{u e} \frac{16 \, {\rm MeV}}{m_u} + C_{d e} \frac{29 \, {\rm MeV}}{m_d} + C_{s e} \frac{49 \, {\rm MeV}}{m_s} \nonumber \\[0.3ex]
&\hspace{1.2em} + C_{c e} \frac{76 \, {\rm MeV}}{m_c} + C_{b e} \frac{74 \, {\rm MeV}}{m_b} + C_{t e} \frac{77 \, {\rm MeV}}{m_t} \, ,
\label{eq:CScontribution}
\end{align}
where numerators arise from the matrix elements given in refs.~\cite{Ellis:2008zy, Junnarkar:2013ac},
\begin{align}
&(m_u + m_d) \langle N | \bar{u} u + \bar{d} d | N \rangle \simeq 90 \, {\rm MeV} \, , \quad \langle N | \bar{u} u - \bar{d} d | N \rangle \simeq 0 \, , \quad m_s \langle N | \bar{s} s | N \rangle \simeq 49 \, {\rm MeV} \, , \nonumber \\[1ex]
&m_c \langle N | \bar{c} c | N \rangle \simeq 76 \, {\rm MeV} \, , \quad m_b \langle N | \bar{b} b | N \rangle \simeq 74 \, {\rm MeV} \, , \quad m_t \langle N | \bar{t} t | N \rangle \simeq 77 \, {\rm MeV} \, , 
\end{align}
with $m_u / m_d \simeq 0.55$.

\subsection{The Barr-Zee diagrams}
\label{sec:diagram(b)}

To calculate the BZ diagram in Fig.~\ref{fig:EDMdiagrams}, it is important to handle the momentum in $F_{\rm DS}^{\phi_j \eta_a}$ with care, as it is integrated as the outer loop momentum in the BZ diagram. 
Additionally, since we have not specified our dark sector at this stage, we cannot have any explicit formula for $F_{\rm DS}^{\phi_j \eta_a}$. 
Therefore, we present the general result before the loop momentum integration. 
Before the calculation, we define the relevant couplings for SM fermions as
\begin{align}
\mathcal{L} \supset \bar{f} \left( g_{f \Phi}^S + i g_{f \Phi}^P \gamma^5 \right) f \Phi + \bar{f} \left( g_{f G}^V \gamma^{\mu} + g_{f G}^A \gamma^{\mu} \gamma^5 \right) f G_{\mu} \, ,
\label{eq:fermioncouplings}
\end{align}
where $\Phi$ and $G_{\mu}$ denote the neutral scalars ($\phi_j, \eta_a$) and neutral gauge bosons ($Z_{\mu}, A_{\mu}$), respectively. 
Comparing with Eq.~\eqref{eq:YukawaandWGenaral}, $g_{f \Phi}^{S, P}$ are related to $\xi_{\phi_j, \eta_a}^f$, and $g_{f G}^{V, A}$ are the usual gauge boson couplings in the SM, as summarized in Table~\ref{tab:fermioncouplings}. 

\begin{table}[t]
\centering
\begin{tabular}{|c|cc||c|cc|}
\hline
 & $\Phi = \phi_j$ & $\Phi = \eta_a$ &  & $G = Z$ & $G = \gamma$ \\ \hline
$g_{f \Phi}^S$ & ${\displaystyle - \frac{m_f}{v_{\rm SM}} \xi_{\phi_j}^f}$ & $0$ & $g_{e G}^V$ & ${\displaystyle \frac{g}{\cos \theta_W} \left( \frac{1}{2} T_3^f - Q_f \sin^2 \theta_W \right)}$ & ${\rm e} Q_f$ \\[2.0ex]
$g_{f \Phi}^P$ & $0$ & ${\displaystyle + \frac{m_f}{v_{\rm SM}} \xi_{\eta_a}^f}$ & $g_{e G}^A$ & ${\displaystyle - \frac{g}{2 \cos \theta_W} T_3^f}$ & $0$ \\[1.0ex] \hline 
\end{tabular}
\caption{Couplings defined in Eq.~\eqref{eq:fermioncouplings}. 
Here, $g$ denotes the gauge coupling of $SU(2)_L$, $\theta_W$ is the weak mixing angle, ${\rm e} = g \sin \theta_W$ is the electromagnetic coupling, $T_3^f = +1/2 \, (-1/2)$ for $f = u \, (d, e)$, and $Q_f$ is the electric charge of $f$.}
\label{tab:fermioncouplings}
\end{table}

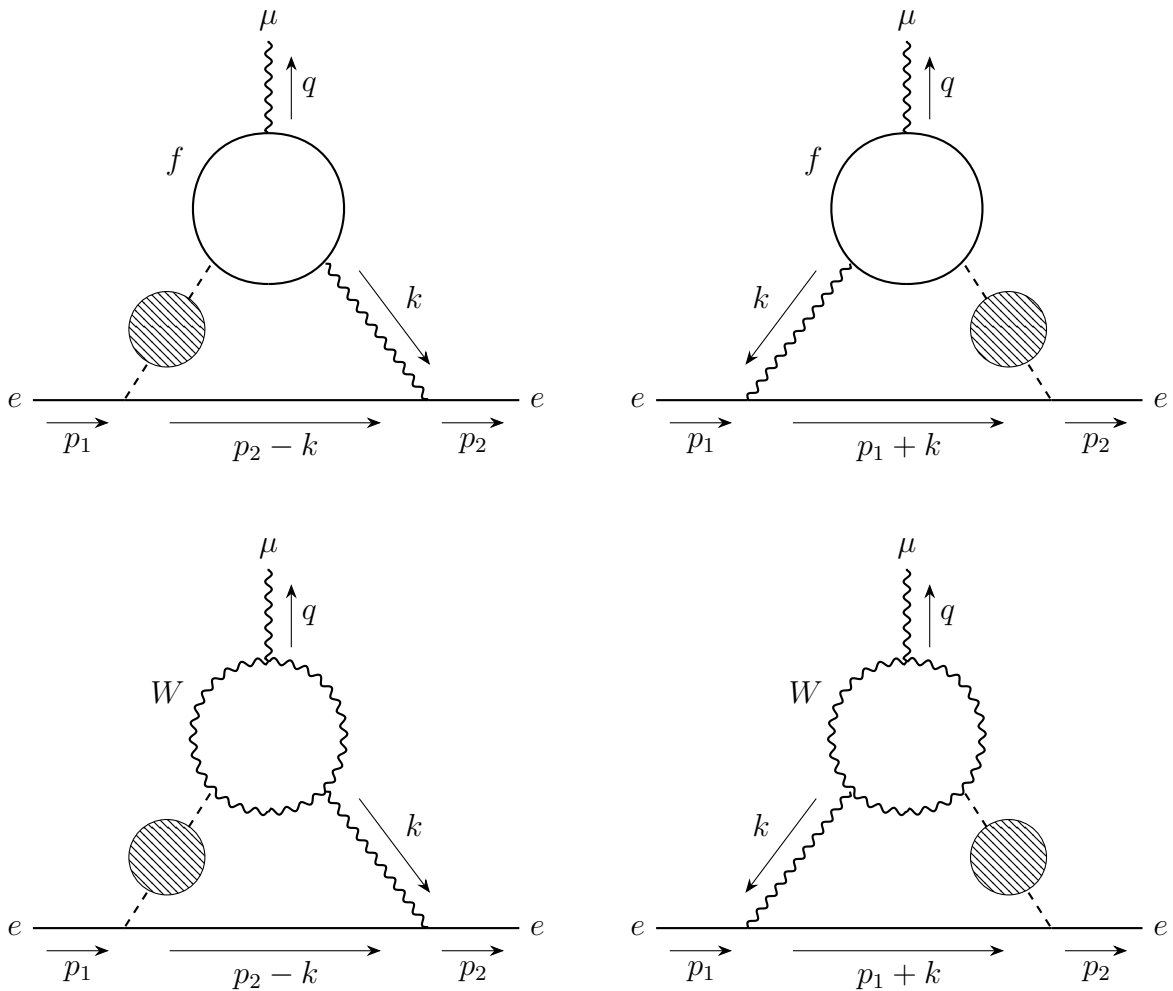
\begin{figure}[t]
\begin{center}
\begin{tikzpicture}
\begin{feynman}[large]
\vertex (a1fL) {\(\mu\)};
\vertex [below=1.5cm of a1fL] (l1fL);
\vertex [below=2.0cm of l1fL] (l2fL);
\vertex [above=0.27cm of l2fL] (t1fL);
\vertex [left=0.75cm of t1fL] (l3fL);
\vertex [right=0.75cm of t1fL] (l4fL);
\vertex [below=0.87cm of l3fL] (tfL);
\vertex [left=0.3cm of tfL] (tl1fL);
\vertex [above=0.4cm of tl1fL] (a2fL);
\vertex [below=0.47cm of tfL] (tl2fL);
\vertex [left=0.85cm of tl2fL] (a4fL);
\vertex [below=0.47cm of a4fL] (tl3fL);
\vertex [left=0.3cm of tl3fL] (a5fL);
\vertex [left=1.2cm of a5fL] (el1fL) {\(e\)};
\vertex [right=4cm of a5fL] (a3fL);
\vertex [right=1.2cm of a3fL] (er1fL) {\(e\)};
\vertex[blob] [left=0.08cm of tfL] (bl1fL) {};
\diagram [medium] {
(a1fL) -- [photon, rmomentum=\(q\)] (l1fL),
(a2fL) -- [scalar] (l3fL),
(a4fL) -- [scalar] (a5fL),
(a3fL) -- [photon, rmomentum'=\(k\)] (l4fL),
(l1fL) -- [half left, looseness=1.7] (l2fL) -- [half left, looseness=1.7, edge label=\(f\), pos=0.6] (l1fL),
(el1fL) -- [momentum'=\(p_1\)] (a5fL) -- [momentum'=\(p_2 - k\)] (a3fL) -- [momentum'=\(p_2\)] (er1fL)
};
\vertex [right=8.4cm of a1fL] (b1fR) {\(\mu\)};
\vertex [below=1.5cm of b1fR] (l5fR);
\vertex [below=2.0cm of l5fR] (l6fR);
\vertex [above=0.27cm of l6fR] (t2fR);
\vertex [left=0.75cm of t2fR] (l7fR);
\vertex [right=0.75cm of t2fR] (l8fR);
\vertex [below=0.87cm of l8fR] (tfR);
\vertex [right=0.3cm of tfR] (tr1fR);
\vertex [above=0.4cm of tr1fR] (b2fR);
\vertex [below=0.47cm of tfR] (tr2fR);
\vertex [right=0.85cm of tr2fR] (b4fR);
\vertex [below=0.47cm of b4fR] (tl3fR);
\vertex [right=0.3cm of tl3fR] (b5fR);
\vertex [right=1.2cm of b5fR] (er2fR) {\(e\)};
\vertex [left=4cm of b5fR] (b3fR);
\vertex [left=1.2cm of b3fR] (el2fR) {\(e\)};
\vertex[blob] [right=0.08cm of tfR] (br1fR) {};
\diagram [medium] {
(b1fR) -- [photon, rmomentum=\(q\)] (l5fR),
(b4fR) -- [scalar] (b5fR),
(b3fR) -- [photon, rmomentum=\(k\)] (l7fR),
(b2fR) -- [scalar] (l8fR),
(l5fR) -- [half left, looseness=1.7] (l6fR) -- [half left, looseness=1.7, edge label=\(f\), pos=0.6] (l5fR),
(el2fR) -- [momentum'=\(p_1\)] (b3fR) -- [momentum'=\(p_1 + k\)] (b5fR) -- [momentum'=\(p_2\)] (er2fR)
};
\vertex [below=7.0cm of a1fL] (a1WL) {\(\mu\)};
\vertex [below=1.5cm of a1WL] (l1WL);
\vertex [below=2.0cm of l1WL] (l2WL);
\vertex [above=0.27cm of l2WL] (t1WL);
\vertex [left=0.75cm of t1WL] (l3WL);
\vertex [right=0.75cm of t1WL] (l4WL);
\vertex [below=0.87cm of l3WL] (tWL);
\vertex [left=0.3cm of tWL] (tl1WL);
\vertex [above=0.4cm of tl1WL] (a2WL);
\vertex [below=0.47cm of tWL] (tl2WL);
\vertex [left=0.85cm of tl2WL] (a4WL);
\vertex [below=0.47cm of a4WL] (tl3WL);
\vertex [left=0.3cm of tl3WL] (a5WL);
\vertex [left=1.2cm of a5WL] (el1WL) {\(e\)};
\vertex [right=4cm of a5WL] (a3WL);
\vertex [right=1.2cm of a3WL] (er1WL) {\(e\)};
\vertex[blob] [left=0.08cm of tWL] (bl1WL) {};
\diagram [medium] {
(a1WL) -- [photon, rmomentum=\(q\)] (l1WL),
(a2WL) -- [scalar] (l3WL),
(a4WL) -- [scalar] (a5WL),
(a3WL) -- [photon, rmomentum'=\(k\)] (l4WL),
(l1WL) -- [photon, half left, looseness=1.7] (l2WL) -- [photon, half left, looseness=1.7, edge label=\(W\), pos=0.6] (l1WL),
(el1WL) -- [momentum'=\(p_1\)] (a5WL) -- [momentum'=\(p_2 - k\)] (a3WL) -- [momentum'=\(p_2\)] (er1WL)
};
\vertex [right=8.4cm of a1WL] (b1WR) {\(\mu\)};
\vertex [below=1.5cm of b1WR] (l5WR);
\vertex [below=2.0cm of l5WR] (l6WR);
\vertex [above=0.27cm of l6WR] (t2WR);
\vertex [left=0.75cm of t2WR] (l7WR);
\vertex [right=0.75cm of t2WR] (l8WR);
\vertex [below=0.87cm of l8WR] (tWR);
\vertex [right=0.3cm of tWR] (tr1WR);
\vertex [above=0.4cm of tr1WR] (b2WR);
\vertex [below=0.47cm of tWR] (tr2WR);
\vertex [right=0.85cm of tr2WR] (b4WR);
\vertex [below=0.47cm of b4WR] (tl3WR);
\vertex [right=0.3cm of tl3WR] (b5WR);
\vertex [right=1.2cm of b5WR] (er2WR) {\(e\)};
\vertex [left=4cm of b5WR] (b3WR);
\vertex [left=1.2cm of b3WR] (el2WR) {\(e\)};
\vertex[blob] [right=0.08cm of tWR] (br1WR) {};
\diagram [medium] {
(b1WR) -- [photon, rmomentum=\(q\)] (l5WR),
(b4WR) -- [scalar] (b5WR),
(b3WR) -- [photon, rmomentum=\(k\)] (l7WR),
(b2WR) -- [scalar] (l8WR),
(l5WR) -- [photon, half left, looseness=1.7] (l6WR) -- [photon, half left, looseness=1.7, edge label=\(W\), pos=0.6] (l5WR),
(el2WR) -- [momentum'=\(p_1\)] (b3WR) -- [momentum'=\(p_1 + k\)] (b5WR) -- [momentum'=\(p_2\)] (er2WR)
};
\end{feynman}
\end{tikzpicture}
\end{center}
\vspace{-0.5cm}
\caption{Focused BZ type diagrams with fermion (top diagrams) and $W$ boson (bottom diagrams) loops.}
\label{fig:outerf}
\end{figure}

The relevant sub-diagrams for the BZ diagrams are shown in Fig.~\ref{fig:outerf}.\footnote{The other contributions are mainly coming from diagrams with the charged scalar. 
As studied in refs.~\cite{Abe:2013qla,Altmannshofer:2020shb}, these contributions are sub-dominant, especially in large $\tan \beta$ case. 
Therefore, we simply ignore its contributions in our work.} 
The top panel diagrams feature a fermion inner loop, while the bottom panel diagrams have the $W$ boson in the inner loop. 
In our calculation, we adopt the Feynman-'t Hooft gauge for the $W$ boson propagator and utilize the results from ref.~\cite{Abe:2013qla}. 
In general, for $W$ boson contributions to the electron EDM, careful consideration of gauge dependence is necessary, as the diagrams shown in Fig.~\ref{fig:outerf} alone do not yield a gauge-independent result. 
A comprehensive gauge-independent analysis of the BZ diagrams with $W$ bosons was conducted in ref.~\cite{Altmannshofer:2020shb}. 
Although numerous diagrams warrant consideration for gauge independence, we focus solely on the diagrams presented in Fig.~\ref{fig:outerf}. 
This is because these diagrams already encompass the leading contributions among the $W$ boson loop diagrams in the Feynman-'t Hooft gauge. 

The sum of four diagrams in Fig.~\ref{fig:outerf} leads to a general result of the electron EDM as
\begin{align}
d_e &= \sum_f d_e^f + d_e^W \, , \label{eq:deBZresGenLO} \\[0.7ex]
d_e^{f, W} &= \sum_{j = 1}^{N_E} \sum_{a = 2}^{N_O} \int_x \int_y \int_k \frac{F_{\rm DS}^{\phi_j \eta_a} ( (k + q)^2, M^2)}{(k^2 - \Delta_{f, W}) (k^2 - m_G^2) (k^2 + 2 k \cdot q - m_{\phi_j}^2) (k^2 + 2 k \cdot q - m_{\eta_a}^2)} \nonumber \\[0.5ex]
&\hspace{1.0em} \times \sum_{\Phi = \phi_j, \eta_a} \left( \frac{A_{\alpha \beta}^{L, \Phi (f, W)} k^{\alpha} k^{\beta} + B_{\alpha}^{L, \Phi (f, W)} k^{\alpha}}{k^2 - 2 k \cdot p_2} + \frac{A_{\alpha \beta}^{R, \Phi (f, W)} k^{\alpha} k^{\beta} + B_{\alpha}^{R, \Phi (f, W)} k^{\alpha}}{k^2 + 2 k \cdot p_1} \right) \, , \label{eq:deBZresGen}
\end{align}
where we have used $p_{1, 2}^2 = m_e^2$ for the outer electron lines, $q = p_1 - p_2$ represents the four-momentum of the external photon, and $\Delta_{f, W}$ is defined as
\begin{align}
\Delta_{f, W} \equiv \frac{m_{f, W}^2}{x (1 - x)} - \frac{2 y}{1 - x} k \cdot q \, . \label{eq:defDeltafW}
\end{align}
In addition, we have defined a shorthand notation for each integral as
\begin{align}
\int_x &\equiv \int_0^1 \! d x \frac{1}{x (1 - x)} \, , ~~ \int_y \equiv \int_0^{1 - x} \! dy \, , ~~ \int_k \equiv \int \! \frac{d^4 k}{(2 \pi)^4} \, , \label{eq:intdef}
\end{align}
where $x$ and $y$ are Feynman parameters arising from the inner loop. The terms $A_{\alpha \beta}^{L, \Phi (f, W)}$, $A_{\alpha \beta}^{R, \Phi (f, W)}$, $B_{\alpha}^{L, \Phi (f, W)}$, and $B_{\alpha}^{R, \Phi (f, W)}$ represent the relevant contributions to the electron EDM from fermion and $W$ boson loop diagrams. 
The explicit expressions for these terms are summarized in appendix~\ref{app:CalcGen}. 

The expressions in Eq.~\eqref{eq:deBZresGen} are general and applicable to any dark sector models. 
Before delving into the numerical calculation, we emphasize the following points:
\begin{itemize}
\item[(1)] The momenta $p_{1, 2}$ correspond to the external electron momenta. 
Thus, terms such as $k \cdot p_i$ and $k \cdot q$ in the expression will contribute sub-dominantly. 
Upon integrating the loop momentum $k$, these terms lead to $p_i \cdot p_j = m_e^2$ or $p_i \cdot q = 0$, as $q \cdot q = 0$. 

\item[(2)] Due to the propagators, Eq.~\eqref{eq:deBZresGen} exhibits a pole around $k^2 \sim v_{\rm SM}^2$ if the neutral scalar masses in the visible sector are of $\mathcal{O}(v_{\rm SM})$. 
Consequently, the dominant contribution to $d_e$ is expected to occur when $k^2 \sim v_{\rm SM}^2$.
\end{itemize}

\subsubsection{The heavy dark sector limit}
\label{sec:HDS}

With the property (2), the general formula can be simplified when the unknown dark sector is significantly heavier than the electroweak scale, $M \gg v_{\rm SM}$. 
By power expanding $F_{\rm DS}^{\phi_j \eta_a} (k^2, M^2)$ in terms of $v_{\rm SM}^2 / M^2 (\sim k^2 / M^2)$, we can simplify $F_{\rm DS}^{\phi_j \eta_a}$ as
\begin{align}
F_{\rm DS}^{\phi_j \eta_a} ((k + q)^2, M^2) = a_0^{\phi_j \eta_a} (M^2) + a_1^{\phi_j \eta_a} (M^2) \frac{(k + q)^2}{M^2} + \mathcal{O} \left( \frac{(k + q)^4}{M^4} \right) \, , \label{eq:HDSexpand}
\end{align}
where $a_{0, 1}^{\phi_j \eta_a} (M^2)$ are independent of $k$, and each order is expected to be suppressed by $\mathcal{O}(v_{\rm SM}^2/M^2)$. 
With this expansion, the electron EDM can be explicitly calculated for the heavy dark sector (HDS). 
The leading results are (see appendix~\ref{app:CalcHDS} for derivation):
\begin{align}
d_e^{f, W} &\simeq \sum_{j = 1}^{N_E} \sum_{a = 2}^{N_O} \sum_{\Phi = \phi_j, \eta_a} \int_x \frac{g_{e G}^V \left( C_E^{f, W} g_{e \Phi}^P - C_O^{f, W} g_{e \Phi}^S \right)}{16 \pi^2 m_{\phi_j}^2} \nonumber \\[0.5ex]
&\hspace{3.0em} \times \left[ \frac{a_0^{\phi_j \eta_a} (M^2)}{m_{\phi_j}^2} j_3^{(1)} (r_{f, W}, r_G, r_{\eta_a}) + \frac{a_1^{\phi_j \eta_a} (M^2)}{M^2}  j_3^{(2)} (r_{f, W}, r_G, r_{\eta_a}) \right] \, , \label{eq:deBZresHDS}
\end{align}
where we omit terms proportional to $m_e^2 / m_{\phi_j}^2 \sim m_e^2 / v_{\rm SM}^2$. 
The integrals of momentum $k$ and the Feynman parameter $y$ have been calculated explicitly. 
Here, $r_{f, W, G, \eta_a}$ are mass squared ratios defined as
\begin{align}
r_{f, W} \equiv \frac{m_{f, W}^2}{m_{\phi_j}^2 x (1 - x)} \, , \qquad r_G \equiv \frac{m_G^2}{m_{\phi_j}^2} \, , \qquad  r_{\eta_a} \equiv \frac{m_{\eta_a}^2}{m_{\phi_j}^2} \, .
\end{align}
The coefficients $C_E^{f, W}$ and $C_O^{f, W}$ encapsulate the information from the inner fermion and $W$ boson loops, and they are defined as~\cite{Abe:2013qla,Nakai:2016atk}:
\begin{align}
C_E^f &= + \frac{N_C^f}{4 \pi^2} g_{f \gamma}^V g_{f G}^V g_{f \Phi}^S m_f \Bigl[ x^2 + (1 - x)^2 \Bigr] \, , \label{eq:CEfyint} \\[1ex]
C_O^f &= - \frac{N_C^f}{4 \pi^2} g_{f \gamma}^V g_{f G}^V g_{f \Phi}^P m_f \, , \label{eq:COfyint} \\[1ex]
C_E^W &= + \frac{\rm e}{8 \pi^2} g_{W W \Phi} g_{W W G} \left\{ \left( 4 - \frac{m_G^2}{m_W^2} \right) - \frac{x (1 - x)}{2} \left[ 6 - \frac{m_G^2}{m_W^2} + \left( 1 - \frac{m_G^2}{2 m_W^2} \right) \frac{m_{\Phi}^2}{m_W^2} \right] \right\} \, , \label{eq:CEWyint} \\[0.3ex]
C_O^W &= 0 \, , \label{eq:COWyint}
\end{align}
where $g_{W W \Phi} = \frac{2 m_W^2}{v_{\rm SM}} \xi_{\phi_j}^W \, (0)$ for CP-even (CP-odd) scalar, and $g_{W W G} = {\rm e} \cos \theta_W / \sin \theta_W \, ({\rm e})$ for $G = Z (\gamma)$. 
The function $j_3^{(1, 2)} (r_{f, W}, r_G, r_{\eta_a})$, arising from the integration of Feynman parameters, is defined by
\begin{align}
j_3^{(n)} (r_{f, W}, r_G, r_{\eta_a}) &\equiv \frac{r_{f, W}^n \ln r_{f, W}}{(r_{f, W} - 1) (r_{f, W} - r_G) (r_{f, W} - r_{\eta_a})} + \frac{r_G^n \ln r_G}{(r_G - 1) (r_G - r_{f, W}) (r_G - r_{\eta_a})} \nonumber \\[0.5ex]
&\hspace{6.0em} + \frac{r_{\eta_a}^n \ln r_{\eta_a}}{(r_{\eta_a} - 1) (r_{\eta_a} - r_{f, W}) (r_{\eta_a} - r_G)} \, .
\label{eq:defj3n}
\end{align}
For the case where the gauge boson is the photon, i.e., $G = \gamma$ ($m_G = 0$), the function $j_3$ becomes:
\begin{align}
j_3^{(n)} (r_{f, W}, 0, r_{\eta_a}) = \frac{r_{f, W}^{n - 1} \ln r_{f, W}}{(r_{f, W} - 1) (r_{f, W} - r_{\eta_a})} + \frac{r_{\eta_a}^{n - 1} \ln r_{\eta_a}}{(r_{\eta_a} - 1) (r_{\eta_a} - r_{f, W})} \, .
\end{align}

\section{Numerical analysis for benchmark models}
\label{sec:NumChecks}

In the previous section, we have derived the formula for the electron EDM in a general CP-violating dark sector.
Let us now consider two specific dark sector models. 
The first model consists of a complex singlet scalar whose potential contains CP violation
(the scalar dark sector model). 
The second model involves a vector-like singlet fermion with a CP-violating Yukawa coupling
and a complex singlet scalar portal to the Higgs sector (the fermion dark sector model).
For both models, we numerically estimate the size of the electron EDM and discuss
the validity of the expression in the HDS limit.

\subsection{Scalar dark sector model}
\label{sec:CPVinscalar}

We introduce a complex singlet scalar denoted as $S$. Its mass term and self-interactions are described by the potential $V_{\rm DS}$. In addition, we allow for interactions between the scalar $S$ and the Higgs doublets of the 2HDM through the potential $V_{H \mathcal{O}}$. The expressions for $V_{\rm DS}$ and $V_{H \mathcal{O}}$ are given by
\begin{align}
V_{\rm DS} &= m_S^2 |S|^2 + \frac{\lambda_S}{2} |S|^4 + \Bigl[ B_S S^2 + D_S S^4 + {\rm h.c.} \Bigr] \, , \label{eq:VSZ4} \\[1ex]
V_{H \mathcal{O}} &= \lambda_{1 S} H_1^{\dagger} H_1 |S|^2 + \lambda_{2 S} H_2^{\dagger} H_2 |S|^2 + \Bigl[ \lambda_{1 2 S} H_1^{\dagger} H_2 S^2 + \lambda_{2 1 S} H_2^{\dagger} H_1 S^2 + {\rm h.c.} \Bigr] \, . \label{eq:VHSZ4}
\end{align}
In our analysis, we parameterize the complex scalar as $S = (s_1 + i s_2) / \sqrt{2}$. 
To simplify and focus on relevant terms in $V_{\rm DS}$ and $V_{H \mathcal{O}}$, we impose a discrete $Z_4$ symmetry,
and its charges are summarized in Table~\ref{tab:Z4charges}. 
According to the charge assignment of $S$, $B_S$ is a soft $Z_4$ breaking term.\footnote{
Due to the remaining $Z_2$ symmetry, $S$ may become a DM candidate. }

\begin{table}[!t]
\centering
\begin{tabular}{|ccc|ccccc|}
\hline
$H_1$ & $H_2$ & $S$ & $Q_L$ & $u_R$ & $d_R$ & $L_L$ & $e_R$ \\ \hline
$i^2$ & $1$ & $i$ & $1$ & $1$ & $z_4^d$ & $1$ & $z_4^e$ \\ \hline
\end{tabular}
\caption{We present an example of $Z_4$ charges for the scalar dark sector model. 
The exact values of $z_4^d$ and $z_4^e$ for the right-handed $d$ quark and electron depend on the specific type of
the 2HDM being considered.}
\label{tab:Z4charges}
\end{table}

There are 6 complex parameters in the scalar potential of the model: $m_{12}^2$ and $\lambda_5$ in Eq.~\eqref{eq:V2HDM} and $B_S$, $D_S$, $\lambda_{1 2 S}$ and $\lambda_{2 1 S}$ in $V_{\rm DS} + V_{H \mathcal{O}}$. 
To explore CP violation, it is advantageous to identify rephasing-invariant phases, as studied in refs.~\cite{Inoue:2014nva, Inoue:2015pza}. 
The details are discussed in appendix~\ref{app:CPVScalarDS}, where we have five invariant phases. For our analysis, the relevant invariant phase in the dark sector is
\begin{align}
\theta_{\rm phys} = \arg \left[ B_S^2 D_S^{*} \right] \, . 
\label{eq:scalarDMphysicalphase}
\end{align}
While $D_S$ is, in principle, a complex parameter, we can select the phase of $S$ to render it real. 
Consequently, in this scenario, the physical phase $\theta_{\rm phys}$ is dictated by the phase of $B_S$, resulting in tree-level CP violation in the mass mixing matrix for the $S$ scalars. In appendix~\ref{app:CPVScalarDS}, another invariant phase $\arg[B_S^2 \lambda_{12S}^{*} \lambda_{21S}^{*}]$ can also induce the same tree-level CP violation once $H_{1,2}$ acquire their VEVs. However, given our assumption of CP violation happening in the dark sector only, we set $\arg[B_S^2 \lambda_{12S}^{*} \lambda_{21S}^{*}] = 0$ to maintain simplicity in the CP violation pattern. Thus, in our analysis, we turn off all the physical phases other than Eq.~\eqref{eq:scalarDMphysicalphase}. 

Our assumption (i) in section~\ref{sec:DS} requires that the scalar field $S$ does not acquire a nonzero VEV. 
Therefore, the CP-violating phase in $B_S$ contributes to the mass matrix of the dark scalars $s_{1,2}$,
resulting in CP-violating mixing. 
Explicitly, the mass matrix for the complex scalar $S$ is given in the basis of $(s_1, s_2)$ as
\begin{align}
M_s^2 = \overline{m}_S^2 \mathbf{1}_{2 \times 2} + \begin{pmatrix}
{\rm Re}(\overline{B}_S) & - {\rm Im}(\overline{B}_S) \\
- {\rm Im}(\overline{B}_S) & - {\rm Re}(\overline{B}_S)
\end{pmatrix} \, ,
\label{eq:M2s}
\end{align}
with $\overline{m}_S^2 \equiv m_S^2 + \frac{1}{2} \lambda_{1 S} v_1^2 + \frac{1}{2} \lambda_{2 S} v_2^2$ and $\overline{B}_S \equiv 2 B_S + (\lambda_{12S} + \lambda_{21S}) v_1 v_2$. 
Note that since the neutral scalars in the 2HDM sector and $s_{1, 2}$ do not mix with each other,
the SM Yukawa couplings and the $W$ boson couplings are not modified from Eq.~\eqref{eq:SMYukawain2HDM} with coefficients $\xi_{h, H, A}^f$ in Table~\ref{tab:xif} and Eq.~\eqref{eq:WWhandWWHin2HDM}. 
As a result, we obtain the following new mass eigenstates $\varphi_{1, 2}$:
\begin{align}
\varphi_1 = s_1 c_{\theta_s} + s_2 s_{\theta_s} \, , \qquad \varphi_2 = - s_1 s_{\theta_s} + s_2 c_{\theta_s} \, ,
\end{align}
where $\theta_s$ is a mixing angle with $m_{\varphi_1}^2 < m_{\varphi_2}^2$, and its value is determined by
\begin{align}
\sin 2 \theta_s &= \frac{{\rm Im}(\overline{B}_S)}{|\overline{B}_S|} \, . \label{eq:sin2ths}
\end{align}
When we assume the portal couplings $\lambda_{1 2 S}$ and $\lambda_{2 1 S}$ to be real,
$\theta_s$ becomes a phase of $B_S$ and corresponds to the physical CP-violating phase.
The mass eigenvalues are given by
\begin{align}
m_{\varphi_1}^2 = \overline{m}_S^2 - |\overline{B}_S| \, , \qquad m_{\varphi_2}^2 = \overline{m}_S^2 + |\overline{B}_S| \, . \label{eq:mvphi2}
\end{align}
Incorporating all the mass eigenstates into the potentials,
we can derive the corresponding neutral scalar cubic and quartic couplings as
\begin{align}
V_{\rm scl} = V_H + V_{H \mathcal{O}} + V_{\rm DS} \supset \, &\frac{v_{\rm SM}}{2} \Bigl( \lambda_{h j k} h^0 \varphi_j \varphi_k + \lambda_{H j k} H^0 \varphi_j \varphi_k + \lambda_{A j k} A^0 \varphi_j \varphi_k \Bigr) \nonumber \\[0.5ex]
&+ \frac{1}{2} \Bigl( \lambda_{h A j j} h^0 A^0 \varphi_j \varphi_j + \lambda_{H A j j} H^0 A^0 \varphi_j \varphi_j \Bigr)
\, ,
\end{align}
where $j, k = 1,2$ and each coupling is summarized in Table~\ref{tab:3sint}. 

\begin{table}[t]
\centering
\begin{tabular}{|c|c|}
\hline
$\lambda_{h 1 1}$ & $- \lambda_{1 S} s_{\alpha} c_{\beta} + \lambda_{2 S} c_{\alpha} s_{\beta} + \left( \lambda_{1 2 S} + \lambda_{2 1 S} \right) c_{\alpha + \beta} c_{2 \theta_s}$ \\
$\lambda_{h 1 2}$ & $- 2 \left( \lambda_{1 2 S} + \lambda_{2 1 S} \right) c_{\alpha + \beta} s_{2 \theta_s}$ \\
$\lambda_{h 2 2}$ & $- \lambda_{1 S} s_{\alpha} c_{\beta} + \lambda_{2 S} c_{\alpha} s_{\beta} - \left( \lambda_{1 2 S} + \lambda_{2 1 S} \right) c_{\alpha + \beta} c_{2 \theta_s}$ \\ \hline
$\lambda_{H 1 1}$ & $\lambda_{1 S} c_{\alpha} c_{\beta} + \lambda_{2 S} s_{\alpha} s_{\beta} + \left( \lambda_{1 2 S} + \lambda_{2 1 S} \right) s_{\alpha + \beta} c_{2 \theta_s}$ \\
$\lambda_{H 1 2}$ & $- 2 \left( \lambda_{1 2 S} + \lambda_{2 1 S} \right) s_{\alpha + \beta} s_{2 \theta_s}$ \\
$\lambda_{H 2 2}$ & $\lambda_{1 S} c_{\alpha} c_{\beta} + \lambda_{2 S} s_{\alpha} s_{\beta} - \left( \lambda_{1 2 S} + \lambda_{2 1 S} \right) s_{\alpha + \beta} c_{2 \theta_s}$ \\ \hline
$\lambda_{A 1 1}$ & $- \left( \lambda_{1 2 S} - \lambda_{2 1 S} \right) s_{2 \theta_s}$ \\
$\lambda_{A 1 2}$ & $- 2 \left( \lambda_{1 2 S} - \lambda_{2 1 S} \right) c_{2 \theta_s}$ \\
$\lambda_{A 2 2}$ & $\left( \lambda_{1 2 S} - \lambda_{2 1 S} \right) s_{2 \theta_s}$ \\ \hline \hline
$\lambda_{h A 1 1}$ & $\left( \lambda_{1 2 S} - \lambda_{2 1 S} \right) s_{\alpha - \beta} s_{2 \theta_s}$ \\
$\lambda_{h A 2 2}$ & $- \left( \lambda_{1 2 S} - \lambda_{2 1 S} \right) s_{\alpha - \beta} s_{2 \theta_s}$ \\ \hline
$\lambda_{H A 1 1}$ & $- \left( \lambda_{1 2 S} - \lambda_{2 1 S} \right) c_{\alpha - \beta} s_{2 \theta_s}$ \\
$\lambda_{H A 2 2}$ & $\left( \lambda_{1 2 S} - \lambda_{2 1 S} \right) c_{\alpha - \beta} s_{2 \theta_s}$ \\ \hline
\end{tabular}
\caption{The relevant neutral scalar cubic and quartic couplings. 
For instance, $\lambda_{h12}$ represents the trilinear coupling involving the mass eigenstate of the SM Higgs and the dark sector scalars $\varphi_1$ and $\varphi_2$.}
\label{tab:3sint}
\end{table}

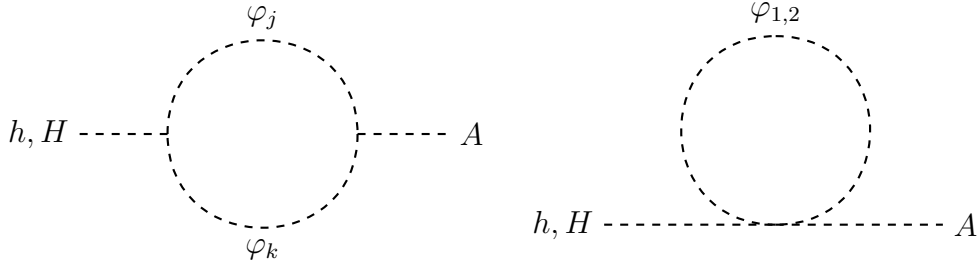
\begin{figure}[t]
\begin{center}
\begin{tikzpicture}
\begin{feynman}[large]
\vertex (a1) {\({h, H}\)};
\vertex [right=1.7cm of a1] (l1);
\vertex [right=2.5cm of l1] (l2);
\vertex [right=1.2cm of l2] (b1) {\(A\)};
\diagram [medium] {
(a1) -- [scalar] (l1),
(l1) -- [scalar, half left, looseness=1.7, edge label=\(\varphi_j\)] (l2),
(l2) -- [scalar, half left, looseness=1.7, edge label=\(\varphi_k\)] (l1),
(l2) -- [scalar] (b1)
};
\vertex [below right=1.7cm of b1] (a2) {\({h, H}\)};
\vertex [right=2.8cm of a2] (l3);
\vertex [above=2.5cm of l3] (l4);
\vertex [right=2.2cm of l3] (b2) {\(A\)};
\diagram [medium] {
(a2) -- [scalar] (l3),
(l3) -- [scalar, half left, looseness=1.7, edge label=\(\varphi_{1, 2}\), pos=1.0] (l4),
(l4) -- [scalar, half left, looseness=1.7] (l3),
(l3) -- [scalar] (b2)
};
\end{feynman}
\end{tikzpicture}
\end{center}
\vspace{-0.5cm}
\caption{One-loop diagrams relevant to the effective $h$-$A$ and $H$-$A$ mixings induced by $\varphi_{1, 2}$
in the scalar dark sector model.}
\label{fig:hAeffmixScl}
\end{figure}

With the cubic and quartic couplings determined, we can now explicitly calculate the one-loop contributions from the dark sector scalars to the two effective CP-violating mixing terms $F_{\rm DS}^{h A}$ and $F_{\rm DS}^{H A}$, as shown in the diagrams of Fig.~\ref{fig:hAeffmixScl},
\begin{align}
F_{\rm DS}^{h A} (k^2) &= - \frac{v_{\rm SM}^2}{64 \pi^2} \lambda_{1 2 S}^- s_{2 \theta_s} \Biggl[ \Biggr. \lambda_{1 2 S}^+ c_{\alpha + \beta} c_{2 \theta_s} \biggl( {\rm DiscB}(k^2, m_{\varphi_1}, m_{\varphi_1}) + {\rm DiscB}(k^2, m_{\varphi_2}, m_{\varphi_2}) \biggr. \nonumber \\[0.5ex]
&\hspace{13.5em} \biggl. - 2 {\rm DiscB}(k^2, m_{\varphi_1}, m_{\varphi_2}) + \frac{m_{\varphi_1}^2 - m_{\varphi_2}^2}{k^2} \ln \frac{m_{\varphi_1}^2}{m_{\varphi_2}^2} \biggr) \nonumber \\[0.5ex]
&\hspace{6.0em} + \lambda_{h S} \left( {\rm DiscB}(k^2, m_{\varphi_1}, m_{\varphi_1}) - {\rm DiscB}(k^2, m_{\varphi_2}, m_{\varphi_2}) - \ln \frac{m_{\varphi_1}^2}{m_{\varphi_2}^2} \right) \Biggl. \Biggr] \nonumber \\[0.5ex]
&\hspace{1.2em} + \frac{\lambda_{1 2 S}^-}{32 \pi^2} s_{\alpha - \beta} s_{2 \theta_s} \left[ \left( m_{\varphi_1}^2 - m_{\varphi_2}^2 \right) \Lambda_{\rm UV}^2 + m_{\varphi_1}^2 \ln \frac{\mu^2}{m_{\varphi_1}^2} - m_{\varphi_2}^2 \ln \frac{\mu^2}{m_{\varphi_2}^2} \right] \, , \label{eq:FDShScl} \\[1.0ex]
F_{\rm DS}^{H A} (k^2) &= - \frac{v_{\rm SM}^2}{64 \pi^2} \lambda_{1 2 S}^- s_{2 \theta_s} \Biggl[ \Biggr. \lambda_{1 2 S}^+ s_{\alpha + \beta} c_{2 \theta_s} \biggl( {\rm DiscB}(k^2, m_{\varphi_1}, m_{\varphi_1}) + {\rm DiscB}(k^2, m_{\varphi_2}, m_{\varphi_2}) \biggr. \nonumber \\[0.5ex]
&\hspace{13.5em} \biggl. - 2 {\rm DiscB}(k^2, m_{\varphi_1}, m_{\varphi_2}) + \frac{m_{\varphi_1}^2 - m_{\varphi_2}^2}{k^2} \ln \frac{m_{\varphi_1}^2}{m_{\varphi_2}^2} \biggr) \nonumber \\[0.5ex]
&\hspace{6.0em} + \lambda_{H S} \left( {\rm DiscB}(k^2, m_{\varphi_1}, m_{\varphi_1}) - {\rm DiscB}(k^2, m_{\varphi_2}, m_{\varphi_2}) - \ln \frac{m_{\varphi_1}^2}{m_{\varphi_2}^2} \right) \Biggl. \Biggr] \nonumber \\[0.5ex]
&\hspace{1.2em} - \frac{\lambda_{1 2 S}^-}{32 \pi^2} c_{\alpha - \beta} s_{2 \theta_s} \left[ \left( m_{\varphi_1}^2 - m_{\varphi_2}^2 \right) \Lambda_{\rm UV}^2 + m_{\varphi_1}^2 \ln \frac{\mu^2}{m_{\varphi_1}^2} - m_{\varphi_2}^2 \ln \frac{\mu^2}{m_{\varphi_2}^2} \right] \, . \label{eq:FDSHScl}
\end{align}
Here, we have defined the shorthand notations $\lambda_{1 2 S}^{\pm} \equiv \lambda_{1 2 S} \pm \lambda_{2 1 S}$,
$\lambda_{h S} \equiv - \lambda_{1 S} s_{\alpha} c_{\beta} + \lambda_{2 S} c_{\alpha} s_{\beta}$ and $\lambda_{H S} \equiv \lambda_{1 S} c_{\alpha} c_{\beta} + \lambda_{2 S} s_{\alpha} s_{\beta}$. 
The function ${\rm DiscB}(k^2, m_1, m_2)$ is defined as
\begin{align}
{\rm DiscB}(k^2, m_1, m_2) &= \frac{\lambda^{1/2} (k^2, m_1^2, m_2^2)}{k^2} \ln \frac{m_1^2 + m_2^2 - k^2 + \lambda^{1/2} (k^2, m_1^2, m_2^2)}{2 m_1 m_2} \, , \label{eq:defDiscB} \\[1ex]
\lambda (k^2, m_1^2, m_2^2) &= k^4 + m_1^4 + m_2^4 - 2 k^2 m_1^2 - 2 k^2 m_2^2 - 2 m_1^2 m_2^2 \, , \label{eq:defKallen}
\end{align}
and $\Lambda_{\rm UV}$ and $\mu$ are a UV divergent parameter and 't Hooft mass parameter, respectively.

To obtain a finite physical prediction, we need to deal with the UV divergent part appropriately. 
For this purpose, let us consider the wave function renormalization for $h$-$A$ and $H$-$A$ self-energies~\cite{Altenkamp:2017ldc, Denner:1991kt}\footnote{To deal with the UV divergent in these self-energies, there is another technique for renormalization, so-called ``CP-odd tadpole renoramlization" proposed in ref.~\cite{Pilaftsis:1998pe} (see also refs.~\cite{Pilaftsis:1998dd,Pilaftsis:1999qt} for its application to the supersymmetric models). 
This methods is basically equivalent to the wave function renormalization, since there is an explicit correspondence in contributions from CP-odd counterterm $T_A$ and renormalization constants $\delta Z_{h A}$, $\delta Z_{A h}$, $\delta Z_{H A}$, $\delta Z_{A H}$ in Eqs.~\eqref{eq:SigmahA} and \eqref{eq:SigmaHA} as
\begin{align}
\frac{T_A}{v_{\rm SM}} s_{\alpha - \beta} \quad &\longleftrightarrow \quad \frac{1}{2} (k^2 - m_h^2) \delta Z_{h A} + \frac{1}{2} (k^2 - m_A^2) \delta Z_{A h} \, , \\[0.5ex]
- \frac{T_A}{v_{\rm SM}} c_{\alpha - \beta} \quad &\longleftrightarrow \quad \frac{1}{2} (k^2 - m_H^2) \delta Z_{H A} + \frac{1}{2} (k^2 - m_A^2) \delta Z_{A H} \, ,
\end{align}
where each left-hand side is obtained by the CP-odd tadpole renormalization. 
We found that the UV divergent parts of each contribution are totally same. 
Note that in this model, we have loop-induced CP-odd tadpole contributions by using $\lambda_{A 1 1}$ and $\lambda_{A 2 2}$ couplings in Table~\ref{tab:3sint}, which are UV divergent ones. 
Refs.~\cite{Krause:2016oke,Denner:2016etu,Altenkamp:2017ldc,Denner:2018opp,Denner:2019vbn,Fontes:2021iue} have discussed about this divergence via Fleischer-Jegerlehner tadpole scheme~\cite{Fleischer:1980ub}.}. 
First, we define the renormalized two-point functions $\hat{\Sigma}_{hA, HA} (k^2)$ as
\begin{align}
\hat{\Sigma}_{h A} (k^2) &\equiv F_{\rm DS}^{h A} (k^2) + \frac{1}{2} (k^2 - m_h^2) \delta Z_{h A} + \frac{1}{2} (k^2 - m_A^2) \delta Z_{A h} \, , \label{eq:SigmahA} \\[0.5ex]
\hat{\Sigma}_{H A} (k^2) &\equiv F_{\rm DS}^{H A} (k^2) + \frac{1}{2} (k^2 - m_H^2) \delta Z_{H A} + \frac{1}{2} (k^2 - m_A^2) \delta Z_{A H} \, . \label{eq:SigmaHA}
\end{align}
In the on-shell subtraction scheme, we can determine $\delta Z_{h A, H A}$ and $\delta Z_{A h, A H}$ by
\begin{align}
{\rm Re} \bigl\{ \hat{\Sigma}_{h A} (m_h^2) \bigr\} = 0 \, , \quad  {\rm Re} \bigl\{ \hat{\Sigma}_{H A} (m_H^2) \bigr\} = 0 \, , \quad  {\rm Re} \bigl\{ \hat{\Sigma}_{h A, H A} (m_A^2) \bigr\} = 0 \, .
\end{align}
By solving these equations, the renormalized two-point functions $\hat{\Sigma}_{h A, H A} (k^2)$ are given by
\begin{align}
\hat{\Sigma}_{h A} (k^2) &= F_{\rm DS}^{h A} (k^2) + \frac{1}{m_h^2 - m_A^2} \Bigl[ \left( m_A^2 - k^2 \right) {\rm Re} \! \left\{ F_{\rm DS}^{h A} (m_h^2) \right\} + \left( k^2 - m_h^2 \right) {\rm Re} \! \left\{ F_{\rm DS}^{h A} (m_A^2) \right\} \Bigr] \, , \label{eq:SigmaDShAScl} \\[0.5ex]
\hat{\Sigma}_{H A} (k^2) &= F_{\rm DS}^{H A} (k^2) + \frac{1}{m_H^2 - m_A^2} \Bigl[ \left( m_A^2 - k^2 \right) {\rm Re} \! \left\{ F_{\rm DS}^{H A} (m_H^2) \right\} + \left( k^2 - m_H^2 \right) {\rm Re} \! \left\{ F_{\rm DS}^{H A} (m_A^2) \right\} \Bigr] \, . \label{eq:SigmaDSHAScl}
\end{align}
One can easily check that the divergent parts (and also other constant terms concerning the momentum $k$)
are canceled, and hence, $\hat{\Sigma}_{h A, H A}$ give finite results. 
For numerical predictions, we need to replace $F_{\rm DS}^{h A} \to \hat{\Sigma}_{h A}$ and $F_{\rm DS}^{H A} \to \hat{\Sigma}_{H A}$ in the expressions that we have obtained in the previous section. 

In the HDS approximation, the dark sector scalar masses $m_{\varphi_1}, m_{\varphi_2}$ are much larger than the masses in the 2HDM. In this case, we can expand $\hat{\Sigma}_{h A (H A)} (k^2)$ as
\begin{align}
\hat{\Sigma}_{h A (H A)} (k^2) = a_0^{h A (H A)} + a_1^{h A (H A)} \frac{k^2}{m_{\varphi_1} m_{\varphi_2}} + \mathcal{O}\left( \frac{k^4}{m_{\varphi_i}^4} \right) \, , \label{eq:HDSexpandScl}
\end{align}
where we choose $M^2$ in Eq.~\eqref{eq:HDSexpand} to be $m_{\varphi_1} m_{\varphi_2}$. 
Each coefficient is given explicitly as
\begin{align}
a_0^{h A} &= \frac{v_{\rm SM}^2 \lambda_{1 2 S}^-}{64 \pi^2} s_{2 \theta_s} \Biggl[ \Biggr. \lambda_{1 2 S}^+ c_{\alpha + \beta} c_{2 \theta_s} \biggl( \biggr. 2 + \left\{ \frac{m_{\varphi_1}^2 - m_{\varphi_2}^2}{m_h^2} + \frac{m_{\varphi_1}^2 - m_{\varphi_2}^2}{m_A^2} - \frac{m_{\varphi_1}^2 + m_{\varphi_2}^2}{m_{\varphi_1}^2 - m_{\varphi_2}^2} \right\} \ln \frac{m_{\varphi_1}^2}{m_{\varphi_2}^2} \nonumber \\[0.5ex]
&\hspace{9.0em} + \frac{1}{m_h^2 - m_A^2} \Bigl[ m_h^2 \mathcal{F}_1 (m_A, m_{\varphi_1}, m_{\varphi_2}) - m_A^2 \mathcal{F}_1 (m_h, m_{\varphi_1}, m_{\varphi_2}) \Bigr] \biggl. \biggr) \nonumber \\[0.5ex]
&\hspace{6.5em} + \frac{\lambda_{h S}}{m_h^2 - m_A^2} \Bigl( m_h^2 \mathcal{F}_2 (m_A, m_{\varphi_1}, m_{\varphi_2}) - m_A^2 \mathcal{F}_2 (m_h, m_{\varphi_1}, m_{\varphi_2}) \Bigr) \Biggl. \Biggr] \, , \label{eq:a0hScl} \\[0.5ex]
a_1^{h A} &= - \frac{v_{\rm SM}^2 \lambda_{1 2 S}^-}{64 \pi^2} s_{2 \theta_s} \Biggl[ \Biggr. \lambda_{1 2 s}^+ c_{\alpha + \beta} c_{2 \theta_s} \biggl( \biggr. \frac{(m_{\varphi_1}^2 + m_{\varphi_2}^2) (m_{\varphi_1}^4 - 8 m_{\varphi_1}^2 m_{\varphi_2}^2 + m_{\varphi_2}^4)}{6 m_{\varphi_1} m_{\varphi_2} (m_{\varphi_1}^2 - m_{\varphi_2}^2)^2} \nonumber \\[0.5ex]
&\hspace{12.0em} + \left\{ \frac{(m_{\varphi_1}^2 - m_{\varphi_2}^2) m_{\varphi_1} m_{\varphi_2}}{m_h^2 m_A^2} + \frac{2 m_{\varphi_1}^3 m_{\varphi_2}^3}{(m_{\varphi_1}^2 - m_{\varphi_2}^2)^3} \right\} \ln \frac{m_{\varphi_1}^2}{m_{\varphi_2}^2} \nonumber \\[0.5ex]
&\hspace{12.0em} - \frac{m_{\varphi_1} m_{\varphi_2}}{m_h^2 - m_A^2} \Bigl[ \mathcal{F}_1 (m_h, m_{\varphi_1}, m_{\varphi_2}) - \mathcal{F}_1 (m_A, m_{\varphi_1}, m_{\varphi_2}) \Bigr] \biggl. \biggr) \nonumber \\[0.5ex] 
&\hspace{3.5em} - \lambda_{h S} \biggl( \biggr. \frac{m_{\varphi_1}^2 - m_{\varphi_2}^2}{6 m_{\varphi_1} m_{\varphi_2}} + \frac{m_{\varphi_1} m_{\varphi_2}}{m_h^2 - m_A^2} \Bigl[ \mathcal{F}_2 (m_h, m_{\varphi_1}, m_{\varphi_2}) - \mathcal{F}_2 (m_A, m_{\varphi_1}, m_{\varphi_2}) \Bigr] \biggl. \biggr) \Biggl. \Biggr] \, , \label{eq:a1hScl} \\[0.5ex]
a_0^{H A} &= \frac{v_{\rm SM}^2 \lambda_{1 2 S}^-}{64 \pi^2} s_{2 \theta_s} \Biggl[ \Biggr. \lambda_{1 2 S}^+ s_{\alpha + \beta} c_{2 \theta_s} \biggl( \biggr. 2 + \left\{ \frac{m_{\varphi_1}^2 - m_{\varphi_2}^2}{m_H^2} + \frac{m_{\varphi_1}^2 - m_{\varphi_2}^2}{m_A^2} - \frac{m_{\varphi_1}^2 + m_{\varphi_2}^2}{m_{\varphi_1}^2 - m_{\varphi_2}^2} \right\} \ln \frac{m_{\varphi_1}^2}{m_{\varphi_2}^2} \nonumber \\[0.5ex]
&\hspace{9.0em} + \frac{1}{m_H^2 - m_A^2} \Bigl[ m_H^2 \mathcal{F}_1 (m_A, m_{\varphi_1}, m_{\varphi_2}) - m_A^2 \mathcal{F}_1 (m_H, m_{\varphi_1}, m_{\varphi_2}) \Bigr] \biggl. \biggr) \nonumber \\[0.5ex]
&\hspace{6.5em} + \frac{\lambda_{H S}}{m_H^2 - m_A^2} \Bigl( m_H^2 \mathcal{F}_2 (m_A, m_{\varphi_1}, m_{\varphi_2}) - m_A^2 \mathcal{F}_2 (m_H, m_{\varphi_1}, m_{\varphi_2}) \Bigr) \Biggl. \Biggr] \, , \label{eq:a0HScl} \\[0.5ex]
a_1^{H A} &= - \frac{v_{\rm SM}^2 \lambda_{1 2 S}^-}{64 \pi^2} s_{2 \theta_s} \Biggl[ \Biggr. \lambda_{1 2 s}^+ s_{\alpha + \beta} c_{2 \theta_s} \biggl( \biggr. \frac{(m_{\varphi_1}^2 + m_{\varphi_2}^2) (m_{\varphi_1}^4 - 8 m_{\varphi_1}^2 m_{\varphi_2}^2 + m_{\varphi_2}^4)}{6 m_{\varphi_1} m_{\varphi_2} (m_{\varphi_1}^2 - m_{\varphi_2}^2)^2} \nonumber \\[0.5ex]
&\hspace{12.0em} + \left\{ \frac{(m_{\varphi_1}^2 - m_{\varphi_2}^2) m_{\varphi_1} m_{\varphi_2}}{m_H^2 m_A^2} + \frac{2 m_{\varphi_1}^3 m_{\varphi_2}^3}{(m_{\varphi_1}^2 - m_{\varphi_2}^2)^3} \right\} \ln \frac{m_{\varphi_1}^2}{m_{\varphi_2}^2} \nonumber \\[0.5ex]
&\hspace{12.0em} - \frac{m_{\varphi_1} m_{\varphi_2}}{m_H^2 - m_A^2} \Bigl[ \mathcal{F}_1 (m_H, m_{\varphi_1}, m_{\varphi_2}) - \mathcal{F}_1 (m_A, m_{\varphi_1}, m_{\varphi_2}) \Bigr] \biggl. \biggr) \nonumber \\[0.5ex] 
&\hspace{3.5em} - \lambda_{H S} \biggl( \biggr. \frac{m_{\varphi_1}^2 - m_{\varphi_2}^2}{6 m_{\varphi_1} m_{\varphi_2}} + \frac{m_{\varphi_1} m_{\varphi_2}}{m_H^2 - m_A^2} \Bigl[ \mathcal{F}_2 (m_H, m_{\varphi_1}, m_{\varphi_2}) - \mathcal{F}_2 (m_A, m_{\varphi_1}, m_{\varphi_2}) \Bigr] \biggl. \biggr) \Biggl. \Biggr] \, , \label{eq:a1HScl}
\end{align}
where we have defined the following functions:
\begin{align}
\mathcal{F}_1 (m, m_{\varphi_1}, m_{\varphi_2}) &\equiv {\rm Re} \! \left\{ {\rm DiscB}(m^2, m_{\varphi_1}, m_{\varphi_1}) \right\} + {\rm Re} \! \left\{ {\rm DiscB}(m^2, m_{\varphi_2}, m_{\varphi_2}) \right\} \nonumber \\[0.3ex]
&\hspace{2.5em} - 2 {\rm Re} \! \left\{ {\rm DiscB}(m^2, m_{\varphi_1}, m_{\varphi_2}) \right\} \, , \label{eq:F1func} \\[0.3ex]
\mathcal{F}_2 (m, m_{\varphi_1}, m_{\varphi_2}) &\equiv {\rm Re} \! \left\{ {\rm DiscB}(m^2, m_{\varphi_1}, m_{\varphi_1}) \right\} - {\rm Re} \! \left\{ {\rm DiscB}(m^2, m_{\varphi_2}, m_{\varphi_2}) \right\} \, . \label{eq:F2func}
\end{align}
Since $a_0^{h A (H A)} = \hat{\Sigma}_{\rm DS}^{h A (H A)} (k^2 \to 0)$, they are used to calculate the coefficient of the four-fermion operator $C_{f f'}$ by inserting $a_0^{h A, H A}$ into Eq.~\eqref{eq:CffpGen}. 
Therefore, we have obtained the electron EDM in the HDS approximation as well as the coefficient $C_S$. 
By applying the limit of the JILA experiment in Eq.~\eqref{eq:deJILA}, one can establish bounds on the parameters for the scalar dark sector model. 

\subsubsection{Numerical results}
\label{sec:resScl}

Let us now perform the numerical analysis for the scalar dark sector model.
We choose the following values for the relevant masses and parameters:
\begin{align}
&m_h = 125.25 ~ {\rm GeV}  , ~~ m_H = 1500 ~ {\rm GeV}  , ~~ m_A = 1550 ~ {\rm GeV}  , ~~ \alpha = - 0.02  , ~~ \tan \beta = 50  , \nonumber \\[0.5ex]
&\theta_s = \frac{\pi}{4}  , ~~ \lambda_{1 S} = 1  , ~~ \lambda_{2 S} = 2  , ~~ \lambda_{1 2 S} = 2  , ~~ \lambda_{2 1 S} = 0.1  . \label{eq:paramScl}
\end{align}
Note that the value of $\theta_s$ corresponds to the maximal choice for the physical CP phase.
With this parameter choice, we have found that the $C_S$ contribution is several orders of magnitude
smaller than the electron EDM $d_e$ from the BZ diagrams. 
Hence, we simply neglect the $C_S$ term in the calculation. 

\begin{figure}[!t]
\centering
\includegraphics[bb=0 0 540 366, width=0.48\linewidth]{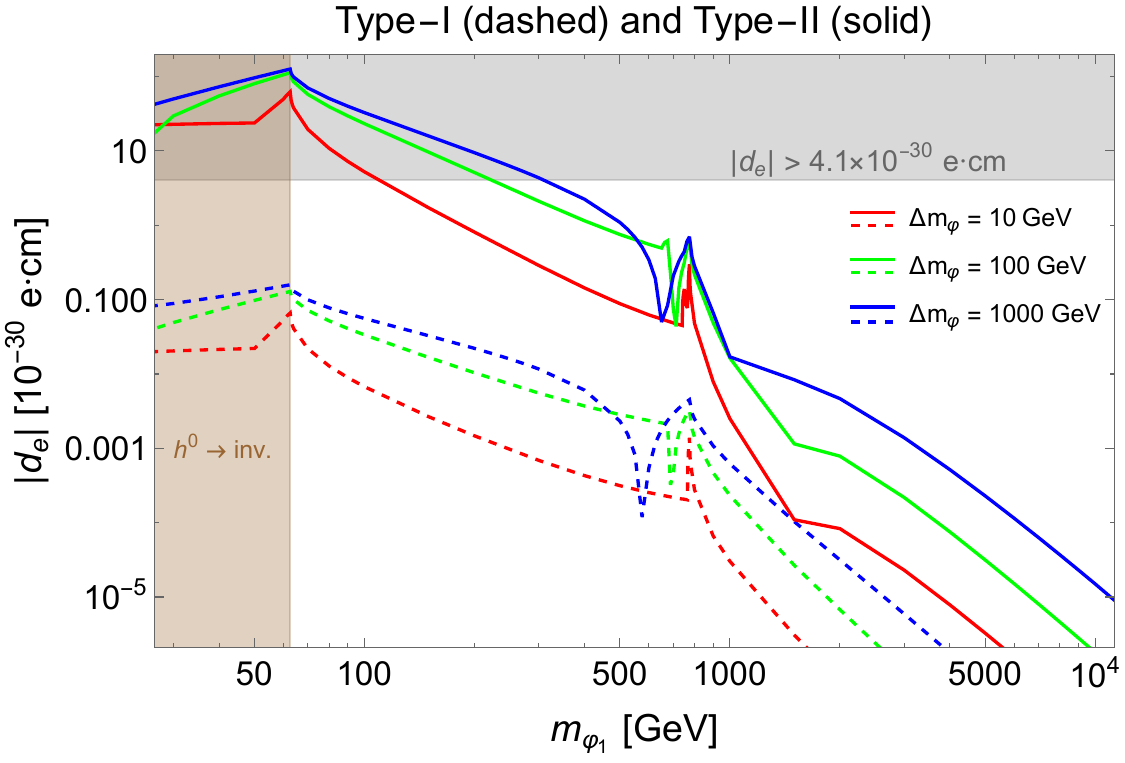}
\hfill
\includegraphics[bb=0 0 540 366, width=0.48\linewidth]{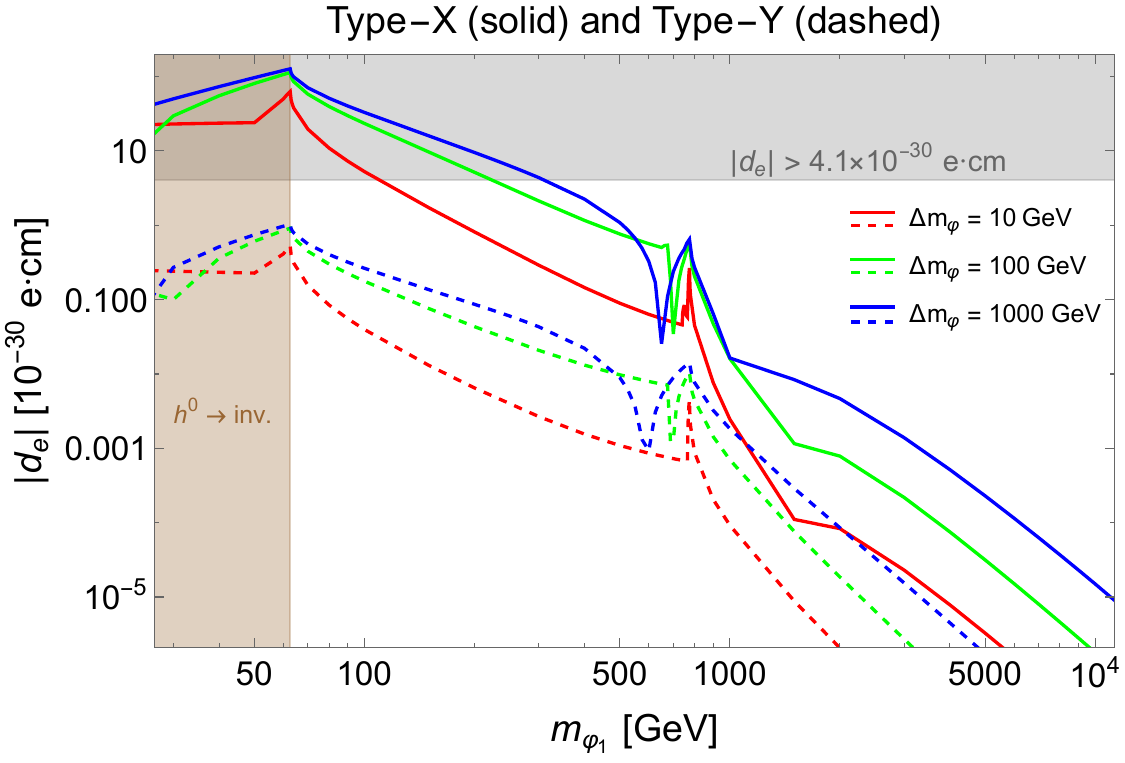}
\caption{Numerical results of the electron EDM in the scalar dark sector model. 
The left panel shows the results of Type-I (dashed lines) and Type-II (solid lines) 2HDMs, while the right panel shows those of Type-X (solid lines) and Type-Y (dashed lines) 2HDMs. 
The color means different choices of $\Delta m_{\varphi} \equiv m_{\varphi_2} - m_{\varphi_1}$: $\Delta m_{\varphi} = 10$ GeV (red), $\Delta m_{\varphi} = 100$ GeV (green) and $\Delta m_{\varphi} = 1000$ GeV (blue). 
The gray shaded region is excluded by the current electron EDM bound from the JILA experiment~\cite{Roussy:2022cmp}, and the brown shaded region is constrained by the SM Higgs invisible decay, $h^0 \to {\rm inv}$~\cite{Biekotter:2022ckj}.}
\label{fig:resScl}
\end{figure}

Fig.~\ref{fig:resScl} shows the size of the electron EDM $d_e$ in terms of $m_{\varphi_1}$, with 3 choices of mass difference for $m_{\varphi_{1, 2}}$: $\Delta m_{\varphi} \equiv m_{\varphi_2} - m_{\varphi_1} = 10$ GeV (red lines), $100$ GeV (green lines) and $1000$ GeV (blue lines). 
The left panel corresponds to the cases of Type-I (dashed lines) and Type-II (solid lines), while the right panel shows those in Type-X (solid lines) and Type-Y (dashed lines). 
The shaded regions are constrained by the JILA experiment for the electron EDM bound~\cite{Roussy:2022cmp} (gray) and the SM Higgs invisible decay $h \to \varphi_i \varphi_j$, which leads to the exclusion of $m_{\varphi_1} < m_h/2$ (brown). 

The value of the electron EDM $d_e$ is obtained from the general expression in Eq.~\eqref{eq:deBZresGen} by omitting $k \cdot q$ and $k \cdot p_{1, 2}$ terms which give small corrections of $\mathcal{O} \left( m_e^2 / v_{\rm SM}^2 \right)$. 
We take account of the BZ diagrams with the 3rd generation fermions and the $W$ boson in the inner loop. 

As expected from Table~\ref{tab:xif}, Type-II and Type-X 2HDMs have enough $\tan \beta$ enhancement from the electron-neutral scalar couplings to reach the current upper bound on $|d_e|$, while Type-I and Type-Y 2HDMs predict several orders of magnitude smaller $|d_e|$, due to the $\tan \beta$ suppression from those couplings. 
We can see from Fig.~\ref{fig:resScl} that Type-Y 2HDM predicts slightly larger $d_e$ than Type-I 2HDM, due to the enhancement from the bottom quark couplings to the neutral scalars. 
In addition, Type-II and Type-X 2HDM cases exhibit similar predictions for $|d_e|$ because the primary contribution to $|d_e|$ arises from the top quark and $W$ boson loop processes which are the same for both cases. 
In the case of a large $\tan \beta$, the value of $\lambda_{1 S}$ becomes irrelevant for $|d_e|$, while the values of $\lambda_{2 S}$ and $\lambda_{1 2 S}^-$ are crucial to enhance $|d_e|$. 
The value of $|d_e|$ is also enhanced by the hierarchy between $m_{\varphi_1}$ and $m_{\varphi_2}$, as observed in Fig.~\ref{fig:resScl}. 
This enhancement occurs because $\hat{\Sigma}_{h A (H A)} (k^2)$ are amplified when $m_{\varphi_2} \gg m_{\varphi_1}$, owing to the logarithmic terms involving the ratio of their masses, which can be achieved by choosing $\overline{m}_S^2 \simeq |\overline{B}_S|$ (see Eq.~\eqref{eq:mvphi2}). 
As a result, we obtain a sizable electron EDM for Type-II and Type-X 2HDMs: for instance, by choosing $(m_{\varphi_1}, m_{\varphi_2}) = (300 ~ {\rm GeV}, 1300 ~ {\rm GeV})$, $|d_e| \sim 4 \times 10^{-30} \, e \, {\rm cm}$ is predicted. 
Even for the case of $\Delta m_{\varphi} = 10$ GeV, a large $|d_e|$ very close to the current upper limit can be obtained for $m_{\varphi_1} \simeq 100$ GeV. 
Note that the behavior of $d_e$ around $m_{\varphi_i} \simeq m_{h, H, A} / 2$ is originated from the function ${\rm Re} \bigl\{ {\rm DiscB}(m_{h, H, A}^2, m_{\varphi_i}, m_{\varphi_i}) \bigr\}$, which has following properties:
\begin{itemize}
\item[$\bullet$] ${\rm Re} \bigl\{ {\rm DiscB}(m^2, m_{\varphi_i}, m_{\varphi_i}) \bigr\}$ is always negative, except for $m_{\varphi_i} = m / 2$. 
\item[$\bullet$] For $m_{\varphi_i} = m / 2$, ${\rm Re} \bigl\{ {\rm DiscB}(m^2, m_{\varphi_i}, m_{\varphi_i}) \bigr\} = 0$ and non-differentiable with respect to $m$ or $m_{\varphi_i}$. 
\item[$\bullet$] ${\rm Re} \bigl\{ {\rm DiscB}(m^2, m_{\varphi_i}, m_{\varphi_i}) \bigr\} \simeq -2$ for $m \ll m_{\varphi_i}$. 
\end{itemize}
For our choice of $\theta_s = \pi / 4$, terms of ${\rm Re} \bigl\{ {\rm DiscB}(m_{h, H, A}^2, m_{\varphi_1}, m_{\varphi_2}) \bigr\}$ are dropped, because these are proportional to $s_{2 \theta_s} c_{2 \theta_s}$ (see Eqs.~\eqref{eq:FDShScl} and \eqref{eq:FDSHScl}).

\subsubsection{Discussion on the HDS approximation}
\label{sec:resHDSapproxScl}

\begin{figure}[t]
\centering
\includegraphics[bb=0 0 540 369, width=0.6\linewidth]{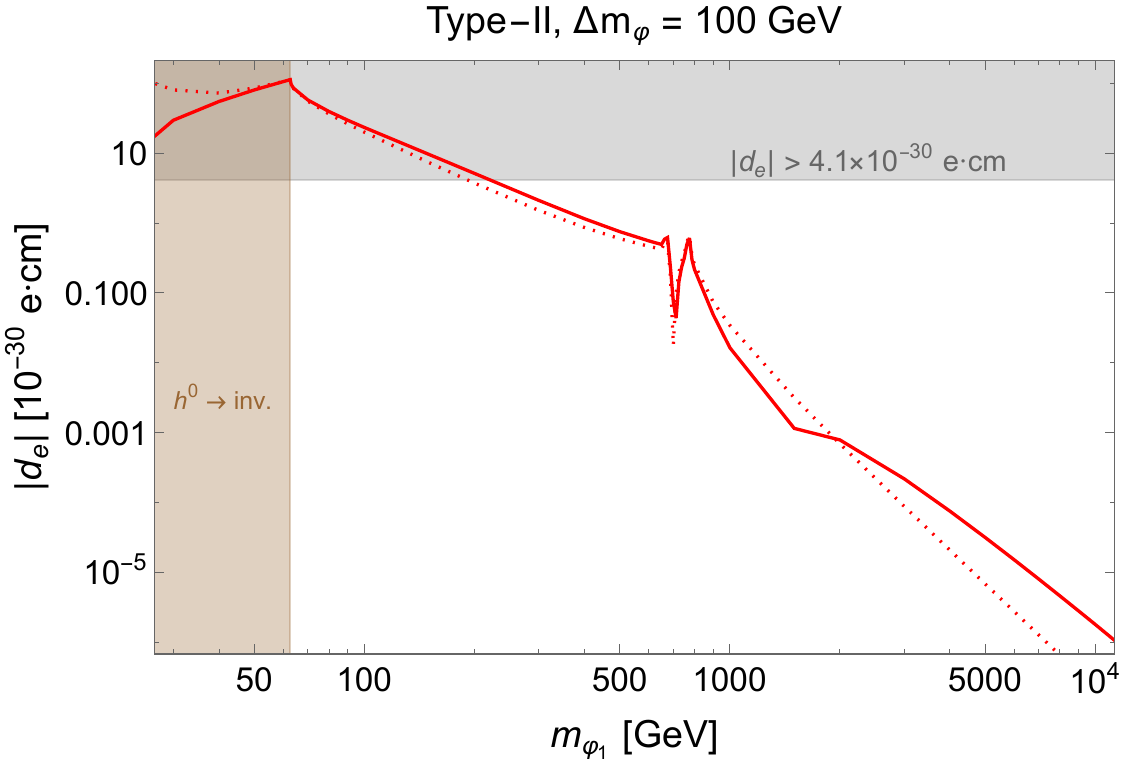}
\caption{Numerical results of the electron EDM with (dotted line) and without (solid line) the HDS approximation in the scalar dark sector model.}
\label{fig:resSclHDS}
\end{figure}

If $m_{\varphi_{1, 2}}$ are much heavier than all neutral scalars of the 2HDM sector, the electron EDM is given by the HDS approximation shown in Eq.~\eqref{eq:deBZresHDS} along with Eqs.~\eqref{eq:a0hScl}-\eqref{eq:a1HScl}. 
Fig.~\ref{fig:resSclHDS} shows the numerical estimates of the electron EDM in Type-II 2HDM for $\Delta m_{\varphi} = 100$ GeV, with (red dotted) and without (solid) the HDS approximation. 
Similar behaviors can be seen for other types and/or other $\Delta m_{\varphi}$ cases. 
We can see from the figure that for large $m_{\varphi_1}$, both curves decrease monotonically. 
However, the values of $|d_e|$ are not close to each other. 
Moreover, even for the mass of $m_{\varphi_1} = \mathcal{O}(v_{\rm SM})$, the HDS approximation looks good. 
To explain these reasons, we first parameterize $d_e$ as
\begin{align}
d_e = d_e^{\rm Org.} + d_e^{\rm Ren.} \, , \label{eq:deparam}
\end{align}
where $d_e^{\rm Org.}$ and $d_e^{\rm Ren.}$ denote the finite part of $F_{\rm DS}^{h A, H A}$ and
the finite part of terms induced by renormalization for $F_{\rm DS}^{h A, H A}$ in Eqs.~\eqref{eq:SigmaDShAScl} and \eqref{eq:SigmaDSHAScl}, whose expressions are summarized in appendix~\ref{app:decomposeEDM}. 
For the HDS approximation result, we add the superscript ``HDS",
\begin{align}
d_e^{\rm HDS} = d_e^{\rm Org., HDS} + d_e^{\rm Ren., HDS} \, . \label{eq:deparamHDS}
\end{align}
Note that $d_e^{\rm Ren.}$ and $d_e^{\rm Ren., HDS}$ are the same at the analytical level. 
Then, the difference between the results with and without the HDS approximation comes from the accuracy of calculations in $d_e^{\rm Org.}$ and $d_e^{\rm Org., HDS}$. 

\begin{figure}[t]
\centering
\includegraphics[bb=0 0 540 366, width=0.48\linewidth]{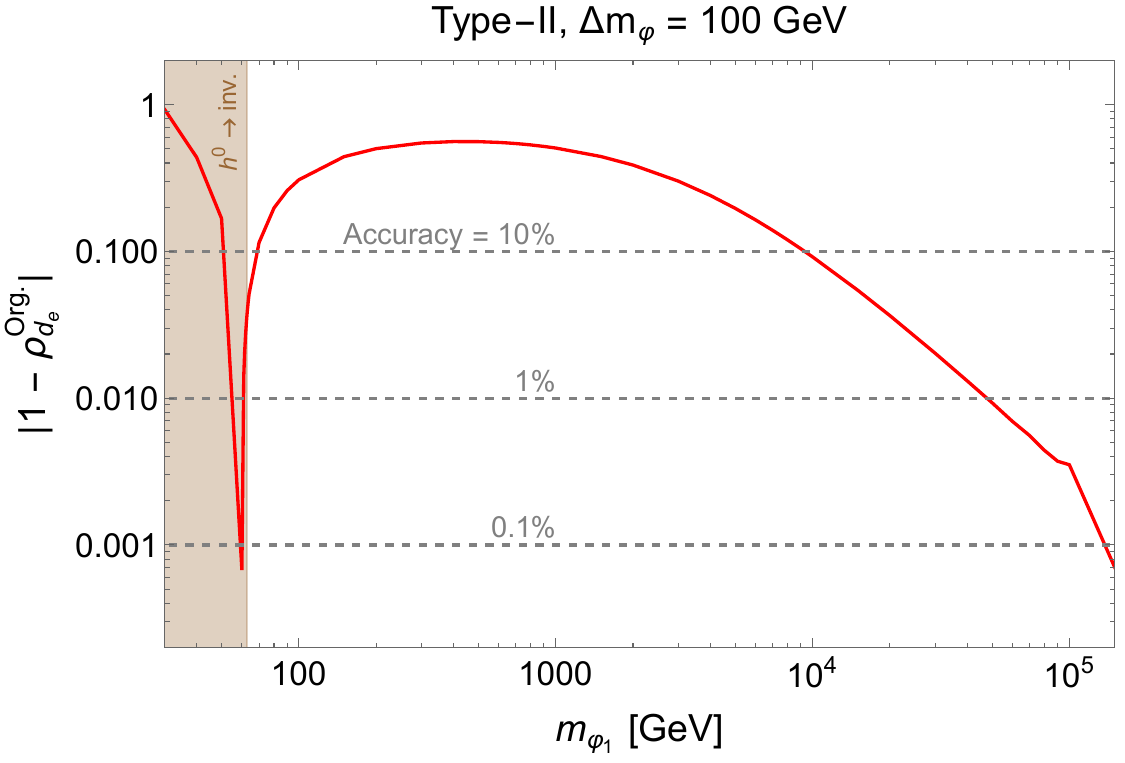}
\hfill
\includegraphics[bb=0 0 540 369, width=0.48\linewidth]{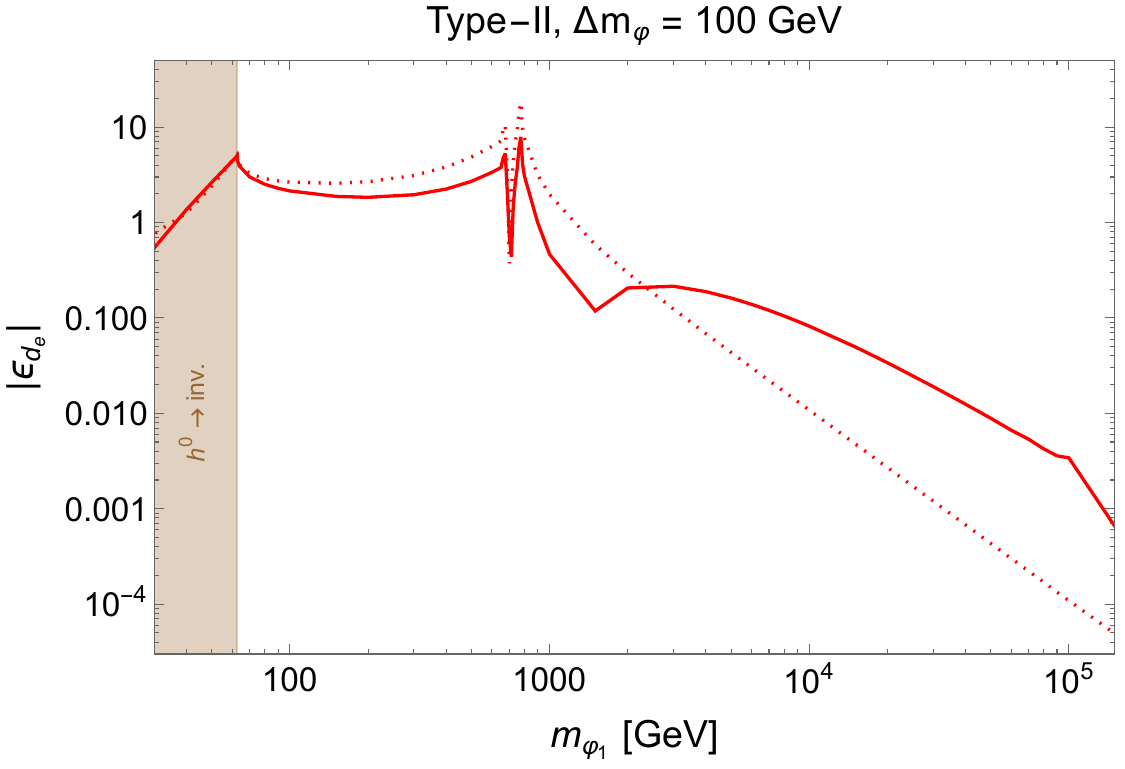}
\caption{The left panel indicates the accuracy of the HDS approximation with $\rho_{d_e}^{\rm Org.} \equiv d_e^{\rm Org., HDS} / d_e^{\rm Org.}$, while the right panel shows the degree of cancellation in $d_e$ (solid line) and $d_e^{\rm HDS}$ (dotted line) with $\epsilon_{d_e} \equiv d_e / d_e^{\rm Org.}$ and $\epsilon_{d_e^{\rm HDS}} \equiv d_e^{\rm HDS} / d_e^{\rm Org., HDS}$. 
Here, we focus on the Type-II 2HDM with $\Delta m_{\varphi} = 100$ GeV.}
\label{fig:AuxresSclHDS}
\end{figure}

To quantify the accuracy of the HDS approximation, we now define 
\begin{align}
\rho_{d_e}^{\rm Org.} \equiv \frac{d_e^{\rm Org., HDS}}{d_e^{\rm Org.}} \, , \label{eq:rhodeDef}
\end{align}
and plot $|1-\rho_{d_e}^{\rm Org.}|$ in the left panel of Fig.~\ref{fig:AuxresSclHDS}.
We can see that the HDS approximation agrees within 1\% when $m_{\varphi_1} \gtrsim 4.5 \times 10^4$ GeV. 
However, as parameterized in Eq.~\eqref{eq:deparam}, the physical prediction is the sum of $d_e^{\rm Org.}$ and $d_e^{\rm Ren.}$, and therefore, a cancellation between them may affect the accuracy of the physical results. 
The right panel of Fig.~\ref{fig:AuxresSclHDS} shows this feature, by defining
\begin{align}
\epsilon_{d_e} \equiv \frac{d_e}{d_e^{\rm Org.}} \, , \qquad \epsilon_{d_e^{\rm HDS}} \equiv \frac{d_e^{\rm HDS}}{d_e^{\rm Org., HDS}} \, . \label{eq:epdeDef}
\end{align}
From this plot, we can understand that $|\epsilon_{d_e}|$ for the HDS approximation is much smaller than that for the result without the HDS approximation for the heavy $m_{\varphi_1}$ region. 
This means that the cancellation in the HDS approximation is strong, and hence, $d_e^{\rm HDS}$ is predicted to be smaller than $d_e$, as one can see from Fig.~\ref{fig:resSclHDS}. 
To improve the accuracy of $d_e$, we need to take more and more higher accuracy in the numerical integration in $d_e^{\rm Org.}$, but it takes lots of calculation costs. 
We would like to emphasize that for such a heavy mass range, the prediction is far below the current upper limit, and hence, we do not need to improve the accuracy of the calculation unless the electron EDM bound is significantly improved. 

Accidentally, the HDS approximation is good even for $m_{\varphi_1} = \mathcal{O}(v_{\rm SM})$. 
This is because around this mass range, the dominant contribution to $d_e$ and $d_e^{\rm HDS}$ is coming from $d_e^{\rm Ren.}$ which is unchanged with and without the HDS approximation. 
In the right panel of Fig.~\ref{fig:AuxresSclHDS}, we can see this feature directly: $|\epsilon_{d_e^{\rm (HDS)}}| > 1$ means that $d_e > d_e^{\rm Org.}$ ($d_e^{\rm HDS} > d_e^{\rm Org., HDS}$), and hence, $d_e \sim d_e^{\rm Ren.}$ ($d_e^{\rm HDS} \sim d_e^{\rm Ren., HDS} = d_e^{\rm Ren.}$).

\subsection{Fermion dark sector model}
\label{sec:CPVinfermion}

Let us consider a dark sector model which contains a vector-like singlet fermion $\psi_{L, R}$ 
and an SM singlet complex scalar $S$. The dark sector Lagrangian and the scalar potential include the following terms:
\begin{align}
\mathcal{L}_{\rm DS} &= - y_{\psi} \overline{\psi}_L \psi_R S - m_{\psi} \overline{\psi}_L \psi_R + {\rm h.c.} \, , \\[0.6ex]
V_{\rm scl} &\supset V_H + m_S^2 |S|^2 + \Bigl[ \mu_{1 S} H_1^{\dagger} H_1 S + \mu_{2 S} H_2^{\dagger} H_2 S + \mu_{1 2 S} H_1^{\dagger} H_2 S + \mu_{2 1 S} H_2^{\dagger} H_1 S + {\rm h.c.} \Bigr] \, , \label{eq:VsFrmDS}
\end{align}
where $V_H$ represents the scalar potential of the 2HDM sector given in Eq.~\eqref{eq:V2HDM}.\footnote{
{For the scalar potential, we have omitted some terms that are allowed by the charge assignment in Table~\ref{tab:Z2charges}, because these terms are irrelevant to our analysis for the electron EDM. 
The full scalar potential and its analysis are shown in appendix~\ref{app:VanalysisFermion}.}}
We assume a $Z_2$ symmetry in which $H_1$ and $\psi_{L, R}$ are odd.
The charge assignment is summarized in Table~\ref{tab:Z2charges}. 
{According to this $Z_2$ assignment, $\mu_{1 2 S}$ and $\mu_{2 1 S}$ terms are soft $Z_2$ breaking terms.} 
It should be noted that $\mu_{1 2 S}$ and $\mu_{2 1 S}$ terms are necessary for the mixing among CP-odd components of $H_{1, 2}$ and $S$ (see appendix~\ref{app:VanalysisFermion} for details). 
Similar to section~\ref{sec:CPVinscalar}, we parameterize $S$ as $S = (s_1^0 + i s_2^0) / \sqrt{2}$ and assume that $S$ does not acquire a nonzero VEV. 

\begin{table}[!t]
\centering
\begin{tabular}{|ccc|ccccc|cc|}
\hline
$H_1$ & $H_2$ & $S$ & $Q_L$ & $u_R$ & $d_R$ & $L_L$ & $e_R$ & $\psi_L$ & $\psi_R$ \\ \hline
$-$ & $+$ & $+$ & $+$ & $+$ & $z_2^d$ & $+$ & $z_2^e$ & $-$ & $-$ \\ \hline
\end{tabular}
\caption{The $Z_2$ charge assignment for the fermion dark sector model where $z_2^d$ and $z_2^e$ depend on the type of 2HDM.}
\label{tab:Z2charges}
\end{table}

The model incorporates numerous complex parameters, encompassing 13 invariant phases as detailed in appendix~\ref{app:CPVFermionDS}. 
Adhering to the presence of CP violation solely within the dark sector, we select the following invariant phase as non-zero, contributing to the electron EDM:
\begin{align}
\theta_{\rm phys} = \arg \left[ \mu_{1 S}^* y_{\psi} \right] \, , \label{eq:fermionDMphysicalphase}
\end{align}
while assuming the other phases to be zero. 
Consequently, if we choose the phase of $S$ to have real $\mu_{1 S}$, then the only CP violation happens exclusively within the Yukawa interaction of the dark sector. 
It is pertinent to note that a chiral phase rotation of the dark fermion $\psi_{L, R}$ has already been applied to render the mass parameter $m_{\psi}$ real. 
Hence, we choose $y_{\psi}$ to be complex and parameterize it by using two nonzero real parameters, $y_{\psi}^{R}$ and $y_{\psi}^{I}$, as $y_{\psi} = y_{\psi}^R + i y_{\psi}^I$. 
Defining $\theta_{\psi} \equiv \arg (y_{\psi})$, we also have $y_{\psi}^R = y_{\psi} \cos \theta_{\psi}$ and $y_{\psi}^I = y_{\psi} \sin \theta_{\psi}$, and $\theta_{\psi}$ corresponds to the physical CP phase. 
Consequently, we have explicit CP violation in the dark sector Yukawa couplings:
\begin{align}
\mathcal{L}_{\slashed{\rm CP}} &= - \frac{\left(y_{\psi}^R s_1^0 - y_{\psi}^I s_2^0 \right)}{\sqrt{2}} \overline{\psi} \psi - \frac{\left(y_{\psi}^I s_1^0  + y_{\psi}^R s_2^0\right)}{\sqrt{2}} \overline{\psi} i \gamma^5 \psi - m_{\psi} \overline{\psi} \psi \, . \label{eq:FDSM-LCPV}
\end{align}

The CP-even scalars $(H_1^0, H_2^0, s_1^0)$ and the CP-odd scalars $(A_1^0, A_2^0, s_2^0)$
are respectively expressed in terms of CP-even mass eigenstates $\phi_{1, 2, 3}^0$
and CP-odd mass eigenstates $\eta_{1, 2, 3}^0$ as
\begin{align}
H_1^0 &= \sum_{j = 1}^3 (R_E)_{1 j} \phi_j^0 \, , \quad H_2^0 = \sum_{j = 1}^3 (R_E)_{2 j} \phi_j^0 \, , \quad s_1^0 = \sum_{j = 1}^3 (R_E)_{3 j} \phi_j^0 \, , \\[0.5ex]
A_1^0 &= \sum_{a = 1}^3 (R_O)_{1 a} \eta_a^0 \, , \quad A_2^0 = \sum_{a = 1}^3 (R_O)_{2 a} \eta_a^0 \, , \quad ~ s_2^0 = \sum_{a = 1}^3 (R_O)_{3 a} \eta_a^0 \, .
\end{align}
Here, we choose $\eta_1^0$ to be the neutral NG mode as mentioned in section~\ref{sec:CPVdiagrams}, and hence, we can parameterize $R_O$ by
\begin{align}
R_O = \begin{pmatrix}
c_{\beta} & - s_{\beta} c_{\theta_O} & s_{\beta} s_{\theta_O} \\
s_{\beta} & c_{\beta} c_{\theta_O} & - c_{\beta} s_{\theta_O} \\
0 & s_{\theta_O} & c_{\theta_O}
\end{pmatrix} \, , \label{eq:ROmat}
\end{align}
where $\theta_O$ is a mixing angle between $A^0$ in the 2HDM sector and $s_2^0$. 
On the other hand, $R_E$ is parameterized by three mixing angles $\alpha, \alpha_{h S}, \alpha_{H S}$ as
\begin{align}
R_E = \begin{pmatrix}
c_{\alpha} c_{\alpha_{h S}} & - s_{\alpha} c_{\alpha_{H S}} - c_{\alpha} s_{\alpha_{h S}} s_{\alpha_{H S}} & - c_{\alpha} s_{\alpha_{h S}} c_{\alpha_{H S}} + s_{\alpha} s_{\alpha_{H S}} \\
s_{\alpha} c_{\alpha_{h S}} & c_{\alpha} c_{\alpha_{H S}} - s_{\alpha} s_{\alpha_{h S}} s_{\alpha_{H S}} & - s_{\alpha} s_{\alpha_{h S}} c_{\alpha_{H S}} - c_{\alpha} s_{\alpha_{H S}} \\
s_{\alpha_{h S}} & c_{\alpha_{h S}} s_{\alpha_{H S}} & c_{\alpha_{h S}} c_{\alpha_{H S}}
\end{pmatrix} \, . \label{eq:REmat}
\end{align}
With these mixing angles, we can rewrite the explicit CP-violating couplings in Eq.~\eqref{eq:FDSM-LCPV}
in terms of the scalar mass eigenstates,
\begin{align}
\mathcal{L}_{\slashed{\rm CP}}^{\rm Yuk}= - \frac{1}{\sqrt{2}} \sum_{j = 1}^3 \overline{\psi} \left( \hat{y}_{\psi, j}^R + i \hat{y}_{\psi, j}^I \gamma^5 \right) \psi \phi_j^0 + \frac{1}{\sqrt{2}} \sum_{a = 2}^3 \overline{\psi} \left( \tilde{y}_{\psi, a}^I - i \tilde{y}_{\psi, a}^R \gamma^5 \right) \psi \eta_a^0 \, , \label{eq:LDSYukawa}
\end{align}
where we have defined $\hat{y}_{\psi, j}^{R, I} \equiv y_{\psi}^{R, I} (R_E)_{3 j}$ and $\tilde{y}_{\psi, a}^{R, I} \equiv y_{\psi}^{R, I} (R_O)_{3 a}$. 
The SM fermion Yukawa couplings and the $W$ boson couplings are also rewritten as
\begin{align}
\mathcal{L}_{\rm Yuk} &\supset - \sum_{f = u, d, e} \frac{m_f}{v_{\rm SM}} \left[ \sum_{j = 1}^3 \xi_{\phi_j}^f \bar{f} f \phi_j - \sum_{a = 2}^3 \xi_{\eta_a}^f \bar{f} i \gamma^5 f \eta_a \right] \, , \\[0.5ex]
\mathcal{L} &\supset \sum_{j = 1}^3 \frac{2 m_W^2}{v_{\rm SM}} \xi_{\phi_j}^W W_{\mu}^+ W^{- \, \mu} \phi_j \, .
\end{align}
Here, $\xi_{\phi_j}^f$ and $\xi_{\eta_a}^f$ are summarized in Table~\ref{tab:SMYukawaFrm}, and $\xi_{\phi_j}^W \equiv c_{\beta} (R_E)_{1 j} + s_{\beta} (R_E)_{2 j}$. 

\begin{table}[t]
\centering
\scalebox{0.9}[0.9]{
\begin{tabular}{|c||cc|cc|cc|}
\hline
& $\xi_{\phi_j}^u$ & $\xi_{\eta_a}^u$ & $\xi_{\phi_j}^d$ & $\xi_{\eta_a}^d$ & $\xi_{\phi_j}^e$ & $\xi_{\eta_a}^e$ \\ \hline
Type-I & $(R_E)_{2 j} / s_{\beta}$ & $(R_O)_{2 a} \cot \beta$ & $(R_E)_{2 j} / s_{\beta}$ & $- (R_O)_{2 a} \cot \beta$ & $(R_E)_{2 j} / s_{\beta}$ & $- (R_O)_{2 a} \cot \beta$ \\
Type-II & $(R_E)_{2 j} / s_{\beta}$ & $(R_O)_{2 a} \cot \beta$ & $(R_E)_{1 j} / c_{\beta}$ & $- (R_O)_{1 a} \tan \beta$ & $(R_E)_{1 j} / c_{\beta}$ & $- (R_O)_{1 a} \tan \beta$ \\
Type-X & $(R_E)_{2 j} / s_{\beta}$ & $(R_O)_{2 a} \cot \beta$ & $(R_E)_{2 j} / s_{\beta}$ & $- (R_O)_{2 a} \cot \beta$ & $(R_E)_{1 j} / c_{\beta}$ & $- (R_O)_{1 a} \tan \beta$ \\
Type-Y & $(R_E)_{2 j} / s_{\beta}$ & $(R_O)_{2 a} \cot \beta$ & $(R_E)_{1 j} / c_{\beta}$ & $- (R_O)_{1 a} \tan \beta$ & $(R_E)_{2 j} / s_{\beta}$ & $- (R_O)_{2 a} \cot \beta$ \\ \hline
\end{tabular}}
\caption{The coefficients $\xi_{\phi_j}^f$ and $\xi_{\eta_a}^f$ for SM fermion Yukawa couplings in the fermion dark sector model.}
\label{tab:SMYukawaFrm}
\end{table}

With new Yukawa couplings for $\psi$ in Eq.~\eqref{eq:LDSYukawa}, the CP-violating contribution to the CP-even scalar $\phi_j$ and CP-odd scalar $\eta_a$ mixing diagrams in Eq.~\eqref{eq:effmix} can be calculated at one-loop by running the vector-like fermion,
\begin{align}
F_{\rm DS}^{\phi_j \eta_a} (k^2) &= N_c^{\psi} \frac{\hat{y}_{\psi, j}^R \tilde{y}_{\psi, a}^I}{8 \pi^2} m_{\psi}^2 \left( \Lambda_{\rm UV}^2 - \int_0^1 \! dx \ln \frac{m_{\psi}^2 - x (1 - x) k^2}{\mu^2} \right) , \label{eq:FDSja} \\
&= N_c^{\psi} \frac{y_{\psi}^R y_{\psi}^I}{8 \pi^2} m_{\psi}^2 (R_E)_{3 j} (R_O)_{3 a} \left( \Lambda_{\rm UV}^2 + 2 + {\rm DiscB}(k^2, m_{\psi}, m_{\psi}) + \ln \frac{\mu^2}{m_{\psi}^2} \right) \, , \nonumber
\end{align}
where $\Lambda_{\rm UV}$ denotes the UV divergent parameter, $\mu$ is the 't Hooft mass parameter and $N_c^{\psi}$ represents some new color factor for the vector-like fermion $\psi$. 
In the present study, we set $N_c^{\psi} = 1$. 

Similar to the scalar dark sector model, we need to properly handle the UV divergent part. 
In the current case, we consider the wave function renormalization for the $\phi_j$-$\eta_a$ self-energy~\cite{Altenkamp:2017ldc, Denner:1991kt}. 
Then, the renormalized two-point function $\hat{\Sigma}_{\phi_j \eta_a} (k^2)$ is defined as
\begin{align}
\hat{\Sigma}_{\phi_j \eta_a} (k^2) \equiv F_{\rm DS}^{\phi_j \eta_a} (k^2) + \frac{1}{2} (k^2 - m_{\phi_j}^2) \delta Z_{\phi_j \eta_a} + \frac{1}{2} (k^2 - m_{\eta_a}^2) \delta Z_{\eta_a \phi_j} \, ,
\end{align}
where no summations over $j$ and $a$ should be understood. 
In the on-shell subtraction scheme, we can determine $\delta Z_{\phi_j \eta_a}$ and $\delta Z_{\eta_a \phi_j}$ by
\begin{align}
{\rm Re} \bigl\{ \hat{\Sigma}_{\phi_j \eta_a} (m_{\phi_j}^2) \bigr\} = 0 \, , \qquad  {\rm Re} \bigl\{ \hat{\Sigma}_{\phi_j \eta_a} (m_{\eta_a}^2) \bigr\} = 0 \, .
\end{align}
By solving these equations, we can obtain the renormalized two-point function $\hat{\Sigma}_{\phi_j \eta_a} (k^2)$ as
\begin{align}
\hat{\Sigma}_{\phi_j \eta_a} (k^2) = \frac{y_{\psi}^R y_{\psi}^I m_{\psi}^2 (R_E)_{3 j} (R_O)_{3 a}}{8 \pi^2 (m_{\phi_j}^2 - m_{\eta_a}^2)} &\Bigl[ \Bigr. (m_{\phi_j}^2 - m_{\eta_a}^2) {\rm DiscB}(k^2, m_{\psi}, m_{\psi}) \nonumber \\[0.5ex]
&+ (m_{\eta_a}^2 - k^2) {\rm Re} \bigl\{ {\rm DiscB}(m_{\phi_j}^2, m_{\psi}, m_{\psi}) \bigr\} \nonumber \\[0.5ex]
&+ (k^2 - m_{\phi_j}^2) {\rm Re} \bigl\{ {\rm DiscB}(m_{\eta_a}^2, m_{\psi}, m_{\psi}) \bigr\} \Bigl. \Bigr] \, , \label{eq:Sighatja}
\end{align}
where the divergent part is properly canceled. 

In the HDS approximation, we can further expand $\hat{\Sigma}_{\phi_j \eta_a} (k^2)$ as
\begin{align}
\hat{\Sigma}_{\phi_j \eta_a} (k^2) = a_0^{\phi_j \eta_a} + a_1^{\phi_j \eta_a} \frac{k^2}{m_{\psi}^2} + \mathcal{O} \left(  \frac{k^4}{m_{\psi}^4} \right) \, . \label{eq:HDSexpandFrm}
\end{align}
Here, $M^2$ in Eq.~\eqref{eq:HDSexpand} corresponds to the vector-like fermion mass $m_{\psi}^2$, and the coefficients are
\begin{align}
a_0^{\phi_j \eta_a} (m_{\psi}) &= \frac{y_{\psi}^R y_{\psi}^I m_{\psi}^2 (R_E)_{3 j} (R_O)_{3 a}}{8 \pi^2 (m_{\phi_j}^2 - m_{\eta_a}^2)} \left[ - 2 (m_{\phi_j}^2 - m_{\eta_a}^2) - m_{\phi_j}^2 {\rm Re} \bigl\{ {\rm DiscB}(m_{\eta_a}^2, m_{\psi}, m_{\psi}) \bigr\} \right. \nonumber \\[0.3ex]
&\hspace{13.0em} \left. + m_{\eta_a}^2 {\rm Re} \bigl\{ {\rm DiscB}(m_{\phi_j}^2, m_{\psi}, m_{\psi}) \bigr\} \right] \, , \label{eq:a0jaFrm} \\[0.5ex]
a_1^{\phi_j \eta_a} (m_{\psi}) &= \frac{y_{\psi}^R y_{\psi}^I m_{\psi}^2 (R_E)_{3 j} (R_O)_{3 a}}{8 \pi^2 (m_{\phi_j}^2 - m_{\eta_a}^2)} \Biggl[ \Biggr. \frac{m_{\phi_j}^2 - m_{\eta_a}^2}{6} - m_{\psi}^2 {\rm Re} \bigl\{ {\rm DiscB}(m_{\phi_j}^2, m_{\psi}, m_{\psi}) \bigr\} \nonumber \\[0.3ex]
&\hspace{13.0em} + m_{\psi}^2 {\rm Re} \bigl\{ {\rm DiscB}(m_{\eta_a}^2, m_{\psi}, m_{\psi}) \bigr\} \Biggl. \Biggr] \, . \label{eq:a1jaFrm}
\end{align}
As in the case of the scalar dark sector model, $a_0^{\phi_j \eta_a} (m_{\psi})$ is used for the calculation of
the four-fermion operator coefficients $C_{f f'}$. 

\subsubsection{Numerical results}
\label{sec:resFrm}

Similar to the scalar dark sector model, to numerically estimate the size of the electron EDM generated in the current model,
we choose the following masses and parameters:
\begin{align}
&m_{\phi_1} = 125.25 ~ {\rm GeV}  , ~ m_{\phi_2} = 1500 ~ {\rm GeV}  , ~ m_{\phi_3} = 1600 ~ {\rm GeV}  , ~ \alpha = \frac{\pi}{3}  , ~ \alpha_{h S} = 0.01  , ~ \alpha_{H S} = \frac{\pi}{6}  , \nonumber \\[0.5ex]
&m_{\eta_2} = 1550 ~ {\rm GeV}  , ~ m_{\eta_3} = 1650 ~ {\rm GeV}  , ~ \tan \beta = 50  , ~ \theta_O = \frac{\pi}{4}  , ~~ y_{\psi}^R = y_{\psi}^I = \frac{1}{\sqrt{2}}  .
\label{eq:parametersFrm}
\end{align}
The choices of $y_{\psi}^{R, I}$ can be obtained by $|y_{\psi}| = 1$ and $\theta_{\psi} = \pi / 4$, and the value of $\theta_{\psi}$ corresponds to the one which maximizes the $d_e$ with the assumption where all parameters in $V_{\rm scl}$ are real. 
We then calculate $|d_e|$ as a function of $m_{\psi}$. 
The $C_S$ contribution is found to be significantly smaller than the electron EDM contribution $d_e$, and hence neglected in the subsequent calculations. 

\begin{figure}[!t]
\centering
\includegraphics[bb=0 0 540 367, width=0.48\linewidth]{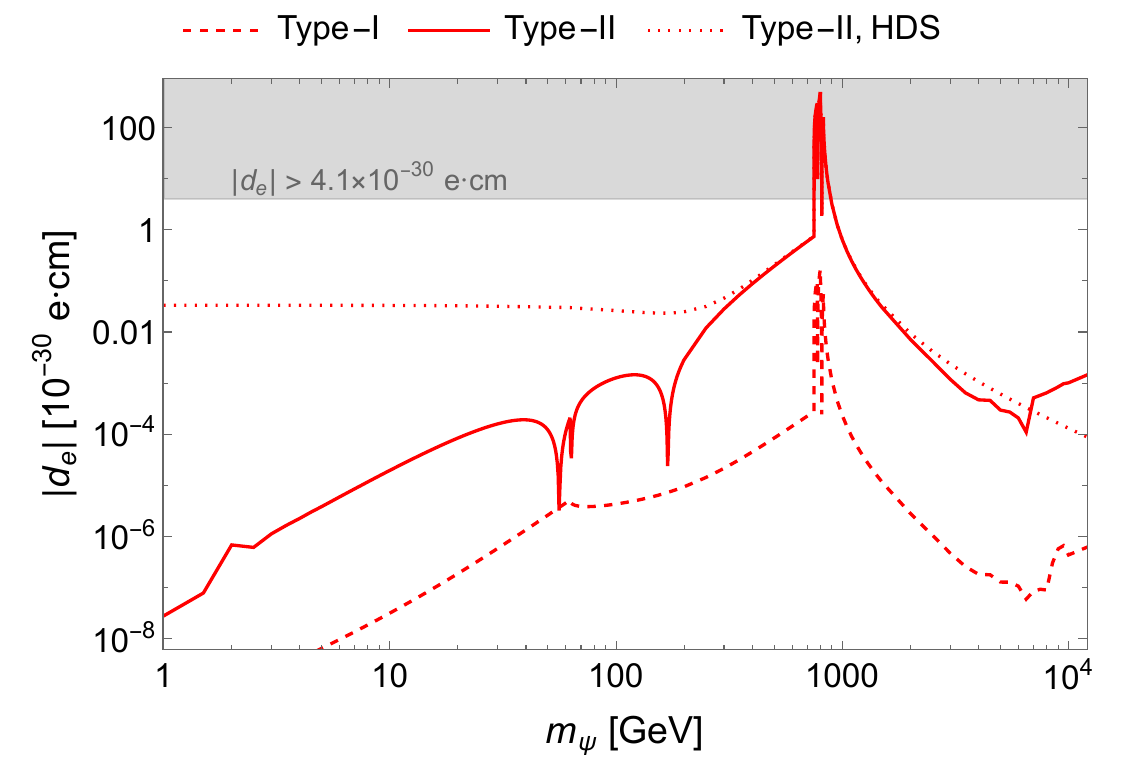}
\hfill
\includegraphics[bb=0 0 540 367, width=0.48\linewidth]{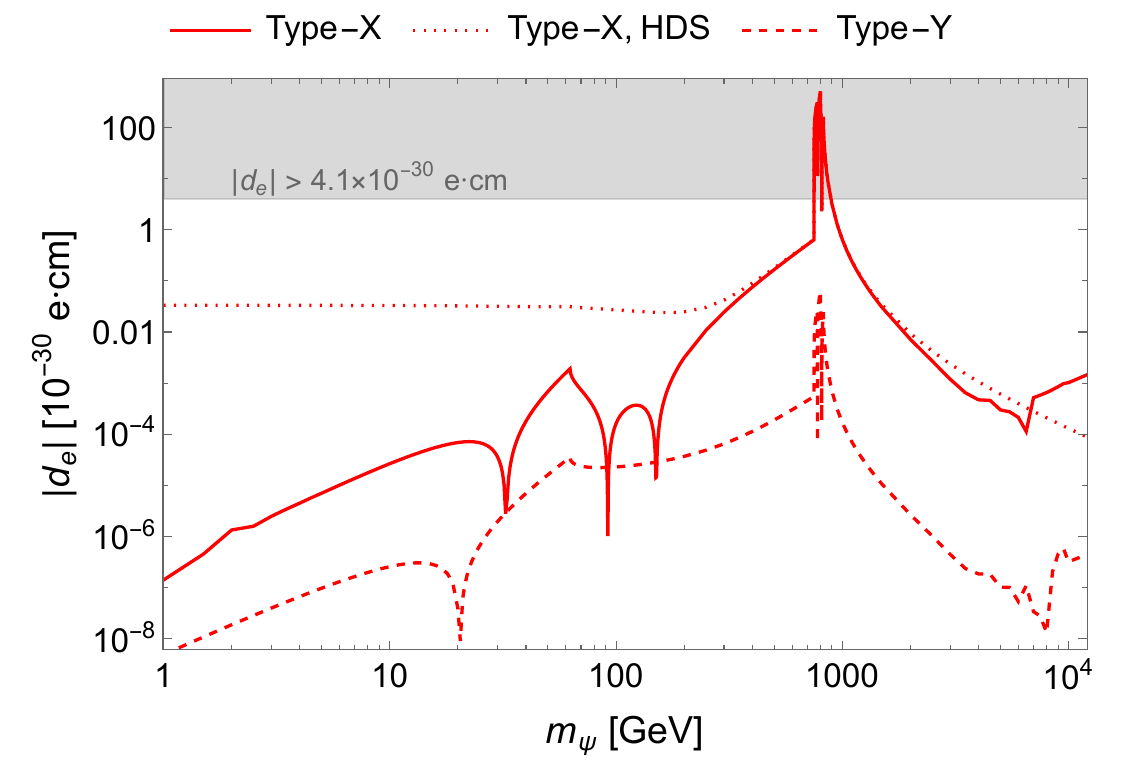}
\caption{Numerical results of the electron EDM in the fermion dark sector model. 
The left panel shows the cases of Type-I (dashed line) and Type-II (solid line) 2HDMs, while the right panel shows those of Type-X (solid line) and Type-Y (dashed line) 2HDMs. 
The results of the HDS approximation for Type-II and Type-X 2HDMs are also shown by dotted lines in each panel. 
The gray-shaded region is excluded by the current electron EDM bound from the JILA experiment~\cite{Roussy:2022cmp}. 
In this plot, we choose the parameters that satisfy the constraint of the SM Higgs invisible decay, $h^0 \to \psi \bar{\psi}$.}
\label{fig:resFrm}
\end{figure}

The left panel of Fig.~\ref{fig:resFrm} presents curves for $|d_e|$ in Type-I (dashed line) and Type-II (solid line) 2HDMs, and the right panel shows those for Type-X (solid line) and Type-Y (dashed line) 2HDMs. 
To get those results, we have used the general expression in Eq.~\eqref{eq:deBZresGen} by omitting $k \cdot q$ and $k \cdot p_{1, 2}$ and calculated contributions from the 3rd generation fermions ($t,b$) and the $W$ boson in the inner loop, as in the case of the scalar dark sector model. 
The gray-shaded region is excluded by the constraint from the JILA experiments~\cite{Roussy:2022cmp}. 
We may have a constraint from the SM Higgs invisible decay, $h^0 \to \psi \bar{\psi}$, especially if the vector-like fermion is light. 
The branching ratio is directly proportional to $|y_{\psi}|^2 \sin^2 \alpha_{h S}$, making $|y_{\psi}|$ and $\alpha_{h S}$ critical parameters. 
Larger values of $|y_{\psi}|$ and/or $\alpha_{h S}$ tend to exclude the lower mass region $m_{\psi}< m_h/2$. 
We have checked that the parameter choice in Eq.~\eqref{eq:parametersFrm} can avoid the constraint. 

The curves of the electron EDM $|d_e|$ in Fig.~\ref{fig:resFrm} show significant complexity, stemming from the $m_{\psi}$ dependence of the renormalized two-point function as given in Eq.~\eqref{eq:Sighatja}. 
For the region $m_{\psi} < 200$ GeV except for $m_{\psi} \simeq m_{\phi_1} / 2 = m_h / 2$, the sharp valleys is due to the sign flipping on $d_e$, while the behavior around $m_{\psi} \simeq m_{\phi_j} / 2, m_{\eta_a} / 2$ is originated from properties of functions ${\rm Re} \bigl\{ {\rm DiscB}(m_{\phi_j}^2, m_{\psi}, m_{\psi}) \bigr\}$ and ${\rm Re} \bigl\{ {\rm DiscB}(m_{\eta_a}^2, m_{\psi}, m_{\psi}) \bigr\}$. 
Notably, the region that $|d_e|$ is peaked corresponds to $m_{\psi} \simeq m_{\phi_j} / 2$ or $m_{\eta_a} / 2$, where $\hat{\Sigma}_{\phi_j \eta_a} (k^2)$ is enhanced. 
As in the case of the scalar dark sector model, Type-II and Type-X 2HDMs can reach the current upper limit in this mass region, while Type-I and Type-Y 2HDMs predict $|d_e|$ that is several orders of magnitude smaller than the limit. 
Moreover, $\hat{\Sigma}_{\phi_j \eta_a} (k^2)$ approaches to zero as $m_{\psi}$ goes to zero or infinity. 
Such a feature is visible in Fig.~\ref{fig:resFrm}. 
Note that for $m_{\psi} \gtrsim 7 \times 10^3$ GeV, the curves of $|d_e|$ change their behaviors, due to the low accuracy in the numerical integration, which is explained below. 

\subsubsection{Discussion on the HDS approximation}
\label{sec:resHDSapproxFrm}

Remarkably, the HDS approximation results which are represented by the dotted curves for the Type-II (left panel) and Type-X (right panel) 2HDMs in Fig.~\ref{fig:resFrm} align well with the results without using the approximation. 
Note that the result for the HDS approximation can be obtained from Eq.~\eqref{eq:deBZresHDS} with Eqs.~\eqref{eq:a0jaFrm} and \eqref{eq:a1jaFrm}. 

Compared with the scalar dark sector model, the HDS approximation is good, due to the simple expression for $\hat{\Sigma}_{\phi_j \eta_a}$ in Eq.~\eqref{eq:Sighatja} (see Eqs.~\eqref{eq:SigmaDShAScl} and \eqref{eq:SigmaDSHAScl} together with Eqs.~\eqref{eq:FDShScl} and \eqref{eq:FDSHScl} for the scalar dark sector model). 
Hence, we may be able to use the HDS approximation results to estimate the $d_e$ prediction in the heavy $m_{\psi}$ limit. 

\begin{figure}[t]
\centering
\includegraphics[bb=0 0 540 365, width=0.48\linewidth]{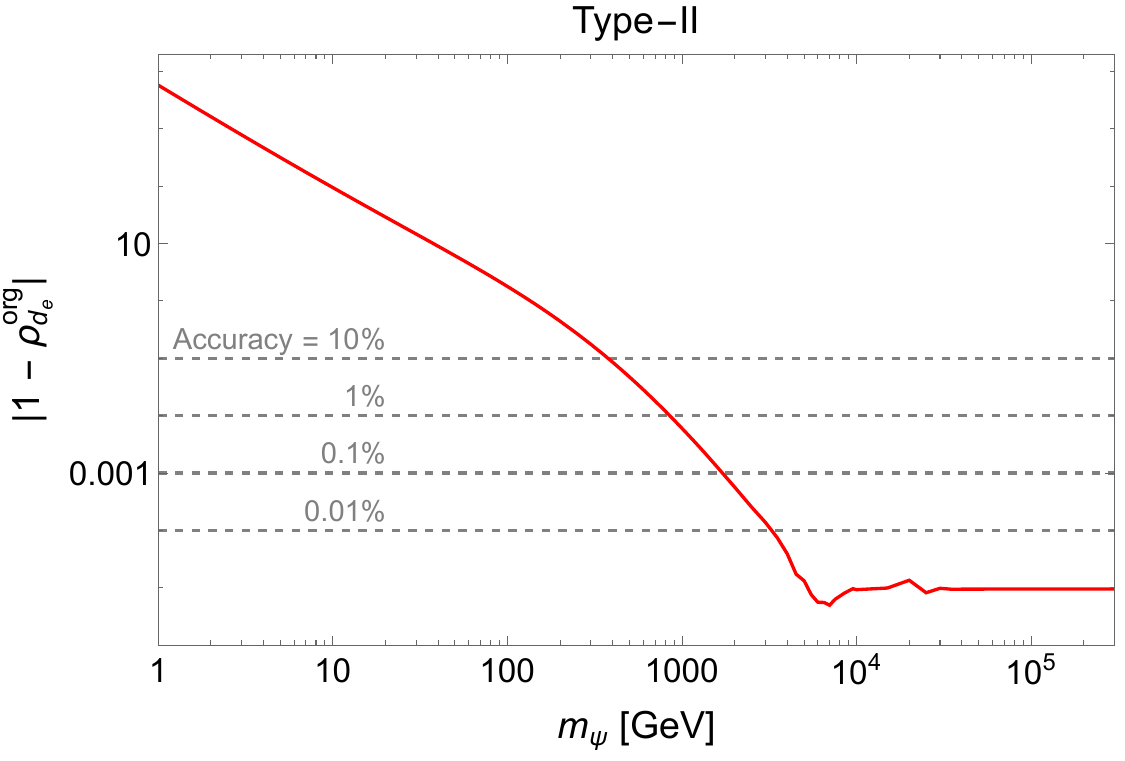}
\hfill
\includegraphics[bb=0 0 540 371, width=0.48\linewidth]{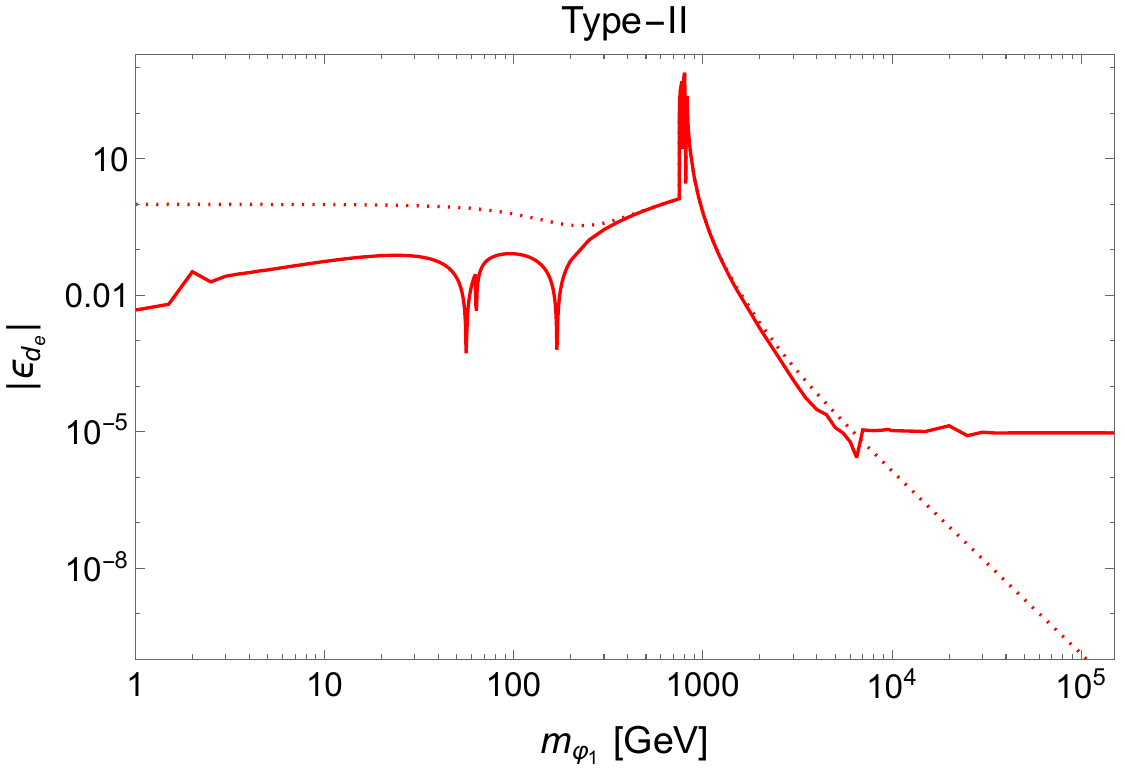}
\caption{Plots corresponding to Fig.~\ref{fig:AuxresSclHDS} for the fermion dark sector model.}
\label{fig:AuxresFrmHDS}
\end{figure}

Fig.~\ref{fig:AuxresFrmHDS} shows $| 1 - \rho_{d_e}^{\rm Org.} |$ and $|\epsilon_{d_e}|$ which are defined in Eqs.~\eqref{eq:rhodeDef} and \eqref{eq:epdeDef}, and the expressions for $d_e^{\rm Org.}$ and $d_e^{\rm Ren.}$ are summarized in appendix~\ref{app:decomposeEDM}. 
From the left panel of Fig.~\ref{fig:AuxresFrmHDS}, we can see that the HDS approximation does not give the correct results for a light $m_{\psi}$, as observed from Fig.~\ref{fig:resFrm}. 
On the other hand, when $m_{\psi} > 1000$ GeV, the accuracy of $d_e^{\rm Org.}$ becomes less than 1\%, and hence, the HDS approximation can be used to estimate the value of the electron EDM. 
However, for $m_{\psi} \gtrsim 7 \times 10^3$ GeV, the accuracy of $d_e^{\rm Org.}$ is saturated, and the curves with and without the HDS approximation deviate from each other (see Fig.~\ref{fig:resFrm}). 
For this mass range, we need a more accurate calculation for numerical integration, which is shown in the right panel of Fig.~\ref{fig:AuxresFrmHDS}. 
For $m_{\psi} \sim 7 \gtrsim 10^3$ GeV, the cancellation between $d_e^{\rm Org.}$ and $d_e^{\rm Ren.}$ is also saturated, at the level of $|\epsilon_{d_e}| \simeq 10^{-5}$, while that between $d_e^{\rm Org., HDS}$ and $d_e^{\rm Ren., HDS}$ decreases monotonically. 
Therefore, even if we obtain a good accuracy for $|1 - \rho_{d_e}^{\rm Org.}|$, the numerical results with and without the HDS approximation deviate for the mass range of $m_{\psi} \gtrsim 7 \times 10^3$ GeV. 
As a result, for such a heavy $m_{\psi}$, the correct prediction can be obtained from the HDS approximation, although the size is far below the current upper limit on $|d_e|$. 

Around $m_{\psi} \simeq m_{\phi_j} / 2$ and $m_{\eta_a} / 2$, it seems that the HDS approximation is quite good in Fig.~\ref{fig:resFrm}. 
The reason is the same as that of the scalar dark sector model: in this mass region, $d_e$ ($d_e^{\rm HDS}$) is dominated by $d_e^{\rm Ren.}$ ($d_e^{\rm Ren., HDS}$), as one can see from the right panel of Fig.~\ref{fig:AuxresFrmHDS}.

\section{Conclusion}
\label{sec:conclusion}

In the present paper, we have discussed the possibility that a CP-violating dark sector coupling to the 2HDM generates a detectable electron EDM. 
The 2HDM was assumed to preserve the CP symmetry at the beginning, otherwise, the model is severely constrained, while the dark sector contains CP violation which is mediated to the 2HDM sector by radiative corrections. 
We have presented a general form of the electron EDM induced by BZ diagrams shown in Fig.~\ref{fig:outerf}, which can be applied to a wide range of dark sector models with CP violation. 

For benchmark scenarios, we considered the scalar dark sector model and the fermion dark sector model. 
In the scalar dark sector model, CP violation arises in the scalar potential involving neutral Higgs and dark sector scalars. 
In the fermion dark sector model, a singlet complex scalar serves as a portal field, connecting a vector-like singlet fermion to the 2HDM through a complex CP-violating Yukawa coupling. 
For both scenarios, we carefully considered the two-point renormalized scalar self-energy functions to ensure finite results under the on-shell subtraction scheme. 
In each model, we found that Type-II and Type-X 2HDM cases exhibit similar predictions for $|d_e|$ and have a chance to reach the current upper limit on the electron EDM. 
Therefore, we can constrain the model parameter space: the mass scale of the dark sector and/or the size of effective mixing for CP-even and CP-odd scalars in the 2HDMs. 
The analytical expression of the electron EDM was given in the Heavy Dark Sector limit, and we have checked that it is valid to use the approximation for the estimation of the size of the electron EDM. 

Since the experimental sensitivity will be improved in the future~\cite{Alarcon:2022ero}, the discovery of the electron EDM or its more stringent limit is expected to be obtained. 
In either case, our results will give important implications for physics beyond the SM with CP violation.

\section*{Acknowledgements}

We would like to thank Michael Ramsey-Musolf for very helpful discussions.
YN is supported by Natural Science Foundation of China under grant No. 12150610465.
JL is supported by Natural Science Foundation of China under grant No. 12075005 and 12235001.

\appendix

\section{The 2HDM}
\label{app:2HDM}

Here, we summarize mass eigenvalues and eigenstates of the 2HDM. 
The Higgs potential in Eq.~\eqref{eq:V2HDM} leads to $2 \times 2$ mass matrices for CP-even and CP-odd neutral scalars as
\begin{align}
M_{\rm even}^2 &= \begin{pmatrix}
m_{12}^2 \tan \beta + \lambda_1 v_1^2 & - m_{12}^2 + \lambda_{345} v_1 v_2 \\
- m_{12}^2 + \lambda_{345} v_1 v_2 & m_{12}^2 \cot \beta + \lambda_2 v_2^2
\end{pmatrix} \, , \label{eq:M2H} \\[1ex]
M_{\rm odd}^2 &= \begin{pmatrix}
m_{12}^2 \tan \beta - \lambda_5 v_2^2 & - m_{12}^2 + \lambda_5 v_1 v_2 \\
- m_{12}^2 + \lambda_5 v_1 v_2 & m_{12}^2 \cot \beta - \lambda_5 v_1^2
\end{pmatrix} \, ,
\label{eq:M2A}
\end{align}
in basis of $(H_1^0, H_2^0)$ and $(A_1^0, A_2^0)$, respectively. For charged scalars, we find
\begin{align}
M_{\pm}^2 = \begin{pmatrix}
m_{12}^2 \tan \beta - \frac{\lambda_4 + \lambda_5}{2} v_2^2 & - m_{12}^2 + \frac{\lambda_4 + \lambda_5}{2} v_1 v_2 \\
- m_{12}^2 + \frac{\lambda_4 + \lambda_5}{2} v_1 v_2 & m_{12}^2 \cot \beta - \frac{\lambda_4 + \lambda_5}{2} v_1^2
\end{pmatrix} \, ,
\label{eq:M2charged}
\end{align}
in basis of $(H_1^+, H_2^+)$. 
Here, $m_{12}^2$ and $\lambda_5$ are considered as real parameters, and we define $\lambda_{345} \equiv \lambda_3 + \lambda_4 + \lambda_5$. 
The minimization conditions for the Higgs potential impose relations among model parameters, 
\begin{align}
m_1^2 &= m_{12}^2 \tan \beta - \frac{1}{2} \left( \lambda_1 v_1^2 + \lambda_{345} v_2^2 \right) \, , \label{eq:mincondm1} \\[0.5ex]
m_2^2 &= m_{12}^2 \cot \beta - \frac{1}{2} \left( \lambda_2 v_2^2 + \lambda_{345} v_1^2 \right) \, . \label{eq:mincondm2}
\end{align}
The above mass matrices are diagonalized by orthogonal matrices, and each mass eigenvalue can be obtained as
\begin{align}
m_h^2 &= \frac{v_{\rm SM}^2}{2} \left[ \left( \lambda_1 c_{\beta}^2 + \lambda_2 s_{\beta}^2 + \lambda_{\beta} \right) - \sqrt{(\lambda_1 c_{\beta}^2 - \lambda_2 s_{\beta}^2 - \lambda_{\beta} c_{2 \beta})^2 + (\lambda_{345} - \lambda_{\beta})^2 s_{2 \beta}^2} \right] \, , \label{eq:mh2} \\[0.5ex]
m_H^2 &= \frac{v_{\rm SM}^2}{2} \left[ \left( \lambda_1 c_{\beta}^2 + \lambda_2 s_{\beta}^2 + \lambda_{\beta} \right) + \sqrt{(\lambda_1 c_{\beta}^2 - \lambda_2 s_{\beta}^2 - \lambda_{\beta} c_{2 \beta})^2 + (\lambda_{345} - \lambda_{\beta})^2 s_{2 \beta}^2} \right] \, , \label{eq:mH2} \\[0.5ex]
m_A^2 &= \left( \lambda_{\beta} - \lambda_5 \right) v_{\rm SM}^2 \, , \label{eq:mA2} \\[0.5ex]
m_{H^{\pm}}^2 &= \left( \lambda_{\beta} - \frac{1}{2} \lambda_4 - \frac{1}{2} \lambda_5 \right) v_{\rm SM}^2 = m_A^2 - \frac{1}{2} \left( \lambda_4 - \lambda_5 \right) v_{\rm SM}^2 \, , \label{eq:mHpm2}
\end{align}
where $m_h$, $m_H$ and $m_A$ are the masses of the SM-like Higgs, heavy CP-even and CP-odd scalars, respectively, and $m_{H^{\pm}}$ is the physical charged scalar mass. 
Here, we have defined
\begin{align}
\lambda_{\beta} \equiv \frac{m_{12}^2}{v_1 v_2} = \frac{1}{s_{\beta} c_{\beta}} \frac{m_{12}^2}{v_{\rm SM}^2} \, . \label{eq:lambdabeta}
\end{align}
Note that each of $M_{\pm}^2$ and $M_{\rm odd}^2$ has one zero eigenvalue, which corresponds to NG modes, $G^{\pm}$ and $G^0$. 
The mass matrix for CP-even neutral scalars $M_{\rm even}^2$ is diagonalized by one mixing angle $\alpha$,
\begin{align}
\sin 2 \alpha = \frac{(\lambda_{345} - \lambda_{\beta}) s_{2 \beta}}{\sqrt{(\lambda_1 c_{\beta}^2 - \lambda_2 s_{\beta}^2 - \lambda_{\beta} c_{2 \beta})^2 + (\lambda_{345} - \lambda_{\beta})^2 s_{2 \beta}^2}} \, , \label{eq:sin2alpha}
\end{align}
while $M_{\pm}^2$ and $M_{\rm odd}^2$ are diagonalized by using $\tan \beta$ defined in Eq.~\eqref{eq:tanbeta}. 
The mass eigenstates are obtained as in Eqs.~\eqref{eq:h0H0} and \eqref{eq:A0G0}, and for charged scalars,
\begin{align}
H_1^{\pm} &= G^{\pm} c_{\beta} - H^{\pm} s_{\beta} \, , \qquad H_2^{\pm} = G^{\pm} s_{\beta} + H^{\pm} c_{\beta} \, , \label{eq:HpmGpm}
\end{align}
where $H^{\pm}$ is a physical charged scalar.

\section{Potential analysis for the fermion dark sector model}
\label{app:VanalysisFermion}

The scalar potential of the fermion dark sector model allowed by the $Z_2$ charge assignments in Table~\ref{tab:Z2charges} is given by
\begin{align}
V_{\rm scl} &= V_H + V_S + V_{H S} \, , \label{eq:appVsclFrm} \\[0.6ex]
V_S &= m_S^2 |S|^2 + \frac{\lambda_S}{2} |S|^4 + \Bigl[ \mu_2^2 S^2 + \mu_3 S |S|^2 + \mu'_3 S^3 + \lambda'_S S^2 |S|^2 + \lambda''_S S^4 + {\rm h.c.} \Bigr] \, , \label{eq:appVSFrm} \\[0.6ex]
V_{H S} &= \Bigl[ \mu_{1 S} H_1^{\dagger} H_1 S + \mu_{2 S} H_2^{\dagger} H_2 S + \mu_{1 2 S} H_1^{\dagger} H_2 S + \mu_{2 1 S} H_2^{\dagger} H_1 S + {\rm h.c.} \Bigr]  \nonumber \\[0.3ex]
&\hspace{1.2em} + \lambda_{1 S} H_1^{\dagger} H_1 |S|^2 + \lambda_{2 S} H_2^{\dagger} H_2 |S|^2 + \Bigl[ \lambda'_{1 S} H_1^{\dagger} H_1 S^2 + \lambda'_{2 S} H_2^{\dagger} H_2 S^2 + {\rm h.c.} \Bigr] \, . \label{eq:appVHSFrm}
\end{align}
We assume that all the parameters in $V_{\rm scl}$ are real, and $S$ does not acquire a nonzero VEV. 
In the potential $V_S$, we can introduce a tadpole term $T_S S + \text{h.c.}$ since $S$ is a singlet under the considered symmetries. 
Nevertheless, this term is removable through a shift in $S$ by appropriately redefining all the relevant couplings. 
The potential leads to the following minimization condition:
\begin{align}
\mu_{1 S} v_1^2 + \mu_{2 S} v_2^2 + \left( \mu_{1 2 S} + \mu_{2 1 S} \right) v_1 v_2 = 0 \, , \label{eq:mincondmu}
\end{align}
with others shown in Eqs.~\eqref{eq:mincondm1} and \eqref{eq:mincondm2}. 
After applying all the minimization conditions, we find mass matrices for CP-even and CP-odd neutral scalars as
\begin{align}
M_{\rm even}^2 &= \begin{pmatrix}
\lambda_{\beta} v_2^2 + \lambda_1 v_1^2 & \left( - \lambda_{\beta} + \lambda_{345} \right) v_1 v_2 & \hat{m}_{1 S}^2 \\
\left( - \lambda_{\beta} + \lambda_{345} \right) v_1 v_2 & \lambda_{\beta} v_1^2 + \lambda_2 v_2^2 & \hat{m}_{2 S}^2 \\
\hat{m}_{1 S}^2 & \hat{m}_{2 S}^2 & m_{S, +}^2
\end{pmatrix} \, , \\[1ex]
M_{\rm odd}^2 &= \begin{pmatrix}
\left( \lambda_{\beta} - \lambda_5 \right) v_2^2 & \left( - \lambda_{\beta} + \lambda_5 \right) v_1 v_2 & \tilde{m}_{1 S}^2 \\
\left( - \lambda_{\beta} + \lambda_5 \right) v_1 v_2 & \left( \lambda_{\beta} - \lambda_5 \right) v_1^2 & \tilde{m}_{2 S}^2 \\
\tilde{m}_{1 S}^2 & \tilde{m}_{2 S}^2 & m_{S, -}^2
\end{pmatrix} \, ,
\end{align}
where we define
\begin{align}
\hat{m}_{1 S}^2 &\equiv - \frac{\sqrt{2} v_2^2}{v_1} \mu_{2 S} - \frac{\mu_{1 2 S} + \mu_{2 1 S}}{\sqrt{2}} v_2 \, , \quad \hat{m}_{2 S}^2 \equiv \sqrt{2} \mu_{2 S} v_2 + \frac{\mu_{1 2 S} + \mu_{2 1 S}}{\sqrt{2}} v_1 \, , \\[0.5ex]
\tilde{m}_{1 S}^2 &\equiv \frac{\mu_{1 2 S} - \mu_{2 1 S}}{\sqrt{2}} v_2 \, , \quad \tilde{m}_{2 S}^2 \equiv - \frac{\mu_{1 2 S} - \mu_{2 1 S}}{\sqrt{2}} v_1 \, , \\[0.5ex]
m_{S, \pm}^2 &\equiv m_S^2 + \frac{\lambda_{1 S}}{2} v_1^2 + \frac{\lambda_{2 S}}{2} v_2^2 \pm \left( \lambda'_{1 S} v_1^2 + \lambda'_{2 S} v_2^2 + 2 \mu_2^2 \right) \, ,
\end{align}
and $\lambda_{\beta}$ is defined in Eq.~\eqref{eq:lambdabeta}. 
Note that terms in the first line of Eq.~\eqref{eq:appVHSFrm} are essential for mixings between neutral scalars in the 2HDM and $S$. 
In particular, if we do not introduce soft $Z_2$ breaking terms $\mu_{1 2 S}$ and $\mu_{2 1 S}$, $\tilde{m}_{1 S}^2 = 0$ and $\tilde{m}_{2 S}^2 = 0$, which results in no mixing between CP-odd components of the 2HDM and $S$. 
Since $S$ is the neutral scalar, the charged scalar mass is the same as that of the 2HDM. See appendix~\ref{app:2HDM}. 

Upon careful examination, it's apparent that the determinant of $M_{\rm odd}^2$ is zero. 
Consequently, one of the eigenvalues derived from the matrix corresponds to the neutral NG boson field. 
By defining one mixing matrix for $M_{\rm odd}^2$ as
\begin{align}
R_{O, 1} = \begin{pmatrix}
c_{\beta} & - s_{\beta} & 0 \\
s_{\beta} & c_{\beta} & 0 \\
0 & 0 & 1
\end{pmatrix} \, ,
\end{align}
we obtain
\begin{align}
R_{O, 1}^T M_{\rm odd}^2 R_{O, 1} = \begin{pmatrix}
0 & 0 & 0 \\
0 & \left( \lambda_{\beta} - \lambda_5 \right) v_{\rm SM}^2 & - \tilde{m}_{1 S}^2 s_{\beta} + \tilde{m}_{2 S}^2 c_{\beta} \\
0 & - \tilde{m}_{1 S}^2 s_{\beta} + \tilde{m}_{2 S}^2 c_{\beta} & m_{S, -}^2
\end{pmatrix} \, .
\end{align}
The remaining part is diagonalized by another mixing matrix,
\begin{align}
R_{O, 2} = \begin{pmatrix}
1 & 0 & 0 \\
0 & c_{\theta_O} & - s_{\theta_O} \\
0 & s_{\theta_O} & c_{\theta_O}
\end{pmatrix} \, ,
\end{align}
with the mixing angle,
\begin{align}
\sin 2 \theta_O = \frac{2}{\Delta m_{\eta}^2} \left( \tilde{m}_{1 S}^2 s_{\beta} - \tilde{m}_{2 S}^2 c_{\beta} \right) \, .
\end{align}
Here, $\Delta m_{\eta}^2$ is the mass-squared difference of two physical CP-odd scalars,
\begin{align}
\Delta m_{\eta}^2 &= m_{\eta_3}^2 - m_{\eta_2}^2 \\[0.3ex]
&= \sqrt{\left\{ \left( \lambda_{\beta} - \lambda_5 \right) v_{\rm SM}^2 - m_{S, -}^2 \right\}^2 + 4 \left( - \tilde{m}_{1 S}^2 s_{\beta} + \tilde{m}_{2 S}^2 c_{\beta} \right)^2} \, .
\end{align}
Then, each mass eigenstate can be obtained as
\begin{align}
&{\rm diag} \left( 0, m_{\eta_2}^2, m_{\eta_3}^2 \right) = R_{O, 2}^T R_{O, 1}^T M_{\rm odd}^2 R_{O, 1} R_{O, 2} = R_O^T M_{\rm odd}^2 R_O \, , \\[0.7ex]
&m_{\eta_2}^2 = \frac{1}{2} \left[ \left( \lambda_{\beta} - \lambda_5 \right) v_{\rm SM}^2 + m_{S, -}^2 - \Delta m_{\eta}^2 \right] \, , \\[0.5ex]
&m_{\eta_3}^2 = \frac{1}{2} \left[ \left( \lambda_{\beta} - \lambda_5 \right) v_{\rm SM}^2 + m_{S, -}^2 + \Delta m_{\eta}^2 \right] \, ,
\end{align}
where $R_O$ is in Eq.~\eqref{eq:ROmat}. 

In contrast to the CP-odd sector, the CP-even sector requires three mixing angles for diagonalization of $M_{\rm even}^2$, as shown in Eq.~\eqref{eq:REmat}. 
The mixing matrix $R_E$ is obtained as
\begin{align}
R_E &= R_{E, 1} R_{E, 2} R_{E, 3} \, , \\[0.7ex]
R_{E, 1} &= \begin{pmatrix}
c_{\alpha} & - s_{\alpha} & 0 \\
s_{\alpha} & c_{\alpha} & 0 \\
0 & 0 & 1
\end{pmatrix} \, , ~ R_{E, 2} = \begin{pmatrix}
c_{\alpha_{h S}} & 0 & - s_{\alpha_{h S}} \\
0 & 1 & 0 \\
s_{\alpha_{h S}} & 0 & c_{\alpha_{h S}}
\end{pmatrix} \, , ~ R_{E, 3} = \begin{pmatrix}
1 & 0 & 0 \\
0 & c_{\alpha_{H S}} & - s_{\alpha_{H S}} \\
0 & s_{\alpha_{H S}} & c_{\alpha_{H S}}
\end{pmatrix} \, .
\end{align}

\section{Rephasing-invariant CP-violating phases}
\label{sec:CPVphase}

We here summarize rephasing-invariant CP-violating (CPV) phases for the scalar dark sector model and the fermion dark sector model we focused on in the present paper. 
Detailed studies have been done in refs.~\cite{Inoue:2014nva, Inoue:2015pza}.

\subsection{CPV phases in the scalar dark sector model}
\label{app:CPVScalarDS}

In the scalar potential, we have six complex parameters: $m_{12}^2, \lambda_5, B_S, D_S, \lambda_{1 2 S}, \lambda_{2 1 S}$, and one of VEVs of $H_{1, 2}$ is generally complex, e.g., $v_1 = v_1^*$ and $v_2 = |v_2| e^{i \xi}$. 
Then, by considering rephases for all complex scalars as
\begin{align}
H_1 \to e^{i \theta_1} H_1 \, , \qquad  H_2 \to e^{i \theta_2} H_2 \, , \qquad S \to e^{i \theta_S} S \, , \label{eq:rephaseScl}
\end{align}
phases of $\theta_2 - \theta_1$ and $\theta_S$ can be absorbed by redefining all complex parameters and VEVs as
\begin{align}
&m_{12}^2 \to m_{12}^2 e^{i (\theta_1 - \theta_2)} \, , \quad \lambda_5 \to \lambda_5 e^{2 i (\theta_1 - \theta_2)} \, , \quad v_1 v_2^* \to v_1 v_2^* e^{i (\theta_1 - \theta_2)} \, ,  \\[0.5ex]
&B_S \to B_S e^{- 2 i \theta_S} \, , \quad D_S \to D_S e^{- 4 i \theta_S} \, , \\[0.5ex]
&\lambda_{1 2 S} \to \lambda_{1 2 S} e^{i (\theta_1 - \theta_2 - 2 \theta_S)} \, , \quad \lambda_{2 1 S} \to \lambda_{2 1 S} e^{i (\theta_2 - \theta_1 - 2 \theta_S)} \, .
\end{align}
As a result, the scalar potential in Eqs.~\eqref{eq:V2HDM}, \eqref{eq:VSZ4} and \eqref{eq:VHSZ4} is invariant under rephasing of Eq.~\eqref{eq:rephaseScl}. 
Then, according to the discussion in refs.~\cite{Inoue:2014nva,Inoue:2015pza}, we have the following five rephasing-invariant CPV phases:
\begin{align}
\delta_{\lambda_5} &= \arg \left[ \lambda_5^* (v_1 v_2^*)^2 \right] , \quad \delta_{m_{12}^2} = \arg \left[ (m_{12}^2)^* v_1 v_2^* \right] , \quad \delta_S = \arg \left[ B_S^2 D_S^* \right]  , \\[0.5ex]
\delta_{\lambda_{1 2 S}} &= \arg \left[ \lambda_{1 2 S}^* \lambda_{2 1 S} (v_1 v_2^*)^2 \right]  , \quad \delta_{\lambda_{2 1 S}} = \arg \left[ B_S^2 \lambda_{1 2 S}^* \lambda_{2 1 S}^* \right]  .
\end{align}
Note that by selecting a CP-conserving scenario within the 2HDM sector, where $m^2_{12} = 0$, the minimization of the scalar potential implies that $\text{Im} \lambda_5 = 0$. 
In the case where the CPV phase only shows up in the dark sector, one could opt to assign real values to $\lambda_{12S}$ and $\lambda_{21S}$, resulting in no invariant phases within the portal sector. 
Consequently, the sole source of CP violation would be $\delta_S$, with $D_S$ assumed to be real and only $B_S$ is complex. 
This configuration leads to the remaining physical CP phase being expressed as Eq.~\eqref{eq:scalarDMphysicalphase}.

\subsection{CPV phases in the fermion dark sector model}
\label{app:CPVFermionDS}

In the relevant Lagrangian of the fermion dark sector model, we have 15 complex parameters: $y_{\psi}, m_{\psi}, m_{12}^2, \lambda_5, \mu_2^2, \mu_3, \mu'_3, \mu_{1 S}, \mu_{2 S}, \mu_{1 2 S}, \mu_{2 1 S}, \lambda'_S, \lambda''_S, \lambda'_{1 S}, \lambda'_{2 S}$. 
Among them, $m_{\psi}$ can be taken to be real by using the chiral phase rotation of $\psi_{L, R}$, as mentioned in the main text. 
As in the case of the scalar dark sector model, rephases for all complex scalars in Eq.~\eqref{eq:rephaseScl} can be absorbed by redefining complex parameters as
\begin{align}
&y_{\psi} \to y_{\psi} e^{- i \theta_S} \, , ~~ m_{12}^2 \to m_{12}^2 e^{i (\theta_1 - \theta_2)} \, , ~~ \lambda_5 \to \lambda_5 e^{2 i (\theta_1 - \theta_2)} \, , ~~ v_1 v_2^* \to v_1 v_2^* e^{i (\theta_1 - \theta_2)} \, , \\[0.5ex]
&\mu_2^2 \to \mu_2^2 e^{- 2 i \theta_S} \, , ~~ \mu_3 \to \mu_3 e^{- i \theta_S} \, , ~~ \mu'_3 \to \mu'_3 e^{- 3 i \theta_S} \, , \\[0.5ex]
&\mu_{1 S} \to \mu_{1 S} e^{- i \theta_S} \, , ~~ \mu_{2 S} \to \mu_{2 S} e^{- i \theta_S} \, , \\[0.5ex]
&\mu_{1 2 S} \to \mu_{1 2 S} e^{i (\theta_1 - \theta_2 - \theta_S)} \, , ~~ \mu_{2 1 S} \to \mu_{2 1 S} e^{i (\theta_2 - \theta_1 - \theta_S)} \, , \\[0.5ex]
&\lambda'_S \to \lambda'_S e^{- 2 i \theta_S} \, , ~~ \lambda''_S \to \lambda''_S e^{- 4 i \theta_S} \, , ~~ \lambda'_{1 S} \to \lambda'_{1 S} e^{- 2 i \theta_S} \, , ~~  \lambda'_{2 S} \to \lambda'_{2 S} e^{- 2 i \theta_S} \, .
\end{align}
Then, we find 13 combinations for rephasing-invariant CPV phases as
\begin{align}
&\delta_{\lambda_5} = \arg \left[ \lambda_5^* (v_1 v_2^*)^2 \right] \, , ~~ \delta_{m_{12}^2} = \arg \left[ (m_{12}^2)^* v_1 v_2^* \right] \\[0.5ex]
&\delta_{\mu_2^2} = \arg \left[ (\mu_2^2)^* y_{\psi}^2 \right] \, , ~~ \delta_{\mu_3} = \arg \left[ \mu_3^* y_{\psi} \right] \, , ~~ \delta_{\mu'_3} = \arg \left[ (\mu'_3)^* y_{\psi}^3 \right] \, , \\[0.5ex]
&\delta_{\lambda'_S} = \arg \left[ (\lambda'_S)^* y_{\psi}^2 \right] \, , ~~ \delta_{\lambda''_S} = \arg \left[ (\lambda''_S)^* y_{\psi}^4 \right] \, , \\[0.5ex]
&\delta_{\mu_{1 S}} = \arg \left[ \mu_{1 S}^* y_{\psi} \right] \, , ~~ \delta_{\mu_{2 S}} = \arg \left[ \mu_{2 S}^* y_{\psi} \right] \, , ~~ \delta_{\mu_{1 2 S}^* \mu_{2 1 S}^*} = \arg \left[ \mu_{1 2 S}^* \mu_{2 1 S}^* y_{\psi}^2 \right] \, , \\[0.5ex]
&\delta_{\mu_{1 2 S}^* \mu_{2 1 S}} = \arg \left[ \mu_{1 2 S}^* \mu_{2 1 S} (v_1 v_2^*)^2 \right] \, , ~~ \delta_{\lambda'_{1 S}} = \arg \left[ (\lambda'_{1 S})^* y_{\psi}^2 \right] \, , ~~ \delta_{\lambda'_{2 S}} = \arg \left[ (\lambda'_{2 S})^* y_{\psi}^2 \right] \, .
\end{align}
Again, if $m_{12}^2 = 0$, the minimization condition implies ${\rm Im} \lambda_5 = 0$ which means $\delta_{\lambda_5} = 0$. 
However, we still have lots of CPV phases in addition to $\theta_{\rm phys} = \delta_{\mu_{1 S}}$.

\section{Details of calculations}
\label{app:calc}

In this appendix, we show the details of the calculations of the BZ diagrams.

\subsection{General dark sector}
\label{app:CalcGen}

Using couplings in Eq.~\eqref{eq:fermioncouplings}, the left diagrams in Fig.~\ref{fig:outerf} have contributions as
\begin{align}
&\int \! \frac{d^4 k}{(2 \pi)^4} i \Gamma_{f, W}^{\mu \nu} \bar{u}(p_2) \left\{ i \left( g_{e G}^V \gamma^{\alpha} + g_{e G}^A \gamma^{\alpha} \gamma^5 \right) \right\} \frac{i (\slg{p}_2 - \slg{k} + m_e)}{(p_2 - k)^2 - m_e^2} \left\{ i \left( g_{e \Phi}^S + i g_{e \Phi}^P \gamma^5 \right) \right\} u(p_1) \nonumber \\[0.5ex]
&\hspace{5.0em}\times \frac{- i g_{\nu \alpha}}{k^2 - m_G^2} \frac{i}{(- k - q)^2 - m_{\eta_a}^2} \left( i F_{\rm DS}^{\phi_j \eta_a} ( (- k - q)^2, M^2) \right) \frac{i}{(- k - q)^2 - m_{\phi_j}^2} \label{eq:outerloopcalc1} \, ,
\end{align}
where effective vertices of photon-gauge boson ($\gamma/Z$)-scalar $\Gamma_{f, W}^{\mu \nu}$ have information about the inner loop with fermion and $W$ boson, respectively, $G$ represents the photon or $Z$ boson, and $\Phi$ is $\phi_j$ or $\eta_a$. 
Note that we should sum up all contributions for the final result. 
For diagrams in Fig.~\ref{fig:outerf}, $\Gamma_{f, W}^{\mu \nu}$ can be parameterized by
\begin{align}
i \Gamma_{f, W}^{\mu \nu} &= i \int_0^1 \! d x \frac{1}{x (1 - x)} \int_0^{1-x} \! dy \frac{C_E^{f, W} \left( k^{\mu} q^{\nu} - k \cdot q \, g^{\mu \nu} \right) + C_O^{f, W} \epsilon^{\mu \nu \rho \sigma} q_{\rho} k_{\sigma}}{k^2 - \Delta_{f, W}} \, . \label{eq:innerL} 
\end{align}
Here, $C_E^{f, W}$ and $C_O^{f, W}$ are obtained from the inner loop calculation as~\cite{Abe:2013qla,Nakai:2016atk}
\begin{align}
C_E^f &= + \frac{N_C^f}{2 \pi^2} g_{f \gamma}^V g_{f G}^V g_{f \Phi}^S m_f (1 - 4 x y) \, , \label{eq:CEf} \\[0.3ex]
C_O^f &= - \frac{N_C^f}{2 \pi^2} g_{f \gamma}^V g_{f G}^V g_{f \Phi}^P m_f \, , \label{eq:COf} \\[0.3ex]
C_E^W &= + \frac{\rm e}{4 \pi^2} g_{W W \Phi} g_{W W G} \left\{ \left( 4 - \frac{m_G^2}{m_W^2} \right) - \left[ 6 - \frac{m_G^2}{m_W^2} + \left( 1 - \frac{m_G^2}{2 m_W^2} \right) \frac{m_{\Phi}^2}{m_W^2} \right] x y \right\} \, , \label{eq:CEW} \\[0.3ex]
C_O^W &= 0 \, , \label{eq:COW}
\end{align}
and $\Delta^{f, W}$ is defined in Eq.~\eqref{eq:defDeltafW}. Similarly, the right diagrams in Fig.~\ref{fig:outerf} are
\begin{align}
&\int \! \frac{d^4 k}{(2 \pi)^4} i \Gamma_{f, W}^{\mu \nu} \bar{u}(p_2) \left\{ i \left( g_{e \Phi}^S + i g_{e \Phi}^P \gamma^5 \right) \right\} \frac{i (\slg{p}_1 + \slg{k} + m_e)}{(p_1 + k)^2 - m_e^2} \left\{ i \left( g_{e G}^V \gamma^{\alpha} + g_{e G}^A \gamma^{\alpha} \gamma^5 \right) \right\} u(p_1) \nonumber \\[0.5ex]
&\hspace{5.0em}\times \frac{- i g_{\nu \alpha}}{k^2 - m_G^2} \frac{i}{(- k - q)^2 - m_{\eta_a}^2} \left( i F_{\rm DS}^{\phi_j \eta_a} ( (- k - q)^2, M^2) \right) \frac{i}{(- k - q)^2 - m_{\phi_j}^2} \label{eq:outerloopcalc2} \, .
\end{align}
Since $\Gamma_{f, W}^{\mu \nu}$ has one loop momentum $k$ in its numerator (see Eq.~\eqref{eq:innerL}), the contributions to the electron EDM can be parameterized as
\begin{align}
\text{Eq.~\eqref{eq:outerloopcalc1}: } &\int_x \int_y \int_k \frac{A_{\alpha \beta}^{L, \Phi (f, W)} k^{\alpha} k^{\beta} + B_{\alpha}^{L, \Phi (f, W)} k^{\alpha}}{(k^2 - \Delta_{f, W}) \cdots (k^2 - 2 k \cdot p_2)} F_{\rm DS}^{\phi_j \eta_a} ((k + q)^2, M^2) \sigma^{\mu \nu} \gamma^5 q_{\nu} \, , \\[0.5ex]
\text{Eq.~\eqref{eq:outerloopcalc2}: } &\int_x \int_y \int_k \frac{A_{\alpha \beta}^{R, \Phi (f, W)} k^{\alpha} k^{\beta} + B_{\alpha}^{R, \Phi (f, W)} k^{\alpha}}{(k^2 - \Delta_{f, W}) \cdots (k^2 + 2 k \cdot p_1)} F_{\rm DS}^{\phi_j \eta_a} ((k + q)^2, M^2) \sigma^{\mu \nu} \gamma^5 q_{\nu} \, ,
\end{align}
where we omit some propagators for simplicity, and $\epsilon^{\mu \nu \alpha \beta} \gamma_{\alpha} \gamma_{\beta} = - i [\gamma^{\mu}, \gamma^{\nu}] \gamma^5 = - 2 \sigma^{\mu \nu} \gamma^5$ is used. 
$A_{\alpha \beta}^{L, \Phi (f, W)}$, $A_{\alpha \beta}^{R, \Phi (f, W)}$, $B_{\alpha}^{L, \Phi (f, W)}$ and $B_{\alpha}^{R, \Phi (f, W)}$ are independent of the loop momentum $k$, but depend on $p_{1, 2}$ and $q$ as well as $x$ and $y$ through $C_{E, O}^{f, W}$. 
These expressions are generally obtained as
\begin{align}
A_{\alpha \beta}^{L, \Phi (f, W)} &= - \frac{2}{\left( 4 m_e^2 - q^2 \right) q^2} \left[ g_{e G}^V \left( C_E^{f, W} g_{e \Phi}^P + C_O^{f, W} g_{e \Phi}^S \right) + i g_{e G}^A \left( C_E^{f, W} g_{e \Phi}^S - C_O^{f, W} g_{e \Phi}^P \right) \right] \nonumber \\[0.3ex]
&\hspace{9.0em} \times \Bigl\{ 2 m_e^2 (p_{1 \alpha} p_{1 \beta} + p_{2 \alpha} p_{2 \beta}) - (2 m_e^2 - q^2) (p_{1 \alpha} p_{2 \beta} + p_{1 \beta} p_{2 \alpha}) \Bigr\} \nonumber \\[0.3ex]
&\hspace{1.2em} + C_O^{f, W} \left( g_{e G}^V g_{e \Phi}^S - i g_{e G}^A g_{e \Phi}^P \right) g_{\alpha \beta} \, , \label{eq:ALfW} \\[0.8ex]
A_{\alpha \beta}^{R, \Phi (f, W)} &= - \frac{2}{\left( 4 m_e^2 - q^2 \right) q^2} \left[ g_{e G}^V \left( C_E^{f, W} g_{e \Phi}^P + C_O^{f, W} g_{e \Phi}^S \right) - i g_{e G}^A \left( C_E^{f, W} g_{e \Phi}^S - C_O^{f, W} g_{e \Phi}^P \right) \right] \nonumber \\[0.3ex]
&\hspace{9.0em} \times \Bigl\{ 2 m_e^2 (p_{1 \alpha} p_{1 \beta} + p_{2 \alpha} p_{2 \beta}) - (2 m_e^2 - q^2) (p_{1 \alpha} p_{2 \beta} + p_{1 \beta} p_{2 \alpha}) \Bigr\} \nonumber \\[0.3ex]
&\hspace{1.2em} + C_O^{f, W} \left( g_{e G}^V g_{e \Phi}^S + i g_{e G}^A g_{e \Phi}^P \right) g_{\alpha \beta} \, , \label{eq:ARfW} \\[0.8ex]
B_{\alpha}^{L, \Phi (f, W)} &= - \frac{4 C_E^{f, W} g_{e G}^V g_{e \Phi}^P m_e^2}{4 m_e^2 - q^2} p_{1 \alpha} + \left[ \frac{2 C_E^{f, W} (2 m_e^2 - q^2) g_{e G}^V g_{e \Phi}^P}{4 m_e^2 - q^2} + 2 i C_E^{f, W} g_{e G}^A g_{e \Phi}^S \right] p_{2 \alpha} \, , \label{eq:BLfW} \\[0.5ex]
B_{\alpha}^{R, \Phi (f, W)} &= - \left[ \frac{2 C_E^{f, W} (2 m_e^2 - q^2) g_{e G}^V g_{e \Phi}^P}{4 m_e^2 - q^2} - 2 i C_E^{f, W} g_{e G}^A g_{e \Phi}^S \right] p_{1 \alpha} + \frac{4 C_E^{f, W} g_{e G}^V g_{e \Phi}^P m_e^2}{4 m_e^2 - q^2} p_{2 \alpha} \, . \label{eq:BRfW}
\end{align}
Note that $A_{\alpha \beta}^{L, \Phi (f, W)}$ and $A_{\alpha \beta}^{R, \Phi (f, W)}$ are symmetric under $\alpha \leftrightarrow \beta$.

\subsection{Heavy dark sector}
\label{app:CalcHDS}

If we expand $F_{\rm DS}^{\phi_j \eta_a}$ by $k^2 / M^2$, as in Eq.~\eqref{eq:HDSexpand}, we can easily perform the loop momentum integration. 
The HDS approximation result is then given by
\begin{align}
d_e = \, & \sum_{j = 1}^{N_E} \sum_{a = 2}^{N_O} \left[ \sum_f \left( d_e^f \right)_{\phi_j \eta_a} + \left( d_e^W \right)_{\phi_j \eta_a} \right] \, , \\[1.0ex]
\left( d_e^{\mathcal{P}} \right)_{\phi_j \eta_a} \simeq \, &\sum_{\Phi = \phi_j, \eta_a} \int_x \int_y \int_{\boldsymbol{z}} \Biggl\{ \Biggr. a_0^{\phi_j \eta_a} (M^2) \left[ \frac{A_{\alpha \beta}^{L, \Phi (\mathcal{P})} + A_{\alpha \beta}^{R, \Phi (\mathcal{P})}}{32 \pi^2 \Delta_5 (\mathcal{P})^2} g^{\alpha \beta} - \frac{f_5^{L, \Phi (\mathcal{P})} + f_5^{R, \Phi (\mathcal{P})}}{8 \pi^2 \Delta_5 (\mathcal{P})^3} \right] \nonumber \\[0.5ex]
&\hspace{6.0em} + \frac{a_1^{\phi_j \eta_a} (M^2)}{M^2} \Biggl[ \Biggr. - \frac{3 (A_{\alpha \beta}^{L, \Phi (\mathcal{P})} + A_{\alpha \beta}^{R, \Phi (\mathcal{P})})}{32 \pi^2 \Delta_5 (\mathcal{P})} g^{\alpha \beta} \nonumber \\[0.3ex]
&\hspace{7.5em} + \frac{\bar{B}^{L, \Phi (\mathcal{P})} \cdot \lambda^L + \bar{B}^{R, \Phi (\mathcal{P})} \cdot \lambda^R}{32 \pi^2 \Delta_5 (\mathcal{P})^2} + \frac{f_5^{L, \Phi (\mathcal{P})} + f_5^{R, \Phi (\mathcal{P})}}{8 \pi^2 \Delta_5 (\mathcal{P})^2} \nonumber \\[0.3ex]
&\hspace{7.5em} + \left( \frac{A_{\alpha \beta}^{L, \Phi (\mathcal{P})} + A_{\alpha \beta}^{R, \Phi (\mathcal{P})}}{32 \pi^2 \Delta_5 (\mathcal{P})^2} g^{\alpha \beta} - \frac{f_5^{L, \Phi (\mathcal{P})} + f_5^{R, \Phi (\mathcal{P})}}{8 \pi^2 \Delta_5 (\mathcal{P})^3} \right) C_p^2 m_e^2 \Biggl. \Biggr] \Biggl. \Biggr\} \, , \label{eq:deBZgenHDS}
\end{align}
where $\mathcal{P} = f$ or $W$, $\mathcal{O}(k^4 / M^4)$ terms are omitted, and we define
\begin{align}
\int_{\boldsymbol{z}} &\equiv \int_0^1 \! d z_1 d z_2 d z_3 d z_4 d z_5 \delta (1 - z_1 - z_2 - z_3 - z_4 - z_5) \, , \label{eq:defintz} \\[0.5ex]
\Delta_5 (f, W) &\equiv \frac{z_1 m_{f, W}^2}{x (1 - x)} + z_2 m_{\phi_j}^2 + z_3 m_{\eta_a}^2 + z_4 m_G^2 + z_5^2 m_e^2 \, , \label{eq:defDelta5} \\[0.5ex]
\bar{B}_{\alpha}^{L, \Phi (f, W)} &\equiv - 2 A_{\alpha \beta}^{L, \Phi (f, W)} (C_q q^{\beta} - C_p p_2^{\beta}) + B_{\alpha}^{L, \Phi (f, W)} \, , \label{eq:BLbarHDS} \\[0.5ex]
\bar{B}_{\alpha}^{R, \Phi (f, W)} &\equiv - 2 A_{\alpha \beta}^{R, \Phi (f, W)} (C_q q^{\beta} + C_p p_1^{\beta}) + B_{\alpha}^{R, \Phi (f, W)} \, , \label{eq:BRbarHDS} \\[0.5ex]
f_5^{L, \Phi (f, W)} &\equiv A_{\alpha \beta}^{L, \Phi (f, W)} (C_q q^{\alpha} - C_p p_2^{\alpha}) (C_q q^{\beta} - C_p p_2^{\beta}) - B_{\alpha}^{L, \Phi (f, W)} (C_q q^{\alpha} - C_p p_2^{\alpha})  \, , \label{eq:fLHDS} \\[0.5ex]
f_5^{R, \Phi (f, W)} &\equiv A_{\alpha \beta}^{R, \Phi (f, W)} (C_q q^{\alpha} + C_p p_1^{\alpha}) (C_q q^{\beta} + C_p p_1^{\beta}) - B_{\alpha}^{R, \Phi (f, W)} (C_q q^{\alpha} + C_p p_1^{\alpha}) \, , \label{eq:fRHDS} \\[0.5ex]
\lambda^L &\equiv 2 (1 - C_q) q + 2 C_p p_2 \, , \label{eq:lamLHDS} \\[0.5ex]
\lambda^R &\equiv 2 (1 - C_q) q - 2 C_p p_1 \, , \label{eq:lamRHDS} \\[0.5ex]
C_q &= \frac{y}{1 - x} z_1 + z_2 + z_3 \, , \quad  C_p = (1 - z_1 - z_2 - z_3 - z_4) \, . \label{eq:defCqCp}
\end{align}
Here, $C_q$ and $C_p$ correspond to the momentum shift originated from the $k \cdot p_i$ and $k \cdot q$ terms in the denominator,
\begin{align}
k \to \tilde{k} = \begin{cases}
k + C_q q - C_p p_2 & \text{for Eq.~\eqref{eq:outerloopcalc1}} \\
k + C_q q + C_p p_1 & \text{for Eq.~\eqref{eq:outerloopcalc2}}
\end{cases} \, .
\end{align}
Note that due to $q^2 = 0$ which leads to $p_{1, 2} \cdot q = 0$, we have $\lambda^L \cdot \lambda^L = \lambda^R \cdot \lambda^R = 4 C_p^2 m_e^2$. 
For the diagram with $\Gamma^{\mu \nu}$ for the inner loop, each term can be calculated as
\begin{align}
(A_{\alpha \beta}^{L, \Phi (f, W)} + A_{\alpha \beta}^{R, \Phi (f, W)}) g^{\alpha \beta} &= - 4 g^V_e \left( C_E^{f, W} g^P_e - C_O^{f, W} g^S_e \right) \, , \label{eq:Agterm} \\[0.5ex]
\bar{B}^{L, \Phi (f, W)} \cdot \lambda^L + \bar{B}^{R, \Phi (f, W)} \cdot \lambda^R &= - 8 C_p^2 C_E^{f, W} g^V_e g^P_e m_e^2 \, , \label{eq:Blamterm} \\[0.5ex]
f_5^{L, \Phi (f, W)} + f_5^{R, \Phi (f, W)} &= - 2 C_p^2 C_E^{f, W} g^V_e g^P_e m_e^2 \, . \label{eq:fterm}
\end{align}
It should be emphasized that the effects of $k \cdot p_i$ and $k \cdot q$ in the denominator of Eqs.~\eqref{eq:outerloopcalc1} and \eqref{eq:outerloopcalc2} appear in the final result as sub-dominant terms: once we set $C_{q, p} = 0$ which means no momentum shift for $k$, Eq.~\eqref{eq:deBZgenHDS} becomes the leading results in Eq.~\eqref{eq:deBZresHDS}. 
This can be understood by the fact that Eqs.~\eqref{eq:Blamterm} and \eqref{eq:fterm} and the last term in Eq.~\eqref{eq:deBZgenHDS} become zero. 
This feature remains even for the general result without the HDS approximation shown in Eq.~\eqref{eq:deBZresGen}.

\subsection{Decomposed expressions for the electron EDM}
\label{app:decomposeEDM}

In Eq.~\eqref{eq:deparam}, we define two parts of contributions to the electron EDM: $F_{\rm DS}^{\phi_j \eta_a}$ and the renormalization part. 
Here, the explicit form of each contribution is shown. 

Let us start with the case of a general dark sector. 
In our setup, we have effective mixing between CP-even ($\phi_j$) and CP-odd ($\eta_a$) scalars via a dark sector CP violation, denoted as $F_{\rm DS}^{\phi_j \eta_a}$. 
If $F_{\rm DS}^{\phi \eta_a}$ has a UV divergent part, as in the two benchmark models, the divergence needs to be treated appropriately. 
In the on-shell subtraction scheme, we can remove it and obtain a renormalized one:
\begin{align}
\hat{\Sigma}_{\phi_j \eta_a} = F_{\rm DS}^{\phi_j \eta_a} + \delta F_{\rm DS}^{\phi_j \eta_a} \, ,
\end{align}
where $\delta F_{\rm DS}^{\phi_j \eta_a}$ corresponds to the term induced by renormalization. 
Then, the UV divergence is canceled, and hence, we find a finite prediction,
\begin{align}
\hat{\Sigma}_{\phi_j \eta_a} = \widetilde{F}_{\rm DS}^{\phi_j \eta_a} + \widetilde{\delta F}_{\rm DS}^{\phi_j \eta_a} \, .
\end{align}
Here, $\widetilde{F}_{\rm DS}^{\phi_j \eta_a}$ and $\widetilde{\delta F}_{\rm DS}^{\phi_j \eta_a}$ do not have any divergent part. 
Note that they also do not have any constant term associated with the momentum passing through this effective mixing. 
Using this notation, we can now find the explicit formulas for $d_e^{\rm Org., Ren.}$ as
\begin{align}
d_e^{\rm Org.} &= \sum_{j = 1}^{N_E} \sum_{a = 2}^{N_O} \int_x \int_y \int_k \frac{\widetilde{F}_{\rm DS}^{\phi_j \eta_a} ( (k + q)^2, M^2)}{(k^2 - \Delta) (k^2 - m_G^2) (k^2 + 2 k \cdot q - m_{\phi_j}^2) (k^2 + 2 k \cdot q - m_{\eta_a}^2)} \nonumber \\[0.5ex]
&\hspace{1.0em} \times \sum_{\Phi = \phi_j, \eta_a} \left( \frac{A_{\alpha \beta}^{L, \Phi} k^{\alpha} k^{\beta} + B_{\alpha}^{L, \Phi} k^{\alpha}}{k^2 - 2 k \cdot p_2} + \frac{A_{\alpha \beta}^{R, \Phi} k^{\alpha} k^{\beta} + B_{\alpha}^{R, \Phi} k^{\alpha}}{k^2 + 2 k \cdot p_1} \right) \, , \label{eq:deBZresOrg} \\[1.0ex]
d_e^{\rm Ren.} &= \sum_{j = 1}^{N_E} \sum_{a = 2}^{N_O} \int_x \int_y \int_k \frac{\widetilde{\delta F}_{\rm DS}^{\phi_j \eta_a} ( (k + q)^2, M^2)}{(k^2 - \Delta) (k^2 - m_G^2) (k^2 + 2 k \cdot q - m_{\phi_j}^2) (k^2 + 2 k \cdot q - m_{\eta_a}^2)} \nonumber \\[0.5ex]
&\hspace{1.0em} \times \sum_{\Phi = \phi_j, \eta_a} \left( \frac{A_{\alpha \beta}^{L, \Phi} k^{\alpha} k^{\beta} + B_{\alpha}^{L, \Phi} k^{\alpha}}{k^2 - 2 k \cdot p_2} + \frac{A_{\alpha \beta}^{R, \Phi} k^{\alpha} k^{\beta} + B_{\alpha}^{R, \Phi} k^{\alpha}}{k^2 + 2 k \cdot p_1} \right) \, , \label{eq:deBZresRen}
\end{align}
where we omit the superscript $f$ and $W$ from the general result in Eq.~\eqref{eq:deBZresGen} for simplicity. 
Note that for $d_e^{\rm Ren.}$, a further calculation can be done as in the case of the HDS limit in section~\ref{sec:HDS}, because the on-shell subtraction scheme gives additional terms which are proportional to the squared momentum and constant (see Eqs.~\eqref{eq:SigmaDShAScl}, \eqref{eq:SigmaDSHAScl} and \eqref{eq:Sighatja}). 
For the HDS approximation, we expand $\widetilde{F}_{\rm DS}^{\phi_j \eta_a} ( (k + q)^2, M^2)$ and $\widetilde{\delta F}_{\rm DS}^{\phi_j \eta_a} ( (k + q)^2, M^2)$ with respect to $k^2 / M^2$. 

For the scalar dark sector model, we have
\begin{align}
\widetilde{F}_{\rm DS}^{h A} (k^2) &= - \frac{v_{\rm SM}^2}{64 \pi^2} \lambda_{1 2 S}^- s_{2 \theta_s} \Biggl[ \Biggr. \lambda_{1 2 S}^+ c_{\alpha + \beta} c_{2 \theta_s} \biggl( \biggr. {\rm DiscB}(k^2, m_{\varphi_1}, m_{\varphi_1}) + {\rm DiscB}(k^2, m_{\varphi_2}, m_{\varphi_2}) \nonumber \\[0.5ex]
&\hspace{13.5em} - 2 {\rm DiscB}(k^2, m_{\varphi_1}, m_{\varphi_2}) + \frac{m_{\varphi_1}^2 - m_{\varphi_2}^2}{k^2} \ln \frac{m_{\varphi_1}^2}{m_{\varphi_2}^2} \biggl. \biggr) \nonumber \\[0.5ex]
&\hspace{6.0em} + \lambda_{h S} \left( {\rm DiscB}(k^2, m_{\varphi_1}, m_{\varphi_1}) - {\rm DiscB}(k^2, m_{\varphi_2}, m_{\varphi_2}) \right) \Biggl. \Biggr] \, , \label{eq:FtilDShScl} \\[1.0ex]
\widetilde{\delta F}_{\rm DS}^{h A} (k^2) &= - \frac{v_{\rm SM}^2}{64 \pi^2} \lambda_{1 2 S}^- s_{2 \theta_s} \biggl[ \biggr. \frac{k^2 - m_A^2}{m_A^2 - m_h^2} \Bigl( \Bigr. \lambda_{1 2 S}^+ c_{\alpha + \beta} c_{2 \theta_s} \overline{\mathcal{F}}_1 (m_h, m_{\varphi_1}, m_{\varphi_2}) \nonumber \\[0.5ex]
&\hspace{13.5em} + \lambda_{h S} \mathcal{F}_2 (m_h, m_{\varphi_1}, m_{\varphi_2}) \Bigl. \Bigr) \nonumber \\[0.7ex]
&\hspace{7.2em} + \frac{m_h^2 - k^2}{m_A^2 - m_h^2} \Bigl( \lambda_{1 2 S}^+ c_{\alpha + \beta} c_{2 \theta_s} \overline{\mathcal{F}}_1 (m_A, m_{\varphi_1}, m_{\varphi_2}) \nonumber \\[0.5ex]
&\hspace{13.5em} + \lambda_{h S} \mathcal{F}_2 (m_A, m_{\varphi_1}, m_{\varphi_2}) \Bigr) \biggl. \biggr] \, , \label{eq:dFtilDShScl} \\[1.0ex]
\widetilde{F}_{\rm DS}^{H A} (k^2) &= - \frac{v_{\rm SM}^2}{64 \pi^2} \lambda_{1 2 S}^- s_{2 \theta_s} \Biggl[ \Biggr. \lambda_{1 2 S}^+ s_{\alpha + \beta} c_{2 \theta_s} \biggl( \biggr. {\rm DiscB}(k^2, m_{\varphi_1}, m_{\varphi_1}) + {\rm DiscB}(k^2, m_{\varphi_2}, m_{\varphi_2}) \nonumber \\[0.5ex]
&\hspace{13.5em} - 2 {\rm DiscB}(k^2, m_{\varphi_1}, m_{\varphi_2}) + \frac{m_{\varphi_1}^2 - m_{\varphi_2}^2}{k^2} \ln \frac{m_{\varphi_1}^2}{m_{\varphi_2}^2} \biggl. \biggr) \nonumber \\[0.5ex]
&\hspace{6.0em} + \lambda_{H S} \left( {\rm DiscB}(k^2, m_{\varphi_1}, m_{\varphi_1}) - {\rm DiscB}(k^2, m_{\varphi_2}, m_{\varphi_2}) \right) \Biggl. \Biggr] \, , \label{eq:FtilDSHScl} \\[1.0ex]
\widetilde{\delta F}_{\rm DS}^{H A} (k^2) &= - \frac{v_{\rm SM}^2}{64 \pi^2} \lambda_{1 2 S}^- s_{2 \theta_s} \biggl[ \biggr. \frac{k^2 - m_A^2}{m_A^2 - m_H^2} \Bigl( \Bigr. \lambda_{1 2 S}^+ s_{\alpha + \beta} c_{2 \theta_s} \overline{\mathcal{F}}_1 (m_H, m_{\varphi_1}, m_{\varphi_2}) \nonumber \\[0.5ex]
&\hspace{13.5em} + \lambda_{H S} \mathcal{F}_2 (m_H, m_{\varphi_1}, m_{\varphi_2}) \Bigl. \Bigr) \nonumber \\[0.7ex]
&\hspace{7.2em} + \frac{m_H^2 - k^2}{m_A^2 - m_H^2} \Bigl( \lambda_{1 2 S}^+ s_{\alpha + \beta} c_{2 \theta_s} \overline{\mathcal{F}}_1 (m_A, m_{\varphi_1}, m_{\varphi_2}) \nonumber \\[0.5ex]
&\hspace{13.5em} + \lambda_{h S} \mathcal{F}_2 (m_A, m_{\varphi_1}, m_{\varphi_2}) \Bigr) \biggl. \biggr] \, , \label{eq:dFtilDSHScl} \\[1.0ex]
\end{align}
where
\begin{align}
\overline{\mathcal{F}}_1 (m, m_{\varphi_1}, m_{\varphi_2}) \equiv \mathcal{F}_1 (m, m_{\varphi_1}, m_{\varphi_2}) + \frac{m_{\varphi_1}^2 - m_{\varphi_1}^2}{m^2} \ln \frac{m_{\varphi_1}^2}{m_{\varphi_2}^2} \, ,
\end{align}
and $\mathcal{F}_1 (m, m_{\varphi_1}, m_{\varphi_2})$ and $\mathcal{F}_2 (m, m_{\varphi_1}, m_{\varphi_2})$ are given in Eqs.~\eqref{eq:F1func} and \eqref{eq:F2func}, respectively. 
For the HDS approximation, we can obtain the expressions by expanding each of them in the heavy $m_{\varphi_{1, 2}}$ limit as
\begin{align}
\widetilde{F}_{\rm DS}^{h A, {\rm HDS}} (k^2) &= - \frac{v_{\rm SM}^2}{64 \pi^2} \lambda_{1 2 S}^- s_{2 \theta_s} \Biggl[ \Biggr. \lambda_{1 2 S}^+ c_{\alpha + \beta} c_{2 \theta_s} \biggl( \biggr. - 2 + \frac{m_{\varphi_1}^2 + m_{\varphi_2}^2}{m_{\varphi_1}^2 - m_{\varphi_2}^2} \ln \frac{m_{\varphi_1}^2}{m_{\varphi_2}^2}\nonumber \\[0.5ex]
&\hspace{8.0em} + k^2 \left\{ \frac{(m_{\varphi_1}^2 + m_{\varphi_2}^2) (m_{\varphi_1}^4 - 8 m_{\varphi_1}^2 m_{\varphi_2}^2 + m_{\varphi_2}^4)}{6 m_{\varphi_1}^2 m_{\varphi_2}^2 (m_{\varphi_1}^2 - m_{\varphi_2}^2)^2} \right. \nonumber \\[0.5ex]
&\hspace{10.0em} \left. + \frac{2 m_{\varphi_1}^2 m_{\varphi_2}^2}{(m_{\varphi_1}^2 - m_{\varphi_2}^2)^3} \ln \frac{m_{\varphi_1}^2}{m_{\varphi_2}^2} \right\} \biggl. \biggr) - \lambda_{h S} \frac{m_{\varphi_1}^2 - m_{\varphi_2}^2}{6 m_{\varphi_1}^2 m_{\varphi_2}^2} k^2 \Biggl. \Biggr] \, , \label{eq:FtilDShHDSScl} \\[1.0ex]
\widetilde{F}_{\rm DS}^{H A, {\rm HDS}} (k^2) &= - \frac{v_{\rm SM}^2}{64 \pi^2} \lambda_{1 2 S}^- s_{2 \theta_s} \Biggl[ \Biggr. \lambda_{1 2 S}^+ s_{\alpha + \beta} c_{2 \theta_s} \biggl( \biggr. - 2 + \frac{m_{\varphi_1}^2 + m_{\varphi_2}^2}{m_{\varphi_1}^2 - m_{\varphi_2}^2} \ln \frac{m_{\varphi_1}^2}{m_{\varphi_2}^2}\nonumber \\[0.5ex]
&\hspace{8.0em} + k^2 \left\{ \frac{(m_{\varphi_1}^2 + m_{\varphi_2}^2) (m_{\varphi_1}^4 - 8 m_{\varphi_1}^2 m_{\varphi_2}^2 + m_{\varphi_2}^4)}{6 m_{\varphi_1}^2 m_{\varphi_2}^2 (m_{\varphi_1}^2 - m_{\varphi_2}^2)^2} \right. \nonumber \\[0.5ex]
&\hspace{10.0em} \left. + \frac{2 m_{\varphi_1}^2 m_{\varphi_2}^2}{(m_{\varphi_1}^2 - m_{\varphi_2}^2)^3} \ln \frac{m_{\varphi_1}^2}{m_{\varphi_2}^2} \right\} \biggl. \biggr) - \lambda_{H S} \frac{m_{\varphi_1}^2 - m_{\varphi_2}^2}{6 m_{\varphi_1}^2 m_{\varphi_2}^2} k^2 \Biggl. \Biggr] \, . \label{eq:FtilDSHHDSScl}
\end{align}
Here, we omit $\mathcal{O}(k^4 / m_{\varphi_i}^4)$ terms, and $\widetilde{\delta F}_{\rm DS}^{h A, {\rm HDS}} (k^2)$ and $\widetilde{\delta F}_{\rm DS}^{H A, {\rm HDS}} (k^2)$ are the same as $\widetilde{\delta F}_{\rm DS}^{h A} (k^2)$ and $\widetilde{\delta F}_{\rm DS}^{H A} (k^2)$. 

For the fermion dark sector model, we find
\begin{align}
\widetilde{F}_{\rm DS}^{\phi_j \eta_a} (k^2) &= \frac{y_{\psi}^R y_{\psi}^I}{8 \pi^2} m_{\psi}^2 (R_E)_{3 j} (R_O)_{3 a} {\rm DiscB}(k^2, m_{\psi}, m_{\psi}) \, , \label{eq:FtilDSjaFrm} \\[1.0ex]
\widetilde{\delta F}_{\rm DS}^{\phi_j \eta_a} (k^2) &= \frac{y_{\psi}^R y_{\psi}^I}{8 \pi^2} m_{\psi}^2 (R_E)_{3 j} (R_O)_{3 a} \biggl[ \biggr. \frac{m_{\eta_a}^2 - k^2}{m_{\phi_j}^2 - m_{\eta_a}^2} {\rm Re} \bigl\{ {\rm DiscB}(m_{\phi_j}^2, m_{\psi}, m_{\psi}) \bigr\} \nonumber \\[0.5ex]
&\hspace{11.5em} + \frac{k^2 - m_{\phi_j}^2}{m_{\phi_j}^2 - m_{\eta_a}^2} {\rm Re} \bigl\{ {\rm DiscB}(m_{\eta_a}^2, m_{\psi}, m_{\psi}) \bigr\} \biggl. \biggr] \, . \label{eq:dFtilDSjaFrm}
\end{align}
For the HDS approximation, they are 
\begin{align}
\widetilde{F}_{\rm DS}^{\phi_j \eta_a, {\rm HDS}} (k^2) &= \frac{y_{\psi}^R y_{\psi}^I}{8 \pi^2} m_{\psi}^2 (R_E)_{3 j} (R_O)_{3 a} \left( - 2 + \frac{k^2}{6 m_{\psi}^2} \right) \, , \label{eq:FtilDSjaHDSFrm}
\end{align}
and $\widetilde{\delta F}_{\rm DS}^{\phi_j \eta_a, {\rm HDS}} (k^2)$ unchanged from $\widetilde{\delta F}_{\rm DS}^{\phi_j \eta_a} (k^2)$.

\bibliographystyle{jhep}
\bibliography{rsc_bib}

\end{document}